\documentclass[useAMS,usenatbib,fleqn]{mn2e}

\usepackage{times}

\usepackage{amsmath}
\usepackage{amssymb}
\usepackage{upgreek}
\usepackage{graphicx}

\renewcommand{\vec}[1]{\ensuremath{\bmath{#1}}}   

\newcommand\aj{AJ}
\newcommand\aap{A\&A}

\newcommand{\apj}{ApJ}
\newcommand{\apjl}{ApJ}

\newcommand\apss{Ap\&SS}

\newcommand{\mnras}{MNRAS}

\newcommand{\pasj}{PASJ}

\title[Shock Driven Flares]{Maser Flares Driven by Isothermal Shock Waves}
\author[M. D. Gray, S. Etoka, A., B. Pimpanuwat and A.M.S. Richards]
{M. D. Gray$^{1,2}$, S. Etoka$^{1}$, B. Pimpanuwat$^{1}$ and A. M. S. Richards$^{1}$\\
$^{1}$Jodrell Bank Centre for Astrophysics, Department of Physics and Astronomy, University of Manchester,
M13 9PL, UK\\
$^{2}$National Astronomical Research Institute of Thailand, 260 Moo 4, T. Donkaew, A. Maerim, Chiangmai 50180, Thailand.}

\begin{document}

\date{}

\pagerange{\pageref{firstpage}--\pageref{lastpage}} \pubyear{2018}

\maketitle

\label{firstpage}

\begin{abstract}
We use 3D computer modelling to investigate the timescales and radiative output from
maser flares generated by the impact of shock-waves on astronomical unit-scale clouds in interstellar and
star-forming regions, and in circumstellar regions in some circumstances. Physical conditions are derived from simple models
of isothermal hydrodynamic (single-fluid) and C-type (ionic and neutral fluid) shock-waves, and based on the ortho-H$_2$O
22-GHz transition. Maser saturation is comprehensively included, and
we find that the most saturated maser inversions are found predominantly in the shocked
material. We study the effect on the intensity, flux density and duration of flares of the following
parameters: the pre-shock level of saturation, the
observer's viewpoint, and the shock speed. Our models are able to reproduce observed flare rise
times of a few times 10 days, specific intensities of up to 10$^5$ times the saturation intensity and
flux densities of order $100(R/d)^2$\,Jy from a source of radius $R$ astronomical units at a distance of $d$ kiloparsec.
We found that flares from C-type shocks are approximately 5 times more
likely to be seen by a randomly placed observer than flares from hydrodynamically shocked clouds of 
similar dimensions. We computed intrinsic beaming
patterns of the maser emission, finding substantial extension of the pattern parallel to the shock front
in the hydrodynamic models. Beaming solid angles for hydrodynamic models can be as small as $1.3\times 10^{-5}$\,sr, 
but are an order of magnitude larger for C-type models.
\end{abstract}


\section{Introduction}
\label{s:intro}

\defcitealias{2018MNRAS.477.2628G}{Paper~1}
\defcitealias{2019MNRAS.486.4216G}{Paper~2}
\defcitealias{2020MNRAS.493.2472G}{Paper~3}
Astrophysical maser flares have been observed from a number of environments, including massive
star-forming regions and the circumstellar envelopes of highly evolved stars. Shock waves are
a potential mechanism for generating flares in these environments.
A maser flare may be loosely defined as a significant brightening of a maser source on a
timescale much shorter than any related to overall structural evolution of the source region.
This paper is the fourth in a series that investigates several plausible mechanisms
for maser flares, and follows earlier works that considered rotation of quasi-spherical clouds
\citep{2018MNRAS.477.2628G} \citepalias{2018MNRAS.477.2628G}, rotation of oblate and prolate
spheroidal clouds \citep{2019MNRAS.486.4216G} \citepalias{2019MNRAS.486.4216G} and changes
in the levels of pumping and background radiation \citep{2020MNRAS.493.2472G}
\citepalias{2020MNRAS.493.2472G}. The radiative mechanisms have been used to recover
physical parameters of maser flares (negative optical depth in the quiescent state and
change in depth during the flare) from observables (the flare variability index and
duty cycle) in two massive star-forming regions, G107.298+5.63 and S255-NIRS3 (Gray, Etoka \&
Pimpanuwat \citeyear{2020MNRAS.498L..11G}). In the present work, we consider in detail the flare
parameters that may be generated by the passage of an idealized isothermal shock wave
through a maser-supporting cloud. Note that for consistency with Papers~1-3
we refer to the gaseous maser-supporting objects
of approximately astronomical unit scale as clouds in the present work. Such objects correspond
approximately to the observational `compact emission centres', or knots, described in
\citet{2022NatAs...6.1068M}. When we need to refer to the much larger objects in which
star formation occurs, we use the term, `molecular cloud'.

Shock waves generally provide a collision-dominated pumping scheme that is
considered typical for many of the known maser transitions of H$_2$O, including
the commonest line at 22\,GHz \citep{1973A&A....26..297D}, which we use for
quantitative examples in the present work. A collisional 
pump is also considered responsible for Class~I methanol masers \citep{1973ApJ...184..763L,1992MNRAS.259..203C},
and operates when the temperature, $T_{\rmn{r}}$, of the local continuum radiation is $<$120\,K, preventing
radiative excitation to the second torsionally excited state. The pump is particularly
effective when $T_{\rmn{r}}<50$\,K, and is exceeded by the gas kinetic temperature \citep{2005MNRAS.362..995V}.

As in \citetalias{2018MNRAS.477.2628G} to \citetalias{2020MNRAS.493.2472G}, we
define the variability index of flares as $F_{\rmn{pk}}/F_{\rmn{qui}}$,
where $F_{\rmn{pk}}$ is the flux density at line centre when the flare is at maximum and $F_{\rmn{qui}}$ is 
the corresponding flux density under quiescent conditions. In the present work, quiescent
conditions imply those before the cloud is impacted by the shock, since there
are a number of relaxation processes that may be important following shock passage
through the cloud.

\subsection{Observational evidence for shock-driven flares}
\label{ss:obsback}

There are many cases of flares observed towards massive star-forming regions, sometimes 
with several maser lines detected simultaneously towards the same source, for example 
\citet{2018MNRAS.478.1077M} (22-GHz H$_2$O, 6.7 and 12.2-GHz methanol and 4 transitions
of OH towards NGC6334I), and \citet{2021MNRAS.502.5658M} (methanol as above and 3 OH transitions
towards G323.459-0.079).
However, here we restrict the discussion to observations that specifically support the view
that shock waves are responsible for the flare. We concentrate on star-forming sources, since
the models used in the present work are more suited to these environments, and the evidence
for shocked clouds, typically of a few au in size, in evolved-star CSEs is limited \citep{2011A&A...525A..56R}. When strong
shocks are invoked for maser pumping in CSEs, it is more for sub-mm H$_2$O masers in
the inner envelope \citep{2020A&A...638A..19B} than for the 22-GHz masers in the more extended
wind acceleration zone.

Discrete H$_2$O maser features at VLBI resolution are much more likely to be
physical clouds than a transient phenomenon based, for example, on random velocity-coherent
paths through a gaseous medium. Evidence for physical clouds comes from both persistence
over many observational epochs, separated by intervals of years, with measured
proper motions, for example \citet{1992ApJ...393..149G}, and from statistical analysis of
position and velocity correlation functions. For example, \citet{2007ASPC..365..196S} found
a two-point velocity increment function that was consistent with incompressible, high
Reynolds number, turbulence at all scales down to a shock dissipation scale, of order 1\,au,
where a steeper power law indicated the onset of rapid energy loss. {\it MERLIN} observations
of the S128 star-forming region also found evidence for disspation of supersonic turbulence
at small scales \citep{2005Ap&SS.295...19R} in an analysis of the surface density of maser
features on the sky: a fractal dimension of 0.38 was derived for features, and 0 for components
within them (uniform distribution), whilst a fractal index of 2.6 is expected for incompressible
turbulence.

 There are many good candidates for a type of flare in which a shock sweeps through a sequence
or chain of maser clouds. For example,
in IRAS16293$-$2422, there have been three periods of strong activity during an observing
programme lasting from 1997 to 2021 \citep{2021MNRAS.507.3285C}. Maser emission during
these flares is about 20 times stronger than in the intervening quiescent interludes. Maser motions in this
source include velocity gradients that are consistent with the passage of a C-type shock of
speed $\sim$15\,km\,s$^{-1}$ through a small number of discrete features in a chain of
overall length approximately 3.5\,au. Flaring in multiple spectral features
implies that several structures are involved. The velocity range that encompasses the flaring
components is modest (-6 to 15)\,km\,s$^{-1}$. The activity cycles are approximately periodic (8\,yr),
probably following the period of a binary orbit. Monitoring of IRAS05358+3543 by \citet{2020ARep...64..839A}
is modelled as the progress of a shock of speed $\sim$15\,km\,s$^{-1}$ through a sequence of au-scale
clouds, given the mean rise (fall) times of 0.3 (0.35)\,yr for flaring in 13 distinct spectral features.
Other broadly similar sources include NGC2071 (in Orion) \citep{2020ARep...64..586A} and
S255IR-SMA1 \citep{2016MNRAS.460..283B}, where shock speeds up to 25\,km\,s$^{-1}$ are apparent.
In G43.8-01, there is a similar pattern of activity cycles lasting years to decades, with
individual flares (9 exceeding 3000\,Jy) within each active phase lasting from months to years
\citep{2019ARep...63..814C}. Ten strong flares in W75N that occured in two distinct cycles of activity were modelled by
\citet{2015AstL...41..517K} as successive excitation of a series of `condensations' by a
shock, with time delays of up to 7 months, implying that these condensations have a scale
of 1.8\,au if the shock is moving at 15\,km\,s$^{-1}$ \citep{2011A&A...527A..48S}. The masers
appear to be associated with a radio jet (VLA~1), and identification of the flaring features
with this continuum source relies on a sequence of interferometric observations
\citep{2011A&A...527A..48S,2014A&A...565L...8S,2013ApJ...767...86K}.

A bipolar outflow scenario for 22-GHz H$_2$O masers is commonly used, 
and sometimes three
main maser clusters are apparent: one near each end of the outflow, and one central cluster
close to the outflow origin, for example in G59.783+0.065 \citep{2021eavw.workE...3M}. Half
of a sample of 36 star-forming regions had either or both of the bipolar and central clusters
\citep{2019A&A...631A..74M}. This three-cluster grouping is
apparent in the flaring high-mass star-forming region, G23.01-0.41, where a slow
(20\,km\,s$^{-1}$) bipolar jet at the base of the outflow develops into a faster (50\,km\,s$^{-1}$)
shock at larger radii \citep{2010A&A...517A..78S}. A powerful flare, with a specific intensity increase
of 200 times, appeared in the central cluster (`C'), associated with 1.3-cm continuum
emission, at the fourth and last VLBI epoch. The masers of cluster~C have a generally higher 
variability than those further from the continuum source. This cluster is modelled by
\citet{2010A&A...517A..78S} as an arc of maser features approximately 200\,au from a protostellar
object from which a shock expands and drives the masers. This is somewhat different from the
more stable clusters at the ends of the outflow that are driven by shocks that result from
the outflow meeting more quiescent gas. Maser behaviour of the generic outflow type also appears 
to extend to star-forming regions of
significantly lower mass, for example a protostellar source with an estimated mass 0.3\,M$_{\sun}$
associated with IRAS16293-2422 in the nearby ($\sim$120\,pc) $\uprho$~Oph molecular cloud
\citep{2016ARep...60..730C,1999PASJ...51..473I}. The activity cycles appear generally shorter
in this low-mass source (0.9-3.4\,yr), but again the strong-emission parts of cycles are punctuated
by flares of shorter duration with individual spectral features that show radial velocity drift, and
is interpreted as a shock moving at modest (15\,km\,s$^{-1}$) speeds through chains of au-scale
maser clouds.

In G43.8-01, all the flaring features occur along an
arc structure of approximate angular size 200\,milliarcsec (560\,au at 2.8\,kpc), possibly a shock from a disc wind, or a bow shock
\citep{2019ARep...63..814C}. A bow-shock structure is also evident in the most northerly
of five H$_2$O maser clusters analysed in the accretion-burst source NGC6334I \citep{2021ApJ...908..175C}.
The bow-shock cluster, known as CM2-W2 appears to be at the northern end of a bipolar outflow. Proper
motions of the masers in CM2-W2 are pointed mostly North, with an average velocity of 112\,km\,s$^{-1}$.

There are sources that behave similarly to the bipolar outflow type, but where the source structure
is different or more complicated.
In S128, five cycles of variability were observed over 37\,yr with intervals of 4-14\,yr, and
again each active portion of a cycle is split into flares of duration typically a few
months \citep{2018ARep...62..609A}. The S128 source is spectrally interesting in that two peaks
separated by 6\,km\,s$^{-1}$ are known to correspond to two sites on an ionization front, separated
by 13''. However, radial velocity drifts that occur in some cycles suggest that the flares
are shock-driven when considered locally to one of the sources, although a different
mechanism is required to link the activity of the widely-separated sites on the
ionization front. Masers associated with IRAS21078+5211
show the pattern of repeated activity cycles, here with a quasi-period of 3.3\,yr, combined
with shorter-timescale flares of individual spectral features \citep{2018ARep...62..200K}. The spatial structure of the maser
clouds in this source at VLBI resolution is of 6 groups that differ considerably in shape
\citep{2013ApJ...769...15X}. A study of radial velocity drift through one sequence of flares
can be interpreted at a shock passing through a chain of au-scale clouds at a speed of $\sim$15\,km\,s$^{-1}$
\citep{2018ARep...62..200K}. In G188.946+0.886, a generally similar pattern of activity cycles
and shorter-term flares is observed \citep{2016AstL...42..652A}. We also note that the 404-d
period associated with 6.7-GHz class~2 methanol masers towards this source by \citet{2011AJ....141..152V}
was not evident in the H$_2$O maser data.

The very strongest flaring sources are sometimes also interpreted via the shock paradigm, but
line-of-sight overlap of clouds is also often cited. We note that the three
sources with the most exceptionally powerful bursts are Orion~KL, W49 and IRAS18316-0602. An
overlap or `local' flare has also been mooted for the strongest flare in IRAS21078+5211 with a
rise time of $<$ 1 month.

The first of these exceptional sources is the very strong
22-GHz H$_2$O maser flare that occured in 2017-2018 towards the W49 region
\citep{2019ARep...63..652V}. This flare has been attributed to the expansion of a shock
originating from a pulsationally unstable protostar into interstellar material of lower
density, including acceleration of maser features at larger radii \citep{1992ApJ...393..149G}.
Some aspects of the flare are consistent with a shock model: the flare appears to come from a
single unresolved VLBI feature that has a linear sixe $\lesssim$10\,au.
However, the W49 flare has at least one feature that is difficult to explain in terms of a shock-generated
flare: it has a very symmetrical light curve that is of exponential form, resulting in a
cusp-like peak, see \citet{2019AstL...45..321V}. The apparent unsaturated state of the
 flaring spectral feature \citep{2019ARep...63..652V}
is not a good discriminator between different flare mechanisms, but is generally indicative
of a high variability index.

The second example is the giant 130-kJy flare in IRAS18316-0602 \citep{2019ARep...63...49V},
observed between 2017 September and 2018 February. The rise and fall profiles of flares in this source
are similar, but significantly asymmetrical. Both rise and fall may be well-fitted by exponentials,
and this probably indicates unsaturated behaviour. In time, the giant flare consists of a broad
component of flux density $\sim$20\,kJy, lasting for the full duration, with two sharp exponential-sided peaks, each lasting
5-10\,d superimposed on this. Both the bright exponential peaks came from a very similar range
of spectral velocity; this and VLBI observations suggest a single cloud dominates the flare.
The mechanism suggested by \citet{2019ARep...63...49V} is an envelope ejected by a multiple protostellar
system impinging on an accretion disc, leading to a powerful system of shocks. However, we note
that the giant flare in this source has also been modelled as line-of-sight overlap of two
maser clouds \citep{2020ARep...64...15A}, owing to its very short rise and decay times. A
rise from 20 to 76\,kJy in 0.5\,d was recorded for the flaring maser feature that is associated
with the radio continuum source VLA1 \citep{2019ApJ...884..140B}. IRAS18316-0602 also contains
44-GHz class~1 methanol masers, but these are $>$1\,arcsec away from the flaring H$_2$O feature.

The final source of exceptional power is Orion~KL, in which three activity cycles have been detected
(1979-1985, 1998-1999 and 2011-2012). The last of these events was studied in detail by
\citet{2014PASJ...66..106H}. The relationship between the flux density and spectral width of the
flaring spectral components indicates largely unsaturated amplification, and, though closely spaced
in frequency, the two dominant spectral features appear to be spatially separated at VLBI
resolution \citep{2014PASJ...66..106H} by about 12\,milliarcsec (sky-plane
linear separation of 5.04\,au). Proper motions with respect to
the driving Source~1 that are close to perpendicular to the long-axes of the maser clouds
support a shock origin for the flares. However, \citet{2014PASJ...66..106H} also suggest line-of-sight
overlap and accidental beaming towards Earth as alternative possibilities. Line-of-sight overlap
has been convincingly put forward as the reason for the previous (1998-99) flare from VLBI
observations \citep{2005ApJ...634..459S}.

In summary, modest H$_2$O maser variability is ubiquitous in star-forming sources, but powerful flares,
reaching flux densities of hundreds to thousands of Jy, are rarer and tend to be
associated with structures comparatively close to the protostellar exciting source.
Typical shock speeds appear to be of order 15-30\,km\,s$^{-1}$. These maser flares are rarely periodic, 
but quasi-periodic variability is common, with typical
intervals of a few yr.

\subsection{Standard Shock Models}
\label{ss:stanshmod}

Two classic works on post-shock H$_2$O maser emission are \citet{1989ApJ...346..983E} and
\citet{1996ApJ...456..250K}. In the former, a J-shock (negligible ambipolar diffusion) that
is dissociative drives into gas with a pre-shock density of order $10^7$\,cm$^{-3}$. In the
post-shock gas, reformation of H$_2$ on grain surfaces leads to a region heated to a fairly stable
temperature $\sim$400\,K that is rich in H$_2$O, and forms the maser zone. The latter work
considers a slower C-type shock, and the boundary velocity between the two models lies in
the range 40-50\,km\,s$^{-1}$. We consider the slower shocks of \citet{1996ApJ...456..250K}
the more likely environment. This is partly for the reasons introduced in \citet{1996ApJ...456..250K}:
400\,K is considerably below the optimum temperature for a collisional H$_2$O pump at 22\,GHz, and
many other collisionally-pumped H$_2$O maser transitions have even higher kinetic temperatures
for optimum inversion, points supported by more recent modelling \citep{2016MNRAS.456..374G}.
However, we also note that, with reference to flares, most of the shock speeds discussed in the observational
material above are in the range 15-30\,km\,s$^{-1}$, speeds consistent with most 22-GHz H$_2$O masers
being excited by C-type shocks.

In a C-type shock, ions and neutrals form two intermingled fluids that are only loosely coupled
by collisions, and a key parameter is $L_{\rmn{in}}$, the ion-neutral coupling length, which is in
turn controlled by the abundance of various charged species. Abundances of these species, based
on cosmic ray ionization \citep{1996ApJ...456..250K} are typically of order $10^{-10}$ with respect to H nuclei at
a number density of $10^7$\,cm$^{-3}$, and detailed plots appear in their Figure~1. Momentum is
transferred from the charged fluid to the neutrals over the shorter length $L_{\rmn{in}}/M_{\rmn{A}}$,
where $M_{\rmn{A}}$ is the Alf\'{e}nic Mach number \citep{1996ApJ...456..250K}. For the range of
pre-shock number densities, $n_0 = 10^7-10^{9.5}$\,cm$^{-3}$, and shock speeds ($15-40$\,km\,s$^{-1}$),
covered in \citet{1996ApJ...456..250K}, and parameterizations of
$M_{\rmn{A}}$, the momentum transfer length, $L_{\rmn{mt}}$ varies between about 0.06\,au, for the highest
$n_0$ and smallest value of $b$ (0.1), used by \citet{1996ApJ...456..250K} to $\sim$85\,au, at
$b=3$ and $n_0 = 10^7$\,cm$^{-3}$. These extremes therefore
range from vastly smaller than, to much greater than, a typical maser cloud scale of
order a few au. These calculations are considered in more detail in Section~\ref{sss:continuous}.
In the shock models by \citet{1996ApJ...456..250K}, H$_2$O is not dissociated
by the shock, but post-shock chemical reactions efficiently enhance the water abundance,
providing the post-shock gas achieves a temperature of at least 400\,K. In a plotted
example (their Fig.~2) the H$_2$O abundance is fairly constant for post-shock distances of $4\times 10^{13}$\,cm
to the full extent of the model at $2\times 10^{14}$\,cm.

If the maser medium, even after being shocked, hosts only unsaturated masers, then it is
relatively straightforward to model the maser emission. However, model turbulence mapping
\citep{2017A&AT...30..173S} compares the expected spectra and maps from a turbulent medium in
both the saturated and unsaturated case. The unsaturated case predicts maps dominated by a 
small number of spatially isolated statistical outliers with very high maser optical depth, whilst saturated
masers tend to form with fractal clustering, and a comparatively small intensity dispersion.
The saturated model compares considerably better with observations.

\subsection{Other Modelling of Maser Flares}
\label{ss:othermods}

\citet{1989ApJ...340L..17D} explained the basis of maser variability by considering changes in the
radiation field affecting the background and pumping, or changes in the effective gain length (velocity-coherent
amplification path).
Such changes due to line-of-sight overlap of slabs and filaments were modelled as the
origin of giant H$_2$O maser outbursts in Orion and W49 \citep{1991ApJ...367..333E}. A model
emphasising the role of J-type shocks in the generation of bright H$_2$O masers, that we do not
consider further here, was proposed by \citet{2013ApJ...773...70H}. A magnetohydrodynamic (MHD) shock origin was suggested
also for OH masers during a flare in W75N \citep{2010MNRAS.404.1121S}.

More recent theoretical work on maser flares has diversified considerably. While the current
authors consider 3D radiative transfer models of au-scale maser clouds, including saturation,
there are several other approaches. For example, an entirely different explanation for maser
flares, Dicke superradiance has been suggested \citep{2017SciA....3E1858R,2019MNRAS.484.1590R}.
\citet{2020A&A...634A..41O} reconstructed 3D source structure from a combination of VLBI data 
and time delays between the flaring of individual maser features. Many-model grids have been used
to derive physical conditions or pumping schemes explaining flares
(for example \citealt{2022MNRAS.512.3215S} for CH$_3$OH masers, and \citealt{2023MNRAS.522.4728M}
for a new maser transition of NH$_3$). An analysis of line widths has been used to link gas
motions to turbulence in the source, and to deduce an absence of homogeneous, spherical maser clouds
\citep{2015AstL...41..517K}. Where the models above solve the radiative trasfer problem, they typically use
approximations, for example the large velocity gradient (LVG) version of the escape-probability method,
or 1D models. 

Another group of models seeks to establish the radiation field, as a function of time and
wavelength, that is generated by time-dependent processes in the central source, for example
colliding binary wind shocks \citep{2014MNRAS.444..620P}, stellar pulsation \citep{2013ApJ...769L..20I}, and 
unsteady accretion \citep{2010ApJ...717L.133A}. An example that computes
spectral energy distributions over a wide range of wavelengths from continuum radiation transfer
solutions is \citet{2021A&A...646A.161S}. 

There is a useful summary of maser models, not necessarily applied to flaring sources, in the
H-atom maser paper by \citet{2020MNRAS.491.2536P}. Mention should be made of the accelerated
lambda iteration code {\sc magritte} \citep{2020MNRAS.492.1812D,2022JOSS....7.3905D} that is fully 3D, and
may soon have the capacity to study saturating masers with full molecular complexity
included.

\section{Radiative Transfer Model}
\label{s:model}

In this section, we briefly describe the model that is used to solve the radiative transfer (RT)
problem for maser radiation, in 3D,  at essentially arbitrary degrees of saturation. We keep this
separate, as far as possible, from the shock-wave physics (see Section~\ref{s:modclouds}) 
that is used to generate input physical conditions for the RT model. We summarize the key points
of the RT model in Section~\ref{ss:keyres}, before entering a more detailed discussion of 
how the model has been upgraded from the version used in \citetalias{2020MNRAS.493.2472G} in
Section~\ref{ss:vbv}. Modifications necessary to model a partially compressed cloud with
various fractions of its volume swept by the shock are deferred until Section~\ref{s:modclouds}.
As in \citetalias{2020MNRAS.493.2472G}, we approximate a time-dependent
model as a series of snapshots that are, in themselves, time-independent. The necessary
assumption that other processes can relax on timescales significantly shorter than the
snapshot interval is perhaps more easily satisfied in the present work because the pumping schemes
are collision-dominated, without the added complexity of external sources of pumping radiation. 
We estimate a suitable minimum snapshot interval in Section~\ref{ss:tscale}.

\subsection{Key Points from Papers~1-3}
\label{ss:keyres}

The motivation for the theory and code in \citetalias{2018MNRAS.477.2628G} was to enable the modelling
of `real-Universe' maser clouds, lacking a specified geometry. Possible problems to be considered
included natural beaming angles and the influence of cloud shape on maser flares, for example in \citetalias{2019MNRAS.486.4216G}.
Model maser clouds in Papers~1-3 and the present work are constructed by DeLaunay triangulation from an original point distribution,
and represent a single cloud. The code can also operate with compound domains comprising many clouds, but
these are not considered further here.
Tetrahedra from the triangulation become the basis for a finite-element solution of the combined radiative transfer 
and statistical equilibrium equations for arbitrarily-saturated masers in a single transition. Such solutions
are 3D generalisations of Schwarzschild-Milne style methods (for example \citealt{1964ApJ...139..397K})
in which radiation integrals, particularly the mean intensity, are eliminated analytically to leave
integral equations in the inversions. On discretization (for example \citet{2006MNRAS.365..779E} 
and \citetalias{2018MNRAS.477.2628G}) these become non-linear algebraic equations for the nodal inversions.
Input radiation to the model comes from a variable background, and spontaneous emission from the maser transition itself is
ignored, as in many maser-focussed studies. We therefore solve a set of coupled non-linear algebraic equations
of the general form
\begin{equation}
\delta'_{i} \! = \! \left\{ \!
                     1 \! + \! \frac{1}{4\mathrm{\upi}^{3/2}} \! \sum_{q=1}^Q
                          \frac{i_{\rmn{BG},q} {\cal A}_q}{l_q^2}
                            \! \times \! \sum_{k=1}^K \zeta_k 
                       \exp \! \left[ \tau_{\rmn{M}} \! \sum_{j=1}^{J(q)}
                           \! \delta'_j \Phi_{j,k}^{q,i}
                          \right]
            \! \right\}^{-1} ,
\label{eq:modsat}
\end{equation}
where $\delta'_{j} = \delta_j /\delta_{0,j}$ is the fractional inversion at node $j$
of the triangulated domain, situated at position $\vec{r}_j$ measured from the domain
origin, and is guaranteed to have a value
$0 \leq \delta'_{j} \leq 1$. The unsaturated inversion at node $j$ is $\delta_{0,j}$, and this is a
function of position, as in \citetalias{2020MNRAS.493.2472G}. The $\delta_{0,j}$ have a global scaling,
to a reference value, usually $\delta_{0,\rmn{max}}$, the largest unsaturated inversion at any node
of the model. Further details appear in Section~2.2 of \citetalias{2020MNRAS.493.2472G}.
At $\vec{r}_i$, a total of $Q$ rays converge: each ray has a background intensity $i_{\rmn{BG},q}$
relative to the saturation intensity, which is the maser intensity that halves the unsaturated inversion,
and is defined symbolically in equation (\ref{eq:isat}). Each ray has
an associated solid angle element equal to the background sky area, ${\cal A}_q$, divided by the 
square of the distance, $l_q$, from the background source to the target node along the ray. Ray $q$ is
bounded by $J(q)$ nodes along its path from domain entry to target node. This set of nodes is decided via
membership of the elements through which ray $q$ passes on the way to the target. 
Ray solid angles are almost equal for every ray converging on the target from a
background source of `celestial sphere' type. The variable velocity modification in Section~\ref{ss:vbv}
introduces a numerical frequency quadrature, achieved via a total of $K$ quadrature abcissae, and 
weights, $\zeta_k$. The overall saturating effect of the model is controlled by the depth 
multiplier, $\tau_{\rmn{M}}$, essentially a measure of the unsaturated inversion,
and the saturation coefficients, $\Phi_{j,k}^{q,i}$, that depend on
target node, $i$, frequency abcissa, $k$, ray, $q$ and bounding node, $j$. Note that equation~\ref{eq:modsat}
contains no radiation integrals: their effect is exerted through the $\Phi_{j,k}^{q,i}$. In its
original form, the analytic elimination of the radiation appeared in \citetalias{2018MNRAS.477.2628G}, where
a restriction to a static medium allowed us to also analytically integrate over the frequency.
A newer derivation of equation~\ref{eq:modsat} appears in Section~\ref{ss:vbv}.

Once a solution of equation~\ref{eq:modsat} has been obtained at all nodes of the model, comparatively
cheap formal solutions of the RT equation may be performed towards an observer, remote compared to
the domain size, and in an arbitrary direction. The formal solutions may then be strightforwardly
converted to synthetic images and spectra. 
Frequency channels in formal solutions used to generate our model spectra and images are allocated,
unless otherwise stated, such that 25 channels cover 7 Doppler widths. For example, H$_2$O molecules
at 670\,K have a Doppler width of 0.784\,km\,s$^{-1}$, so one channel has a width of 0.22\,km\,s$^{-1}$.

\subsection{Variable bulk velocity}
\label{ss:vbv}

The model used in the present work is, in many ways, substantially simpler than that used
in \citetalias{2020MNRAS.493.2472G}: there are no driving functions to be considered, and
uniform density, pseudo-spherical, clouds seem a reasonable approximation to use for the
initial state, prior to shock impact. However, a necessary complication is to include a
velocity field within the cloud. The aim here is to introduce this velocity variation, whilst maintaining an
essential tenet of previous code versions: that saturation can be represented by an array of
pre-computed coefficients, the $\Phi_{j,k}^{q,i}$ in equation~\ref{eq:modsat}, that 
remain constant throughout the iterative procedure that
calculates the maser inversions at all nodes of the model (hereafter nodal inversions). The  
nodes are points within the computational model that are vertices of one or more of the
tetrahedral elements generated by triangulation of a 3D structure (see, for example
Fig.~1 of Paper~1).

The theory introduced in Section~2 of \citetalias{2018MNRAS.477.2628G} starts out with
equations that are general enough to consider variation in both the internal velocity
of the cloud and in the Doppler line width (through variation in kinetic temperature or
microturbulent speed). In that work, we subsequently made some more restrictive assumptions
in Section~3.1 that included a negligible internal velocity field. Here, we return to 
equation (11) of \citetalias{2018MNRAS.477.2628G},
the formal solution of the radiative transfer equation for the specific
intensity as a multiple of the saturation intensity, which we reproduce here,
\begin{equation}
i_{\bar{\nu}}(\tau) = i_{\rmn{BG}}(\Omega) \exp \left\{
    \int_{\tau_0}^\tau \!\!\!\! d\tau' \delta'(\tau') \eta(\tau')
    e^{-(\bar{\nu} - \hat{\vec{n}} \cdot \vec{u}(\tau'))^2 \eta^2(\tau')}
                                          \right\}
\label{eq:formsol0}
\end{equation}
with the notational change that we use the variable-density inversions, $\delta'(\tau')$, 
as in \citetalias{2020MNRAS.493.2472G}, rather than the original uniform density $\Delta(\tau')$.
This new inversion scaling is carried through to the
gain coefficient in the radiative transfer part of the problem, changing the optical depth scale,
$\tau'$, in equation (\ref{eq:formsol0}).

Symbols used in equation (\ref{eq:formsol0}), together with
key stages in its discretization are set out in detail in
Appendix~\ref{a:vvd}, showing the construction of the saturation coefficients,
$\Phi_{j,k}^{q,i}$, where the various indices are as introduced in Section~\ref{ss:keyres}
If the Gauss-Hermite quadrature
method is used, with weights $\zeta_k$ and abcissae $\varpi_k$, then the discretized
mean intensity that has been developed via a frequency and solid-angle average of
equation (\ref{eq:formsol0}) is the final equation of Appendix~\ref{a:vvd}, equation~\ref{eq:formsol4}.
When this result, representing the mean intensity of the maser at the target node, is substituted into
the following expression for the fractional population inversion:
\begin{equation}
\delta'_{i} = [1 + \bar{j}(\vec{r})]^{-1} ,
\end{equation}
we recover equation~\ref{eq:modsat}, an example from a set
of non-linear algebraic equations in the nodal inversions. Compared with the model in \citetalias{2020MNRAS.493.2472G},
the chief added complexity is the additional index $k$, corresponding to the frequency abcissae, which
raises considerably the memory requirement for the array that stores the coefficients.

\section{Model Clouds}
\label{s:modclouds}

\subsection{Compressed clouds}
\label{ss:compcld}

\begin{figure}
\includegraphics[bb=0 0 460 400 ,scale=0.38,angle=0]{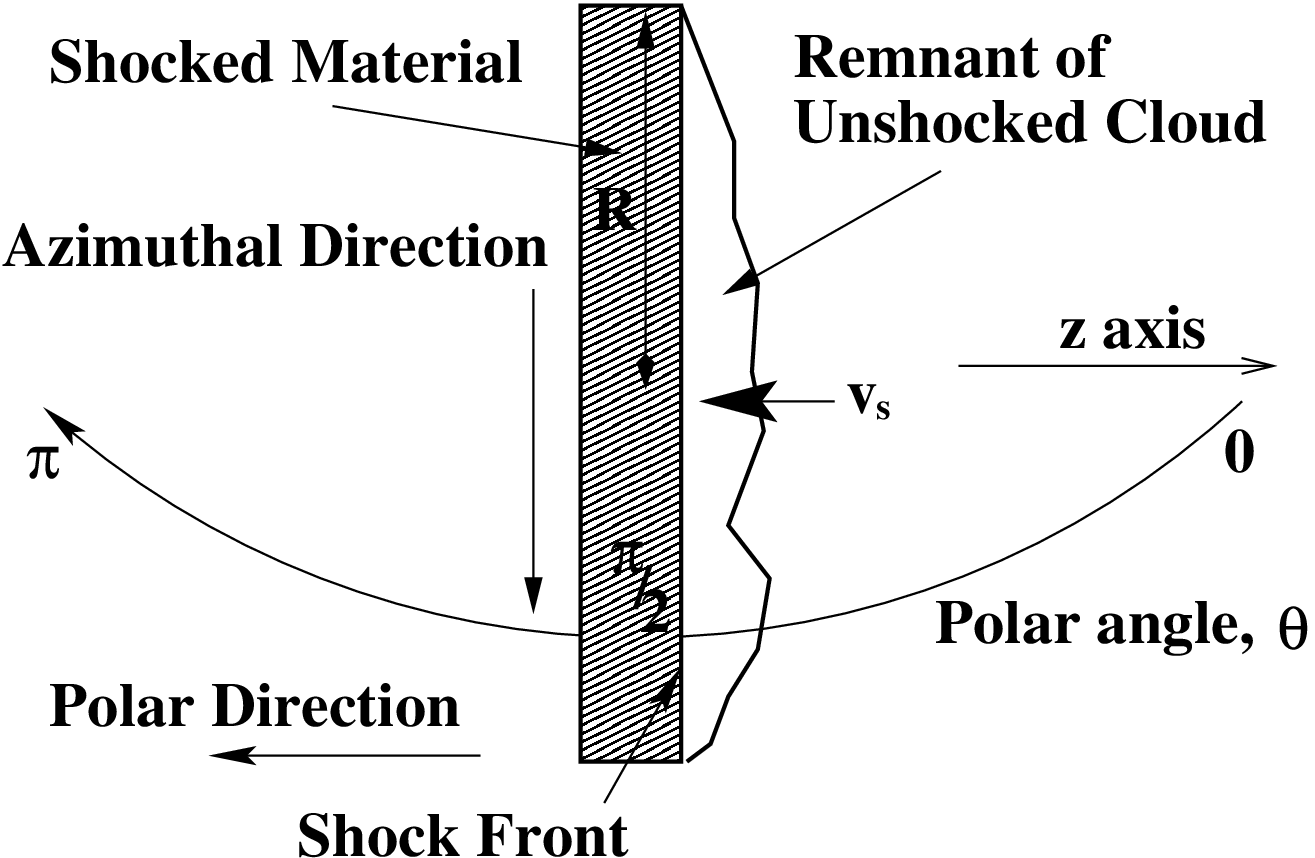}
\caption{A cartoon of the situation near the end of shock passage through a cloud, viewed
parallel to the shock front. In this
frame, a small remnant of the unshocked cloud approaches the shock front, which is at rest,
at speed $v_{\rmn{s}}$. The shocked material is approximated as a short cylinder, shown here edge-on
as a shaded rectangle. Polar and azimuthal directions are marked, noting that the latter
follows the curved edge of the cylinder.}
\label{f:viewpoint}
\end{figure}

For those models requiring a structural change to the cloud due to shock impact,
compressed clouds were constructed from original pseudo-spherical point distributions, and
then compressing the $z$-coordinate for a fraction of the cloud by an amount corresponding
to the shock velocity. The algorithm used first calculates $\Delta z$, the maximum separation
of any pair of nodes along the $z$ axis of the domain. The $z$-position of the shock was
then computed as $z_0 = z_{\rmn{min}} + f_{\rmn{s}} \Delta z$, where $z_{\rmn{min}}$ is the $z$-coordinate of the node
with the most negative $z$-position and $f_{\rmn{s}}$ is the fraction of the cloud shocked (by 
linear distance, rather than volume). For all nodes with a $z$ position such that
$z < z_0$, a modified $z$-position was computed from
\begin{equation}
z' = (z - z_0) / x,
\end{equation}
where $x$ is the compression factor imposed by the shock. The partially compressed domain was then triangulated,
and density and velocity corrections applied to the file of physical conditions.
Densities in the part of the domain with $z < z_0$ were then adjusted to $ x n_0$, where $n_0$ is 
the density in the unshocked material, and the $z$-component of the velocity was set equal to
the shock speed for all nodes in the same part of the domain. Shocked fractions generally
run from zero to 1.0 in steps of 0.05, giving 21 points covering the change in cloud structure
and approximating to the rise time of the flare, which has an approximate range of 30-300\,d.
Figure~\ref{f:viewpoint} shows a sketch of a cloud with a large compressed fraction, in which
the shocked material is represented as a short cylinder.

Models corresponding to C-type shocks (see Section~\ref{sss:continuous}) retained the original
pseudo-spherical point distribution, with the shock affecting only the number density of
the maser molecule. This quantity had pre-shock and post-shock values, and an intermediate
zone in $z$, where the abundance of the maser molecule varied linearly between the two extremes.
The overall number density is assumed to vary negligibly over the range of distance where the
number density of the maser molecule rises rapidly (\citealt{1996ApJ...456..250K}, Fig.~2). Even
at the greater distance corresponding to the hottest part of the shock, compression is still
said to be less than a factor of 2.

\subsection{Simple Shock Models}
\label{ss:simple}

The purely geometric considerations introduced in Section~\ref{ss:compcld} are now modified in
ways that make the present work rather specific to collisionally-pumped transitions of H$_2$O, and
especially the 22-GHz maser transition. There are a small number of model types, set out
in Table~\ref{t:shockmods}, and these are introduced
below in order of increasing sophistication. All of these models are treated as isothermal, and
we justify this approximation on the following basis.
In the hydrodynamic type, the most important parameter that governs the behaviour of the
post-shock gas is the distance $L_{\rmn{cool}}$, over which the initially shock-heated
gas returns to its pre-shock temperature. 
For shocks propagating into molecular gas, H$_2$O itself is an important
coolant if its abundance is high enough, and the temperature reaches at least 250\,K
\citep{2016JPhB...49d4002T}. We consider the cooling time in \citet{2011piim.book.....D},
equation (35.33):
\begin{equation}
t_{\rmn{cool}} = (f+2) n k T / (2\Lambda) = 4 n k T / \Lambda ,
\label{eq:tcool}
\end{equation}
where the most right-hand version in equation (\ref{eq:tcool}) assigns $f=6$ degrees of freedom
for H$_2$O at temperatures $\lesssim$1500\,K. In equation (\ref{eq:tcool}), $n$, $T$ and $\Lambda$
are respectively the number density, kinetic temperature and cooling function in the
post-shock gas. The cooling function for H$_2$O in \citet{2016JPhB...49d4002T} is in
a form per unit solid angle, and per molecule. Multiplication by the number density
of H$_2$O and by $4\mathrm{\upi}$ then places this function in the more usual
erg\,s$^{-1}$\,cm$^{-3}$. We then find that only the fractional number density of H$_2$O
is needed, and using the numerical value of 10$^{-13}$ for the Tennyson et al. cooling
function, the cooling time at 1500\,K is $6\,600 / f_{-4}$\,s, where $f_{-4}$ is the water
abundance with respect to the total number density in multiples of 10$^{-4}$. The actual
cooling time is arguably shorter, since we have not included any other cooling species,
though we estimate the contribution of H$_2$ at densities above 10$^{6}$\,cm$^{-3}$ to be
considerably lower than that of H$_2$O according to data in \citet{2016MNRAS.457.3732C}.
Models never assume maser-zone H$_2$O abundances below $f_{-4}=0.1$, yielding a
maximum cooling time of 66\,000\,s, or 18.3\,hr. Converting this to a
length via the equation,
\begin{equation}
L_{\rmn{cool}} = (v_{\rmn{s}} / 4) t_{\rmn{cool}},
\label{eq:lcool}
\end{equation}
we find that the cooling distance is never significantly larger than $1.7\times 10^5$\,km for
a shock speed of 10\,km\,s$^{-1}$. As this is vastly
smaller than typical inter-nodal spacings in our model domains, we are justified in treating this
type of shock, including its cooling zone, as a localized disturbance of infinitesimal thickness.
We also always use the approximation of a plane-shock front propagating through a pseudo-spherical cloud, and
acknowledge that this ignores the possibilities of both large-scale curvature of the shock
front and smaller-scale distortions that may arise from Rayleigh-Taylor and similar instabilities.

This type of hydrodynamic shock has a jump in physical conditions almost
immediately behind the shock front, and we approximate the compression factor
by $x = M_0^2$, where $M_0$ is the Mach number, given by $v_{\rmn{s}}/v_{\rmn{iso}}$, the ratio of the
shock speed to the isothermal sound speed. This model type corresponds to
strong coupling between the ion and neutral fluids, as might be the case for
magnetic fields that are abnormally weak compared to the predictions of the relation
defining $b$ \citep{1996ApJ...456..250K}, where $b$ is in the range 0.1-3,
or to a higher than typical fractional ionization.
Typical values of $n_{\rmn{H}}$, the number density of H-nuclei, lie in the range $10^6-10^{8.5}$\,cm$^{-3}$
in the pre-shock gas.

In the very simplest model, we use the compression factor above, and make the naive assumption
that the unsaturated inversion itself follows the overall density compression. This amounts to allowing 
the shock to generate a fresh supply of H$_2$O molecules. This simplest model is implemented as Model~0.
Models of the form Model~$N$, where $0 < N \leq 3$, adopt a constant fractional abundance of the maser 
species, and consider the consequences of the increased post-shock density on the maser pumping scheme, based
on the more sophisticated analysis described below. We defer an additional discussion of the C-type models
($3 < N \leq 6$), where the abundance of the maser molecule is variable, to Section~\ref{sss:continuous}. 
The important parameters of all models are listed in Table~\ref{t:shockmods}.

In all our shock model types, an important computational parameter is the optical depth 
multiplier, $\tau_{\rmn{M}}$, a dimensionless representation of the inversion in a specific transition
of the maser species. The parameter $\tau_{\rmn{M}}$ relates the inversion to the dimensioned 
cloud size, $R$.
Specification of $\tau_{\rmn{M}}$ and $R$
leads directly to the dimensioned unsaturated inversion, $\Delta^0$, (in cm$^{-3}$) in the unshocked gas, since
\begin{equation}
\tau_{\rmn{M}} = \frac{g_{\rmn{u}} \lambda_0^3 R \Delta^0 A}{8 \mathrm{\upi}^{3/2} W}
  \rightarrow
         \frac{\Delta_{\rmn{cc}} R_{\rmn{AU}}}{2.8 \sqrt{T_{400}}} ,
\label{eq:taum}
\end{equation}
which is derived by setting the scaled radius of the original unshocked cloud to $\tau_{\rmn{M}}$, so 
that $\tau_{\rmn{M}} = R \gamma_0$, where $\gamma_0$ is a gain coefficient. Other parameters
include the velocity width $W$, a constant in the isothermal model, the transition wavelength
$\lambda_0$ and Einstein A-value, $A$, and the upper-state statistical weight, $g_{\rmn{u}}$. The
second expression on the right-hand side of equation (\ref{eq:taum}) uses the parameters of the
22-GHz transition of ortho-H$_2$O with the scaled parameters $T_{400}$, the kinetic temperature
in units of 400\,K, $R_{\rmn{AU}}$, the cloud radius in astronomical units and $\Delta_{\rmn{cc}}$, the
22-GHz inversion in cm$^{-3}$. 

\begin{figure}
  \includegraphics[bb=60 50 460 300, scale=0.72,angle=0]{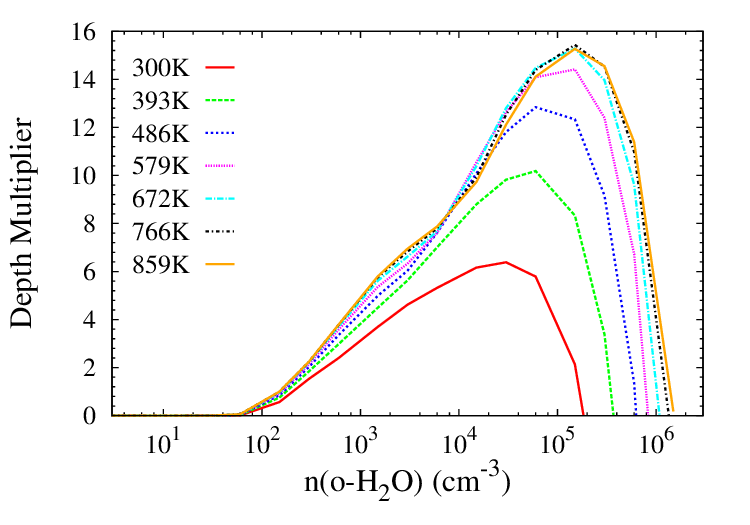}
  \caption{Model optical depth multiplier in the 22-GHz maser transition, $\tau_{\rmn{M}}$ as 
a function of $n_{\rmn{o-H_2O}}$, the
number density of ortho-H$_2$O at kinetic temperatures of 300, 393, 486, 579, 672,
766 and 859\,K, as used in \citet{2016MNRAS.456..374G}, 
and shown in the key. The fractional abundance of o-H$_2$O with respect to H$_2$
is $3\times 10^{-5}$. Values of $\tau_{\rmn{M}} < 0$, where the 22-GHz transition is in absorption,
are not plotted.}
\label{f:tauvn}
\end{figure}

Data in \citet{2016MNRAS.456..374G} show that the relation between the inversion and the overall
number density of the maser species is complicated, and only in a naive model could we take the inversion
to be simply a fixed fraction of the species number density. Assuming that the maser species is 
ortho-H$_2$O, then while $\tau_{\rmn{M}}$ is simply related to the inversion via equation (\ref{eq:taum}), we need
to establish a more complicated relation between $\tau_{\rmn{M}}$ and $n_{\rmn{o-H_2O}}$, the number density
of ortho-H$_2$O. We do this in the following way:
The  maser depth in \citet{2016MNRAS.456..374G} is related to the mean gain coefficient, $\langle \gamma \rangle$ via
$\tau = z \langle \gamma \rangle$, where $z$ is the slab thickness, and $\langle \gamma \rangle$ is related
to an inversion {\it per unit bandwidth at line centre}, $\Delta^{0,\nu}$ , via the equation from \citet{1997MNRAS.285..303Y}
\begin{equation}
\langle \gamma \rangle = \Delta^{0,\nu} A \lambda_0^2 / ( 8\upi) .
\end{equation}
Using the value of the gaussian molecular response at line centre, we then find
$\Delta^0 = \mathrm{\upi}^{1/2} W \Delta^{0,\nu} / \lambda_0$, and the optical depth in the
3D model is 
\begin{equation}
\tau_{\rmn{M}} = g_{\rmn{u}} R \tau / z ,
\label{eq:relate_taus}
\end{equation}
which reduces to $\tau_{\rmn{M}} = 0.871 \tau R_{\rmn{AU}}$ for the 22-GHz transition. It is now possible to
draw a line at constant temperature through the data underlying the 22-GHz
panel of Fig.~5 of \citet{2016MNRAS.456..374G}, yielding directly $\tau$ as a function of
$n_{\rmn{o-H_2O}}$. This relation can then be converted to $\tau_{\rmn{M}}$ as a function of $n_{\rmn{o-H_2O}}$
through equation (\ref{eq:relate_taus}). We plot the relation between $\tau_{\rmn{M}}$ and $n_{\rmn{o-H_2O}}$,
for a number of kinetic temperatures, in Fig.~\ref{f:tauvn}. The curves in this figure
incorporate, in principle, all the complexity of the pumping scheme for the 22-GHz transition,
and it is apparent that, while $\tau_{\rmn{M}}$ rises at modest values of $n_{\rmn{o-H_2O}}$, there is also
a decaying part of each curve, dominated by collisional quenching of the inversion at high
density. It is also apparent from Fig.~\ref{f:tauvn} that for kinetic temperatures above
$\sim$500\,K, the curves are not strong functions of temperature, particularly on the
low-density side of the peak.

The practical procedure for the use of Fig.~\ref{f:tauvn} is to take the value of $\tau_{\rmn{M}}$ for
the unshocked cloud and continue it to an intercept with the curve of approriate temperature,
reading off the corresponding $n_{\rmn{o-H_2O}}$ from the $x$ axis. The shock compression factor $x$
is then applied to $n_{\rmn{o-H_2O}}$, noting that this is taken to be a constant fraction of the
number density of H$_2$ in Models~1-3, but varies in the C-type models, 4-6 that use
Fig.~\ref{f:dense_taumvn} instead (see Section~\ref{sss:continuous}). 
The value of $n_{\rmn{o-H_2O}}$ in the 
compressed gas is then continued to the correct curve, in order to obtain $\tau_{\rmn{M}}$ in the shocked gas.
All models except Model~0 apply the practical procedure described above, to limit the post-shock
inversion (or $\tau_{\rmn{M}}$) in line with expectations of high-density collisional quenching.

\subsubsection{Continuous model}
\label{sss:continuous}

Our remaining models, 4-6 in Table~\ref{t:shockmods}, draw heavily on the continuous (C-type) shocks described 
in \citet{1996ApJ...456..250K}, where physical variables change
smoothly across the shocked portion of the cloud. The only physical parameter that
changes rapidly (over $\simeq$0.5\,au) in these models is the H$_2$O abundance, and we
treat this as the `shock front' in our models. We do
not consider details of chemistry in the post-shock gas, and rely on the results in 
\citet{1996ApJ...456..250K} for H$_2$O abundances (relative to the H$_2$ number density
unless otherwise stated) as necessary. In continuous models,
the compression factor, $x$, is achieved only after a long process of momentum transfer
from the ionized to the neutral fluid, and $x$ is related
to the shock speed via the Alfv\'{e}nic
Mach number, $M_{\rmn{A}}$. For an isothermal shock where the isothermal Mach number, $M_{\rmn{ISO}}$, is much larger
than $M_{\rmn{A}}$, and both Mach numbers are $>$1, the ultimate compression factor is $x=\sqrt{2} M_{\rmn{A}}$ (for
example \citealt{2011piim.book.....D}), for a magnetic field in the plane of the shock.
The Alfv\'{e}nic speed in km\,s$^{-1}$ is given by
\begin{equation}
v_{\rmn{A}} = 1.8 b
\label{eq:vasimp}
\end{equation}
\citep{1996ApJ...456..250K} where $b$ is a constant in the approximate range $0.1-3.0$. However,
$b\gtrsim 1$ is generally required by the requirement of $M_{\rmn{ISO}} > M_{\rmn{A}}$. The magnetic flux density
in the pre-shock gas is 
\begin{equation}
B = b ( n_{\rmn{H}} / \mathrm{cm^{-3}} )^{1/2} \;\;\; \mathrm{\umu G}
\label{eq:magfluxpre}
\end{equation}
where $n_{\rmn{H}}$ is the pre-shock number density of hydrogen nuclei. For our models 4-6, equation (\ref{eq:magfluxpre})
leads, respectively, to $B=10,14.1,20$\,mG.

\begin{figure}
  \includegraphics[bb=60 50 460 300, scale=0.72,angle=0]{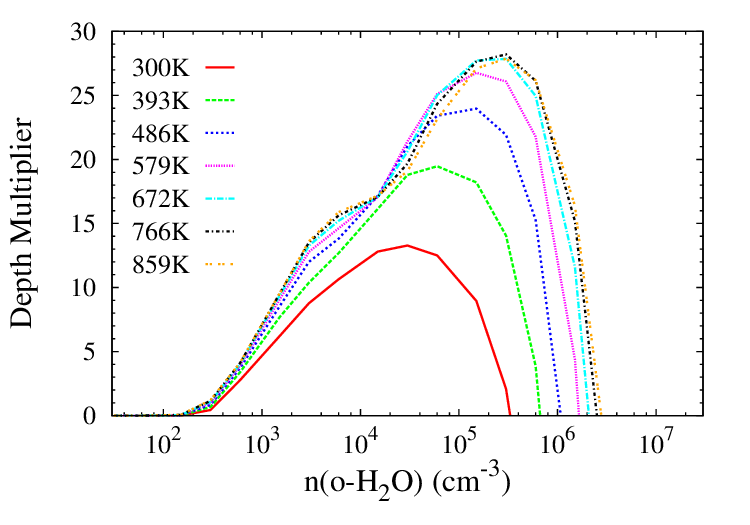}
  \caption{As for Fig.~\ref{f:tauvn}, but for a higher fractional abundance of ortho-H$_2$O,
equal to $3\times 10^{-4}$ with respect to H$_2$. The parameter $\tau_{\rmn M}$, the depth multiplier, is a measure of the available
population inversion in the 22-GHz transition.}
\label{f:dense_taumvn}
\end{figure}

Momentum is transferred from ions to neutral particles over the distance $L_{\rmn{mt}}$. 
We consider extremes for this distance via the equation
\begin{equation}
L_{\rmn{mt}} = 1.8 b L_{\rmn{in}} / v_{\rmn{s}}(\mathrm{km\,s^{-1}}),
\label{eq:lengthmt}
\end{equation}
where we have used equation (\ref{eq:vasimp}), and we derive values of the ion-neutral coupling length, $L_{\rmn{in}}$
from Fig.~1 of \citet{1996ApJ...456..250K}. Equation~\ref{eq:lengthmt} yields the shortest
$L_{\rmn{mt}}$ for the maximum pre-shock density and highest shock speed. We use the shock models
that include populations of large charged (PAH-type) molecules, and small dust grains: the
alternative yields values of $L_{\rmn{in}}$ and $L_{\rmn{mt}}$ that are so large that physical conditions
vary so little across a cloud of a few au in diameter that these models are not useful for the generation
of flares. With the PAHs and grains, the densest pre-shock models in \citet{1996ApJ...456..250K}
($n_{\rmn{H}} = 2 \times 10^{9.5}$\,cm$^{-3}$) correspond in their Fig.~1 to $L_{\rmn{in}} \sim 10^{14.3}$\,cm. With this distance,
$b=0.1$, the smallest value considered, and a shock speed of 40\,km\,s$^{-1}$, the upper limit for an unambiguous C-shock,
we obtain from equation (\ref{eq:lengthmt}), the smallest reasonable momentum
transfer length of $L_{\rmn{mt}} = 9.0 \times 10^{11}$\,cm (0.06\,au). This distance is so
short compared to the cloud scale that one could reasonably treat the entire shock as a thin disturbance, and use 
a model similar to the hydrodynamic types discussed above.
By contrast, the lowest density used by \citet{1996ApJ...456..250K}, where $n_{\rmn{H}}$ is equal
to $2 \times 10^7$\,cm$^{-3}$, corresponds to the much larger $L_{\rmn{in}}$ of $10^{15.55}$\,cm (236\,au), and
the largest reasonable value of $L_{\rmn{mt}}$ (with $b=3$ and a 15\,km\,s$^{-1}$ shock) of 85\,au. The
C-shock paradigm therefore spans a range of momentum-transfer scales from a regime small
enough to be treated as a discontinuity, at the resolution of our maser models, to scales much larger than
typical flare-supporting clouds (0.5-2\,au in AGB stars, 6-9\,au in red supergiants:
\citealt{2012A&A...546A..16R}, and $\sim$1\,au in star-forming regions: \citealt{2005ApJ...634..468U} and
further examples in Section~\ref{ss:obsback}). We resolve this issue in part by noting that at
$n(\mathrm{H_2})=10^{9.5}$\,cm$^{-3}$ and an ortho-H$_2$O fraction of $3\times 10^{-4}$ (modest for the
models in \citet{1996ApJ...456..250K}) $n_{\rmn{o-H_2O}} =9.5\times 10^5$\,cm$^{-3}$, and the
22-GHz inversion is already entering the quenching zone
on the right-hand side of Fig.~\ref{f:dense_taumvn} for temperatures $>400$\,K. This figure is an analogue of Fig.~\ref{f:tauvn}
for an order of magnitude higher H$_2$O abundance. To obtain strong inversions, we adopt
$n(\mathrm{H_2})$ close to $10^8$\,cm$^{-3}$, and here we find $L_{\rmn{mt}}=3.9$\,au for a 35\,km\,s$^{-1}$ shock, favouring a model where
all quantities except the H$_2$O abundance change only slowly within an au-scale cloud. This view is reinforced
by examining Fig.~2 of \citet{1996ApJ...456..250K}. We note that significantly larger H$_2$O maser cloud
sizes have been estimated from observations. for example 0.6-14.5\,au in S140-IRS \citep{2010MNRAS.404..120A}, but these
are not necessarily associated with flares. 

With the above considerations, we adopt the following C-type model: the passage of the H$_2$O
abundance front causes no deformation of the cloud structure and leaves a constant overall
number density. The velocity of neutral species is close to constant, but we allow a small velocity gradient,
in line with Fig.~2 of \citet{1996ApJ...456..250K}, of 10 per cent of the shock velocity
over $0.5 L_{\rmn{mt}}$. The kinetic temperature is set to a constant value of 672.41\,K.
Obviously, this is incorrect for the pre-abundance-front gas, but for a maser model the value
here is irrelevant, since the H$_2$O abundance is approximately 5000 times smaller in this
material, when compared to the post-abundance front gas. After passage of
the abundance front, we justify the constant number density on the grounds that the generated inversion
does not depend strongly on $T_{\rmn{K}}$ for temperatures $>400$\,K (see Fig.~\ref{f:dense_taumvn}), and
any temperature-dependent effect is dwarfed by that due to the abundance rise in o-H$_2$O by
over three orders of magnitude. This abundance rise is assumed to occur exponentially over a transition zone
of thickness $0.5 (35/v_{\rmn{s}} \mathrm{km\,s^{-1}} ) $\,au. In \citet{1996ApJ...456..250K} the
abundance of H$_2$O is expressed relative to oxygen atoms not bound in CO. However, since
the same work assumes that CO is the only reservoir of gas-phase carbon, it is straightforward
to convert this to an abundance relative to H$_2$. Conversion of free oxygen to H$_2$O is
essentially complete for $v_{\rmn{s}} > 15$\,km\,s$^{-1}$, see Fig.~4 of
\citet{1996ApJ...456..250K}, leading to abundances as large as
$8.5 \times 10^{-4}$. In our models, we use the more conservative value of $4.0 \times 10^{-4}$
so that, assuming a 3:1 abundance ratio of ortho- to para-H$_2$O, our abundance of
o-H$_2$O is $3.0 \times 10^{-4}$. Our figure is consistent with the post-shock H$_2$O abundance 
of $3.5\times10^{-4}$ found by \citet{2000ApJ...539L..87M} for Orion~KL.

\subsection{Applicable Timescales}
\label{ss:tscale}

The basic timescale that governs the models used here is the shock crossing time of the original
cloud, or $t_{\rmn{s}} = 2 r_0 / v_{\rmn{s}}$, where $r_0$ is the radius of the original pseudo-spherical cloud
and $v_{\rmn{s}}$ is the shock speed. It is convenient to measure the cloud radius in au, and the shock
speed in multiples of 10\,km\,s$^{-1}$, and in these units, the crossing time in days is
\begin{equation}
t_{\rmn{s}} = 346 \left( \frac{r_0}{1 \mathrm{au}} \right) \left( \frac{v_{\rmn{s}}}{10 \mathrm{km\,s^{-1}}} \right)^{-1} \;
      \mathrm{d}.
\label{eq:tcross}
\end{equation}
As the light-crossing time is of order 30\,000 times shorter, we can assume, with greater
confidence than in \citetalias{2020MNRAS.493.2472G}, that radiation transfer is set up
on a much shorter timescale than $t_{\rmn{s}}$, even if the pumping scheme of the maser depends on
some transitions that are substantially optically thick. As a justification, we use, as in
\citetalias{2020MNRAS.493.2472G}, estimates of optical depths of 350-1000, for an au-scale cloud,
 in the `sink' transition, at 53.1\,$\umu$m, from the o-H$_2$O spin species that is an important
part of the pump for the 22-GHz maser transition, see for example \citet{1973A&A....26..297D}. 
Even at the upper end of this optical depth range, the photon diffusion time, of order
the optical depth multiplied by the light-crossing time, is still 30 times shorter than
$t_{\rmn{s}}$, and is approximately $12$\,d. Since both timescales include a crossing time, use
of a larger cloud does not increase the ratio of the diffusion time to $t_{\rmn{s}}$ unless the
optical depth also increases substantially. If we appeal to effective scattering to reduce
the diffusion time, as in \citetalias{2020MNRAS.493.2472G}, the model can at least be a
starting point for discussion for timescales as small as
$1.2 r_{\rmn{AU}}$\,d, where $r_{\rmn{AU}}$ is the cloud radius in astronomical units.

In all cases where the shock can be counted as a thin disturbance within the approximations
of our model, the timescale $t_{\rmn{s}}$ from equation (\ref{eq:tcross}) is the only one that governs the
rise of the flare. However, in a continuous shock, the momentum transfer time and the associated
ion-neutral coupling time are also important, to the extent that they may far exceed the
time that is considered suitable for the initial rise of a flare. To the best of the knowledge
of the current authors, there is no formal definition that separates times appropriate for
a spectacular flare from those that represent some gentler form of variability. From the
observations considered in Section~\ref{s:intro}, an approximate upper limit for what
might be considered a flare is perhaps of order a few years. Therefore, 
in our C-shock models, which cannot be considered thin, there may be substantial post-flare
evolution as the cloud becomes cooled, compressed and the neutral velocity approaches $v_{\rmn{s}}$.
The initial enhancement of H$_2$O in the C-shock models can be estimated by dividing the thickness of
the transition zone from Section~\ref{sss:continuous} by the shock speed. The result is
\begin{equation}
t_{\rmn{H_2O}} = 24.7 [ 35 / ( v_{\rmn s} \;\;  \mathrm{km\,s^{-1}} )]^2 \;\;\; \mathrm{d} .
\label{eq:th2o}
\end{equation}

Our models are generally considered complete when either the shock front (in hydrodynamic
models) or the H$_2$O abundance front plus its transition zone have passed completely through
the cloud. Therefore, models have a duration of approximately $t_{\rmn{s}}$. The time taken for the radiation
flux density to rise from the pre-shock level to its maximum may differ somewhat from $t_{\rmn{s}}$. We
do not attempt to model in detail the further evolution and decay of the flare, but we
make the following observations here. For hydrodynamic models, if the pre-shock cloud was 
in approximate pressure equilibrium with its surroundings, then
the shocked cloud will be considerably over-pressured because of its enhanced density. If
we need to model the decay of a flare of this type, we use a simple exponential recovery of the
overpressured gas towards the original pressure, of the form $e^{-t/\tau_{\rmn d}}$, where
$\tau_{\rmn d}$ is a dynamical time. For $\tau_{\rmn d}$, we use,
\begin{equation}
\tau_{\rmn d} = L_z / c_{\rmn s},
\label{eq:taudyn}
\end{equation}
where $c_{\rmn s}$ is sound speed in the shocked gas, and $L_z$ is the $z$-axis thickness of the 
shocked cloud once shock passage is complete (shocked fraction = 1.0). It is most unlikely 
that the compressed cloud will relax back to anything resembling its original shape.

If the flare decay is very uncertain in the hydrodynamic case, it is much more
so for the C-type shocks. The compressive evolution, over a time $>L_{\rmn{mt}}/v_{\rmn s}$ introduced
above, may enhance the flare for mild density increases, but could also destroy
the inversion through quenching as $x$ approaches its ultimate value. Cooling to
temperatures $<400$\,K is also detrimental to the 22-GHz inversion (see Fig.~\ref{f:dense_taumvn}), and
such cooling accompanies compression in the models by \citet{1996ApJ...456..250K}. A final
dissolution of the cloud may then occur on a timescale found by substituting $v_{\rmn A}$ for
$c_{\rmn s}$ in equation (\ref{eq:taudyn}), where $v_{\rmn A}$ is the post-shock Alf\'{e}n speed. However,
it is quite possible that the magnetic field,
oriented in the plane of the shock front, may inhibit to some extent any
relaxation flow in the $z$ direction.

\subsection{Notes on Pumping Variability}
\label{ss:pumpv}

All our models treat the gas as isothermal.
Variability therefore depends entirely on changes in density and/or abundance. We also expect that
none of the shocks we consider will dissociate H$_2$O, so that the fractional abundance
of water will remain approximately constant throughout the cloud in the hydrodynamic models,
and will rise quickly to a constant value in the C-type models. The original cloud in
a hydrodynamic model is therefore already a potential maser, since no external radiation is needed
to drive the standard `collisional' pumping mechanism for the 22-GHz transition of 
H$_2$O. The original cloud could even be an observable VLBI maser feature, though
probably rather a weak one. With this in mind, the kinetic temperature may
change significantly across the shock, but the pre-shock value is not very important
because the unshocked gas provides only a very small fraction of the maser flux density
during a flare.

The appearance of the shock may increase the brightness of the maser in two ways,
one of which depends on the viewpoint of the observer, and the other which does not.
The view-independent effect of the shock is that it may increase the available
inversion through a combination of increased $n_{\mathrm{o-H_2O}}$ in the shocked material,
and an increased efficiency of the pumping mechanism. This increased efficiency is based
in \citet{1989ApJ...346..983E} on the enhanced escape probability for line pumping
radiation in directions perpendicular to the maser propagation direction in cylindrical
and filamentary masers, relative to spheres. The combined effect may be
estimated from, for example, Fig.5 of \citet{2016MNRAS.456..374G}, where a compression
of a factor of 10 at a kinetic temperature of $\sim$700\,K could shift the number
density from $10^4-10^5$ o-H$_2$O\,cm$^{-3}$ (for the standard fractional abundance, corresponding
number densities of H$_2$ are $3.3\times 10^{8}$ and $3.3 \times 10^{9}$\,cm$^{-3}$). This
number density increase raises the maser depth in that model from approximately 7 to 17, a
factor of considerably less than that in the number density itself. This supports the view that we
should not simply use the number density as a proxy for the initial inversion. Over a given length,
if the maser remained unsaturated, a single ray of the maser would become brighter by a factor of
$e^{10}$, or 22\,000 based on the raw number density of o-H$_2$O, but only by $e^{17/7}$ (11.34) based
on the maser depth that depends only on the number density of inverted o-H$_2$O molecules,
modelled as nodal inversions. As the shock advances, more of the cloud becomes
enhanced in o-H$_2$O, and angle-averaged quantities, such as the flux-density would
also be expected to increase.

The pre-shock number density is crucial: too high and the post-shock o-H$_2$O number density may be past
the optimum masing conditions, corresponding to the peaks in Fig.~\ref{f:tauvn} and Fig.~\ref{f:dense_taumvn}, 
and into the quenching zones on the right-hand sides of these curves, where the 22-GHz
transition falls rapidly into absorption. If the initial number density is too low, the maser may brighten, but is
unlikely to be considered observationally spectacular.

The viewpoint of the observer is also very important in a shock model. Although the specific intensities
and flux densities received by an observer depend in a complicated way on the passage of many
competing rays through a cloud, an important parameter is the velocity-coherent column density of
molecules of the maser species along a given ray, and this is directly affected by the shock.
Perpendicular to the shock front, the amount of material along a ray is approximately the same as in the original cloud,
independently of the distance the shock has penetrated into the cloud. However,
the velocity coherence is reduced by the velocity change due to the shock.
By contrast, rays moving parallel to the shock front within the post-shock gas experience 
little change in velocity coherence, while the overall column density of the maser species rises
due to the higher post-shock number density. Providing this post-shock density is
not high enough to quench the inversion, we therefore have a basic expectation that the brightest
rays will be emitted towards an observer viewing the cloud parallel to the shock front.

\subsection{Parameter Ranges}
\label{ss:parang}

The most important parameters of the models considered in the present work are displayed
in Table~\ref{t:shockmods}. The standard isothermal kinetic temperature
used was 672.41\,K, corresponding approximately to the temperature that
generated the maximum maser depth in \citet{2016MNRAS.456..374G}.
Subscripts on a basic model number correspond to a final, or target,
value of $\tau_{\rmn M}$. Generally, we increase $\tau_{\rmn M}$ from a starting value of 0.1 
(optically thin and unsaturated) to the target
value in steps of 0.1. Model~0 is special in this respect, because the solution at every
value of $\tau_{\rmn M}$ is a valid model, rather than just a numerical staging point on the way
to the target value. The maximum $\tau_{\rmn M}$ used in Model~0 was 5.0, corresponding to 
significant saturation, and a large value of the inversion.

All hydrodynamic models, including Model~0, use an abundance ratio of o-H$_2$O to H$_2$ of
$3.0 \times 10^{-5}$ to agree with the models in \citet{2016MNRAS.456..374G}. For Models~1-3,
the target $\tau_M$ values correspond to pre-shock number densities of ortho-H$_2$O between 
60 and 3.75$\times$10$^{4}$\,cm$^{-3}$. The compression factors of 9, 16 and 25 are applied
to these pre-shock number densities, but are applied to the inversion only in Model~0.
Post-shock inversions in Model~1 to Model~3 are limited via consideration of Fig.~\ref{f:tauvn}. The maximum
value of the pre-shock o-H$_2$O number density (or of $\tau_{\rmn M}$) corresponds approximately
to the boundary of the quenching zone in the post-shock gas, beyond which models are
uninteresting from the point of view of generating flares.

C-type models (4-6) have a subscript that specifies the shock velocity (from 15-40\,km\,s$^{-1}$).
Within one model, the post-shock o-H$_2$O number density follows immediately from the
pre-shock H$_2$ number density and our standard fractional abundance for these models
of $3.0 \times 10^{-4}$. The pre-shock abundance of o-H$_2$O is small ($1/5000$ of
the post-shock value). As the H$_2$ number density is considered unchanged by the passage
of the H$_2$O abundance front, there is just one target value of $\tau_{\rmn M}$, based upon
the post-shock abundance of o-H$_2$O. The $b$ parameter from equation (\ref{eq:vasimp}) is 1.0.

\begin{table*}
 \centering
 \begin{minipage}{175mm}
  \caption{Parameters of shock models.}
  \hskip-0.6cm
  \begin{tabular}{@{}lrrrrrrrrr@{}}
  \hline
   Model  & Comp. & $v_{\rmn s}$ & $\tau_{\rmn M}$ & $\Delta^0_{\rmn{pre}}$ &
     $\Delta^0_{\rmn{post}}$ & $n_{\mathrm{o-H_2O}}$(pre) & $n_{\mathrm{o-H_2O}}$(post) & $B$(pre) & n(H$_2$) (pre) \\
        & ($x$) & km\,s$^{-1}$ &  & cm$^{-3}$ & cm$^{-3}$ & cm$^{-3}$ & cm$^{-3}$ & mG & cm$^{-3}$ \\
 \hline
 0  & 9  &  7.50 & 1.0-5.0  &  3.63-18.15 & 32.67-163.35 &       
X\footnote{Directly proportional to inversions in column~5}           &    
Y\footnote{Directly proportional to inversions in column~6}              &   0.0 &  $(5.13-25.7)\!\times\!10^6$ \\
 1$_{\rmn t}$  & 9         &  4.60 &  1.0-10.0 &  3.63-36.3  & 20.13-54.27  & (1.54-102)$\!\times\!10^2$ & (1.39-91.8)$\!\times\!10^3$ &  0.0  & (5.13-340)$\!\times\!10^6$  \\
 2$_{\rmn t}$  &16         &  6.14 &  1.0-10.0 &  3.63-36.3  & 23.33-53.03  & (1.54-102)$\!\times\!10^2$ & (2.46-163)$\!\times\!10^3$ &  0.0   & (5.13-340)$\!\times\!10^6$  \\
 3$_{\rmn t}$  &25         &  7.67 &  1.0-10.0 &  3.63-36.3  & 26.24-51.44  & (1.54-102)$\!\times\!10^2$ & (3.85-255)$\!\times\!10^3$ &  0.0    & (5.13-340)$\!\times\!10^6$  \\
 4$_{\rmn v}$  &11.8-31.4  & 15-40 &  $\sim$0.0& $\sim$0.0   & 61.98        & $3.0$                   & $1.50\!\times\!10^4$ &  10.0        & $5.0\!\times\!10^7$ \\
 5$_{\rmn v}$  &11.8-31.4  & 15-40 &  $\sim$0.0& $\sim$0.0   & 82.52        & $6.0$                   & $3.00\!\times\!10^4$ &  14.1        & $1.0\!\times\!10^8$ \\
 6$_{\rmn v}$  &11.8-31.4  & 15-40 &  $\sim$0.0& $\sim$0.0   & 90.80        & $12.0$                  & $6.00\!\times\!10^4$ &  19.9        & $2.0\!\times\!10^8$ \\
\hline
\end{tabular}
\label{t:shockmods}
Columns numbered from left to right are: (1) model number, (2) compression factor, (3) shock speed, (4) maser depth
of the unshocked cloud, (5) \& (6) inversion number density in pre- and post-shocked gas as marked, (7) \& (8)
number density of ortho-H$_2$O in the pre- and post-shock gas as marked, (9) pre-shock magnetic flux density, (10) pre-shock
number density of H$_{2}$.
Subscript $\rmn{t}$ specifies a model depth from $\{1,2,3,4,5,6,8,10 \}$. Subscript $\rmn{v}$ specifies a shock speed in
km\,s$^{-1}$ from 15 to 40, inclusive, in steps of 5. Note that figures in column~10 are unchanged after passage of the H$_2$O
abundance front and compression factors from column~2 should be applied only to obtain ultimate H$_2$ number densities after
momentum transfer to the neutrals.
\end{minipage}
\end{table*}

\section{Results}
\label{s:results}

\subsection{Important Results from Papers~1-3}
\label{ss:imporpoi}

We summarise the observational characteristics of flares from previously studied mechanisms as follows:
\begin{enumerate}
\item Rotation of non-spherical clouds, studied in \citetalias{2019MNRAS.486.4216G}, can have 
variability indices in the range of thousands if the observer's line of sight is near optimal, but 
indices of tens are typical for a randomly chosen line of sight. 
For a pseudo-spherical cloud, see \citetalias{2018MNRAS.477.2628G},
the variability index is only of order 3. Periodicity is unlikely owing to stability considerations.
\item Radiatively driven flares, considered in
\citetalias{2020MNRAS.493.2472G}, are generally more powerful, with variability indices due
to pumping variations typically in the range of thousands to tens of thousands. Extreme
cases, corresponding to small, unsaturated, initial maser depth
(negative optical depth) and large depth change
produce variability indices exceeding 10$^5$ (10$^7$) for oblate (prolate) clouds with
an optimal line of sight. Periodicity in this mechanism naturally follows that of any
source of pumping radiation, and a similar comment may be made regarding variability of background
radiation.
\item Flares due to variations in the level of the background radiation
have somewhat lower variability indices, typically hundreds to thousands, and are also
limited by the variability index of the background itself. This type of flaring also
has a characteristically long duty cycle.
\item Cloud overlap in the line of sight can cause flares with a variability index
of $>$50 from observations (in W~Hya, \citealt{2012A&A...546A..16R}). 
Timescales are approximately $D_{\rmn{AU}}/v_{10}$ in years, where $D_{\rmn{AU}}$ is the cloud diameter
and $v_{10}$ is the velocity component perpendicular to the line of sight in km\,s$^{-1}$.
Such flares are unlikely to be periodic unless there is some favourable cyclic or orbital arrangement.
\end{enumerate}

From the list above, another useful discriminator between the various flare
processes is the ability to generate periodic and quasi-periodic flares. There are many
other possible styles of maser variability \citep{2004MNRAS.355..553G}. The duty
cycle is also a strong discriminator between flares generated by variable pumping and
variable background radiation \citepalias{2020MNRAS.493.2472G}. For the shock-driven
flares studied in the present work, we consider periodicity unlikely for 
individual clouds, because of the
fundamental structural change inflicted by the shock passage. 
However, the geometry of flaring sources could produce quasi-periodic flares if a 
shock passes through multiple regions each containing many clouds. Periodic flares may
also be generated from
the circumstellar envelopes (CSEs) of evolved stars, where periodic
pulsation shocks sweep a distribution of clouds that has statistically similar properties from one
pulsation to the next, provided that there is a continuous supply of outflowing gas from
the stellar photosphere that can renew the cloud population.

The maser beaming angle may also be useful in distinguishing shock-driven flares from other types. A
beaming angle related to the diameter-to-length ratio of the maser was introduced by
\citet{1989ApJ...346..983E}, where the H$_2$O masers in star-formaing regions are modelled as long, thin cylinders.
In evolved star atmospheres, approximately spherical maser clouds were distinguished from the
shock-flattened variety on the basis of the frequency dependence of the observed angular
size of maser features \citep{2011A&A...525A..56R}. Unshocked clouds have an apparent angular
size that is smaller than that of the whole cloud (they are `amplification bounded') and the
apparent size depends on the intrinsic beaming angle of the maser. Moreover, the beaming angle
depends on chords of amplification through the cloud, and amplification is in turn frequency
dependent. By contrast, a shocked cloud, if it presents a fairly flat face to the observer, 
presents its full cross section at any frequency (it is matter, or size, bounded).

All solutions are computed from equation~\ref{eq:modsat}, noting that the coefficients for each
target node, ray, ray-bordering node, and frequency abcissa do not change during iteration, and
may therefore be pre-computed, given sufficient memory. A numerical integration over the frequencies
is necessary in the present work, whilst an anlytical form could be used in Papers~1-3. A solution
takes the form of a list of saturated inversions, one for each model node. The inversions are on the $\delta'$ scale, that
is measured relative to the unsaturated inversion in the individual node, so that $1.0$
means unsaturated and $0.0$ implies ultimate saturation.
A saturated inversion of 0.5 at a node corresponds to the case where the mean intensity there is $\bar{J} = I_{\rmn s}$,
where the saturation intensity, $I_{\rmn s}$ is defined as 
\begin{equation}
I_{\rmn s} = 2 h c \Gamma / [ (1 + g_i / g_j) A \lambda_0^3] ,
\label{eq:isat}
\end{equation}
and the mean intensity is an average over both frequency and solid angle. The rest wavelength
of the maser transition and its Einstein A-value are $\lambda_0$ and $A$ respectively, and the
loss rate, $\Gamma$, 
is taken to be independent of pre- or post-shock conditions. For the 22-GHz H$_2$O transition,
the statistical weights are $g_i=13$ and $g_j=11$. The loss rate has previously been defined
in equation (A5) of \citetalias{2020MNRAS.493.2472G}, where $Z$ is used for the loss-rate. The
change to $\Gamma$ in the present work is purely a notational change.
The loss rate is not trivial to
calculate, and involves summing a sub-set of the larger Einstein A-values and rate-coefficients
out of the upper energy level of the chosen maser transition.
Details may be found in Appendix~B. The overall method of calculating $\Gamma$ is the same as that
used in \citetalias{2020MNRAS.493.2472G}.

\subsection{Comparison with Previous Versions}
\label{ss:res_compare}

To demonstrate that results of the new code converge with those of the old,
we compare a run of the old code, which necessarily
has zero velocity (meaning all model nodes are stationary with respect to each other),
against two runs of the new code, as detailed in Section~\ref{s:model}. In the runs
of the new model, the 250-node domain had an internal sub-thermal velocity that varied
linearly along the $z$-axis, such that, in moving from a node with a scaled $z$-position 
of -1 to one at +1, the velocity increased from $-v$ to $+v$, where $v$ is one of the values
in the first column of Table~\ref{t:vtest}. Input parameters for the old and new models
were otherwise the same. All of these models were
run to a depth-multiplier of $\tau_{\rmn M}=30$. The saturated
inversions on the $\delta'$ scale at three randomly chosen nodes, numbered 9, 67 and 225, for three models
are shown in Table~\ref{t:vtest}. 
\begin{table}
  \caption{Saturated inversions at three randomly chosen nodes in $\tau_{\rmn M}=30.0$ models
for velocties as shown.}
  \begin{tabular}{@{}lcccr@{}}
  \hline
   Velocity  & Node 9            & Node 67    & Node 225    \\
 km\,s$^{-1}$ &                   &            &             \\
 \hline
0.02         &  0.021911         & 0.019843   & 0.031152     \\
0.01         &  0.020539         & 0.018703   & 0.030100     \\
0.00         &  0.020514         & 0.017699   & 0.027527     \\

\hline
\end{tabular}
\label{t:vtest}
\end{table}
It is not entirely trivial to run jobs at velocities that are as close to zero as those
in Table~\ref{t:vtest}, because the expressions for the saturation coefficients in equation (\ref{eq:ykint})
become inaccurate as the $\beta$ denominators approach zero. We therefore use a version of the
coefficient $\Phi_{j,k}^{q,i}$ Taylor-expanded to second order in small $\beta_{qj}$ in this situation,
the modified formula being
\begin{align}
\Phi_{j,k}^{q,i} & \xrightarrow[\beta_{qj} \rightarrow 0]{} s_{j} e^{-(\alpha_{qj} + \varpi_k)^2 }
   \left\{ A_j (1 - \beta_{qj} (\alpha_{qj} + \varpi_k) \right. \nonumber \\
               & \left.
       + (B_j/2) \left[
                    1 - (\beta_{qj}/2) (4\alpha_{qj} + 4\varpi_k + \beta_{qj})
                 \right]
   \right\} .
\label{eq:approxphi}
\end{align}
It was found that a transition value of $\beta_{qj} = 10^{-7}$ gives a smooth transition between
the full formula and the approximation in equation (\ref{eq:approxphi}).

\subsection{Nodal Solutions}
\label{ss:res_nodalsol}

Computation of nodal solutions follows the methods used in 
\citetalias{2019MNRAS.486.4216G} and \citetalias{2020MNRAS.493.2472G}, with the
same non-linear equation solver: the Orthomin(K) algorithm \citep{2001ApMCo.124..351C} with 
order $K=2$. Unless otherwise stated, depth multipliers were increased in steps of 0.1 from $\tau_{\rmn M}=0.1$
to 30.0 for each domain. Complete velocity redistribution, provided by largely isotropic
infra-red radiation \citep{1974ApJ...190...27G,1988rmgm.conf..339F,1994A&A...282..213F,mybook} was
assumed throughout. The example here uses Model~0 from Table~\ref{t:shockmods}.

\begin{figure*}
  \includegraphics[scale=1.18,angle=0]{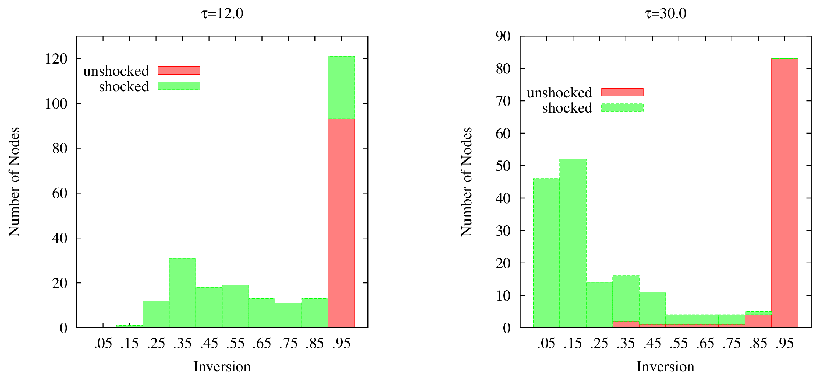}
  \caption{Inversion distributions amongst 10 bins for depth multipliers of $\tau_{\rmn M}=12$
(left panel) and 30 (right panel) for a cloud in which a shock of compression factor 9
has passed through half the $z$-extent of the material.}
\label{f:histogs}
\end{figure*}

The new factor introduced by variable velocity is the extent to which saturation might
be biased to nodes in either the pre- or post-shock zones of the domain. We therefore
plot in Fig.~\ref{f:histogs} histograms of the distribution of nodal populations
for ten bins, marking the contribution of the two zones in each case. In the left-hand
panel, we show a case of moderate saturation, and on the right, for the most
saturated solution ($\tau_{\rmn M}=30.0$) for the same model, in which the shock has penetrated
half way through the $z$-extent of the cloud.
We note that half the $z$-extent is not the same as half the nodes, and that this
model has more than half its nodes in the shocked region.

A strong result is that highly saturated nodes (remaining inversion $<0.5$) are
concentrated in the shocked part of the domain. This result is not confined to the
example shown here and is more extreme in less saturated models (for example the left-hand
panel compared to the right-hand panel in Fig.~\ref{f:histogs}).

\subsection{Sample Images and Light Curves}
\label{ss:samples}

\begin{figure*}
  \includegraphics[scale=1.30,angle=0]{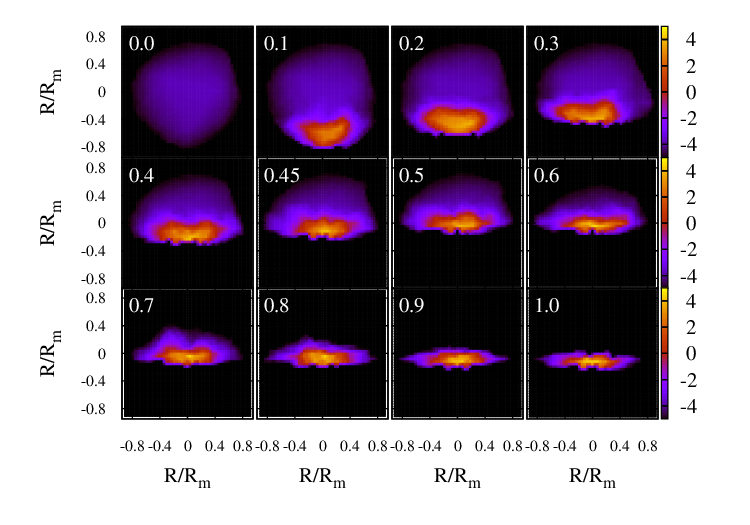}
  \caption{A sequence of images showing the maser brightness distribution varying as the shock
progresses through the originally pseudo-spherical cloud. The white number at the top-left of each
panel is the fraction, $f$, of the $z$ axis extent of the cloud that has passed through the shock. The
$z$ axis itself points vertically from the bottom towards the top of each panel. The colour
bars to the right of each row show the base-10 logarithm of the specific intensity in units of the
saturation intensity of the maser. Note that there is a frame change at $f=0.5$: at earlier times, the
shock is shown moving in the positive $z$ direction into the cloud. at later times, the shock is shown
stationary at $z=0$, while the remnants of the cloud flow into it in the negaitve $z$-direction.
This shift is purely for the viewer's convenience.}
\label{f:frames}
\end{figure*}

\begin{figure}
  \includegraphics[bb=60 50 460 300, scale=0.72,angle=0]{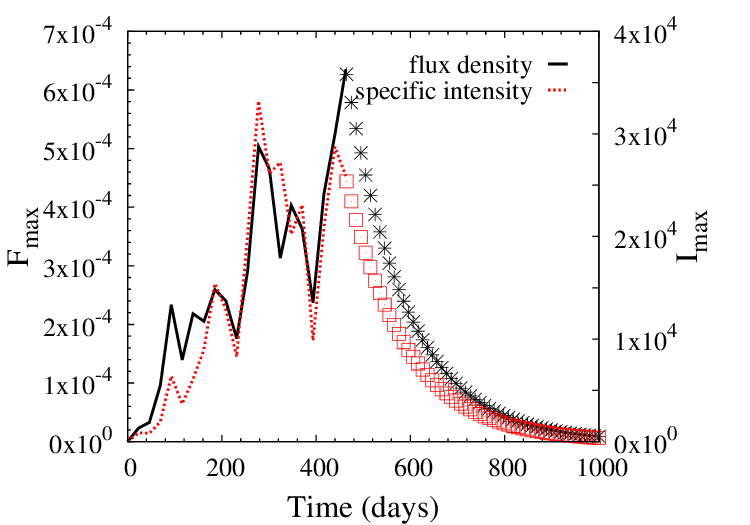}
  \caption{Light curve of the same model as in Fig.~\ref{f:light}, but with
linear $y$-axes and added exponential decays (curve sections plotted in symbols).
The $x$-axis has been converted to time, based on a shock-crossing time for the
original cloud of 463\,d.}
\label{f:linlight}
\end{figure}

\begin{figure}
  \includegraphics[bb=60 50 460 300, scale=0.72,angle=0]{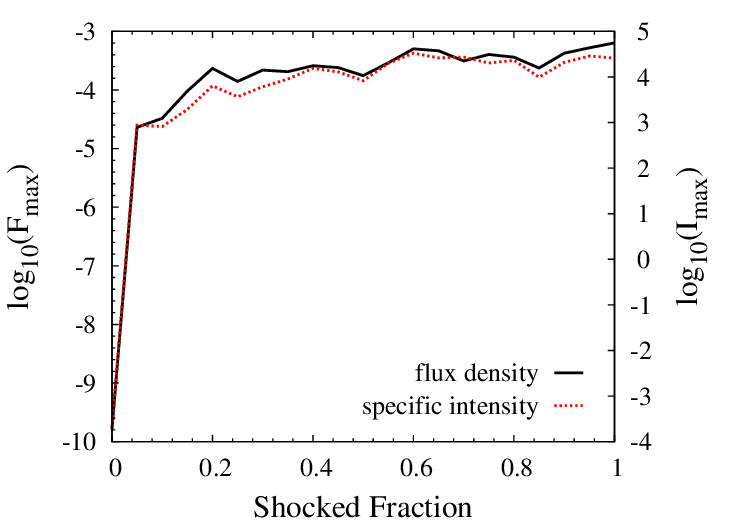}
  \caption{Logarithmic light curves for the maximum flux density in any spectral channel
(black, solid line and left-hand $y$-axis scale), and the maximum brightness in any image pixel
(red, dotted line and right-hand $y$-axis scale). The brightness is scaled to the saturation intensity
of the maser transition. The flux density scale results from using a background brightness
of 10$^{-6}$ with respect to the saturation intensity and placing the observer at a distance
of 1000 domain units, yielding a background level of $\upi \times 10^{-12}$.}
\label{f:light}
\end{figure}

We select a low depth multiplier of $\tau_{\rmn M}=3$, corresponding to rather weak saturation
in the unshocked part of the cloud, and plot a sequence of modelled images (Fig.~\ref{f:frames})
and light curves (Fig.~\ref{f:linlight} and Fig.~\ref{f:light}). Distances along the $z$-axis
are represented as fractions of the original $z$-axis extent of the cloud in Fig.~\ref{f:frames}
and Fig.~\ref{f:light}, ranging from 0.0 to 1.0 in steps of 0.1 for the images and 0.05 for the 
light curve. In Fig.~\ref{f:linlight}, the $z$-axis distance is represented as a time from
initial shock impact, based on a shock-crossing time of 463\,d for the original cloud. The $y$-axis
values in Fig.~\ref{f:linlight} are linear, whilst those in Fig.~\ref{f:light} are logarithmic to
show the initial rapid rise to good effect. Fig.~\ref{f:linlight} also shows exponential decays, 
based on a relaxation time
of 126\,d, in accord with equation (\ref{eq:taudyn}) with $L$ equal to $R/9$ and sound-wave crossing.
In fact, both Fig.~\ref{f:linlight} and Fig.~\ref{f:light} show two versions of the light curve: one is
based on the maximum brightness, $I_{\rmn{max}}$, found in any pixel of the images in
Fig.~\ref{f:frames}, in multiples of the saturation intensity, 
and comparable to the highest specific intensity found in an
interferometer image. The other is based on the maximum flux density in a simulated
single-dish spectrum, $F_{\rmn{max}}$, and corresponds to an integral over the solid angle of the source.
The flux density scaling is explained in the caption of Fig.~\ref{f:light}.
If the maser source is amplification bounded \citep{1992ApJ...394..221E}, then these two representations are related
through the relation $F_{\rmn{max}}/I_{\rmn{max}} = \Omega_{\rmn M} l^2 /d^2$, where $\Omega_{\rmn M}$ is the beaming solid angle
of the maser, $l$ is the intrinsic size of the maser source, and $d$, its distance from the observer.

The compression factor is 9, and in this model (Model~0), it is the same for
the unsaturated inversion. The shock velocity is 7.5\,km\,s$^{-1}$ in the $z$ direction.
With this velocity, the shock would pass completely through a cloud of original radius 1\,au in
463\,d from equation (\ref{eq:tcross}). From Fig.~\ref{f:light}, it is easy to see that a very rapid
rise in the light curve, to within an order of magnitude of the eventual maximum, occurs within
about 10 per cent of the total crossing time, or 46\,d. Shock passage can therefore generate
flares with rise times of order 1 month, as has been observed from some sources, for
example G25.65+1.05 \citep{2018ARep...62..584S}. There is then 
a substantial plateau, filling the rest of the crossing time. The eventual maximum in the brightness
occurs when 60 per cent of the cloud has been shocked. The flux density peaks only after
the entire cloud has been shocked. 

In terms of the specific intensity, the flare is spectacular: the maximum brightness in the central
frequency channel in the original cloud is $2.1 \times 10^{-4}$ of the saturation intensity; the
eventual maximum is 33\,000 times the saturation intensity, yielding a brightness gain of $1.57 \times 10^8$
over the unshocked cloud. To convert to an absolute brightness temperature, we calculated the
saturation intensity for the 22-GHz transition from equation (\ref{eq:isat}), following the methods 
in Appendix~A of \citetalias{2020MNRAS.493.2472G} and using the parameter values in Apprendix~B of the present work.
We then equated our value of $I_{\rmn s}$ to the specific intensity from a black body in the Rayleigh-Jeans limit
in order to obtain the brightness temperature corresponding to $I_{\rmn s}$, which we found to be
$T_{B,\rmn{sat}}=2.18 \times 10^{8} I_{\rmn s}/ \Omega_{\rmn M}$\,K. With the 33\,000 gain
over this temperature, $T_{\rmn B}$ at the peak of the flare is $7.20 \times 10^{12} /\Omega_{\rmn M}$\,K. A
beaming solid angle, $\Omega_{\rmn M}$ is still required to compare $T_{\rmn B}$
to the brightest 22-GHz H$_2$O masers known, for example $T_{\rmn B} > 10^{16}$\,K
\citep{2000aprs.conf..109K} or $8 \times 10^{17}$\,K \citep{1982SvAL....8...86S} for a flare in the
Orion source.

A great advantage of a 3D model is that it is straightforward to compute an
absolute single-dish flux density once the saturation intensity is known. Our version of
equation (A6) of \citetalias{2020MNRAS.493.2472G} for the 22-GHz H$_2$O transition and
the current number density in the shocked gas is
\begin{equation}
F_{\rmn{kpc}} = 73.1 f_{\nu} (R_{\rmn{AU}} / d_{\rmn{kpc}})^2 \;\mathrm{kJy} ,
\label{eq:fkpc}
\end{equation}
where $f_{\nu}$ is the flux density with the scaling on the left-hand $y$-axis in Fig.~\ref{f:light},
$R_{\rmn{AU}}$ is the original cloud radius in astronomical units and $d_{\rmn{kpc}}$, its distance in kpc.
Since the peak value of $f_{\nu}$ attained in Fig.~\ref{f:light} is $6.351 \times 10^{-4}$, an
au-scale cloud at a distance of 1\,kpc would yield a flux density of 46.4\,Jy. The shock mechanism
is therefore capable of lifting a single cloud from complete obscurity to easily
observable.

Another notable feature of Fig.~\ref{f:frames} is that during the flare the area contributing
very high brightness rays to the observer quickly becomes much smaller than the original cloud.
To quantify this, in the bottom right panel, where $f=1.0$,
the fraction of the area that contributes half the total flux density in
the brightest spectral channel is only $5.51 \times 10^{-3}$. Another way of putting
this is that half the flux density is emitted by just 2 of the 363 rays, or pixels,
allocated in the formal solution. Since
the overall emission comes from an area of approximately $\upi$ square astronomical units, half
the emission then emanates from an area of $(2/363)\upi$\,(au)$^2$, with a corresponding linear
scale of 0.13\,au. This is similar in size to the compact halo structure detected by {\it RadioAstron}
towards Cepheus~A \citep{2018ApJ...856...60S}. Spectrally, in the model from the last frame
of Fig.~\ref{f:frames}, the central channel contains 0.7473 of the total flux density, and this
is typical of models with a substantial fraction of the cloud compressed.

\subsection{Density and Saturation}
\label{ss:res_density}

We begin by studying the effect of saturation, as imposed by the significant density increase
in the shocked part of the cloud. We assume here a close-to-optimum observer's viewpoint, where
the line of sight from cloud to observer is parallel to the shock front. However, we now also consider
the more sophisticated models 1-6 (see Table~\ref{t:shockmods}) where the post-shock inversion is limited, and does not simply 
follow the post-shock density as it does in the simplest model (Model~0) that has been discussed in
Section~\ref{ss:samples}.

\begin{figure}
  \includegraphics[bb=60 50 460 300, scale=0.72,angle=0]{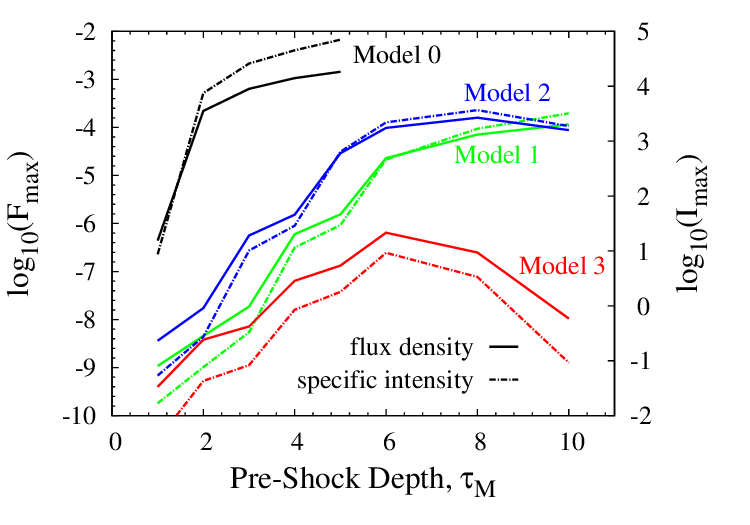}
  \caption{Flux density in the brightest spectral channel (left-hand $y$-axis and solid lines) and
the specific intensity in the brightest map pixel (right-hand $y$-axis and chained lines). Pairs of graphs
are for Model~0 to Model~3 as marked; colour coding is black (Model~0), green (Model~1),
blue (Model~2) and red (Model~3). Scalings are as for
Fig.~\ref{f:light}. Values are taken for the brightest epoch found in the light curve (see main
text) and may differ between the flux density and specific intensity graphs for a particular model.}
\label{f:fluxandepth}
\end{figure}

\begin{figure}
  \includegraphics[bb=60 50 460 300, scale=0.72,angle=0]{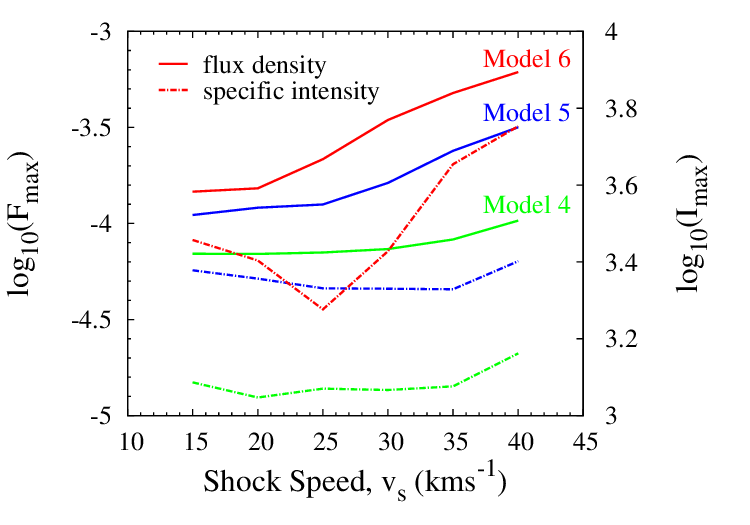}
  \caption{Flux density in the brightest spectral channel (left-hand $y$-axis and solid lines) and 
specific intensity in the brightest pixel (right-hand $y$-axis and chained lines) as a
function of shock speed for the C-type shock models, Model~4-6. There is one pair of lines
for each model. Colour codes are: green, Model~4, $n(\rmn{H_2})=5\times 10^7$\,cm$^{-3}$; blue, Model~5,
$n(\rmn{H_2})=10^8$\,cm$^{-3}$; red, Model~6, $n(\rmn{H_2})=2\times 10^8$\,cm$^{-3}$. Plotted flux density and intensity
values are the highest found in the model, regardless of epoch, as in Fig.~\ref{f:fluxandepth}.}
\label{f:mhdfluxdep}
\end{figure}

In Models~0-3, we use the model optical depth in the unshocked gas as a proxy for the number density of
inverted molecules, that is the difference of the upper-state and lower-state populations
of the maser transition, divided by their respective statistical weights. The model optical
depth and the number density of molecules of the maser species are related
through the graph in Fig.~\ref{f:tauvn}. In Models~1-3, the graphs from Fig.~\ref{f:tauvn} and
the shock compression factor then determine the inverted number density in the post-shock gas 
and the likelihood of saturation in the resulting maser.
In Models~4-6, graphs from Fig.~\ref{f:dense_taumvn} and the shock velocity
serve much the same function. Therefore, unlike Model~0,
each of the other models can contribute only one point per subscripted version to
the graphs in Fig.~\ref{f:fluxandepth}, for Models~1-3, or to Fig.~\ref{f:mhdfluxdep} (for Models~4-6).
For Model~0, where the inversion follows the overall density, the single model can provide data at
all the depths sampled in the upper curves of Fig.~\ref{f:fluxandepth}.

In all cases,
the maximum flux density and maximum specific intensity achieved follow qualitatively similar
functional forms. As expected, these are approximately exponential in the unsaturated regime,
and linear in the saturated regime. Where the physical conditions in the post-shock gas
are sufficient to cause collisional quenching,
maser output begins to decline. The dividing line between the exponential
and linear regimes is not clearly predictable on the basis of the
pre-shock cloud alone, being close to $\tau_{\rmn M} = 6$ for the curves corresponding to
Models~1-3, but at only $2-3$ for Model~0. The continued rise in
the curves for Model~0 suggests a less complete saturation in this case. In Model~0
and Model~1, where the compression factor in number density
is 9, all curves peak at the final epoch, so the points plotted on the
black and green curves in Fig.~\ref{f:fluxandepth} correspond to
100 per cent compression of the cloud. However, for Model~2 (compression factor 16) both
flux density and specific intensity reach their maximum values before the cloud is
fully compressed. Therefore, the points plotted on the blue curve correspond to
shocked fractions of 75-80 per cent, dependent on $\tau_{\rmn M}$, and specific intensities to 65-85 per cent.
Corresponding fractions for Model~3 (compression factor 25) are 80-95 per cent shocked for
flux density points and 65-90 per cent for specific intensity. We note that whilst Model~2 
achieves slightly higher output in both flux density and specific intensity compared to Model~1,
the output of Model~3 is much lower, especially at higher $\tau_{\rmn M}$. This is at least in part
due to the earlier onset of quenching, but almost certainly also involves a viewpoint effect
(see Section~\ref{ss:res_viewpoint} below).

The Model~0 curves are stopped at a pre-shock maser depth of 5.0: it is not reasonable to continue
to higher depth, as further points would almost certainly correspond to fractional H$_2$O abundances beyond
the likely maximum of approximately $4 \times 10^{-4}$ \citep{2021A&A...648A..24V}. If the model is also considered
for evolved stars, there is a little more leeway in this respect, with fractional abundances closer to
$10^{-3}$ reported for W~Hya \citep{1996A&A...315L.241B} and VY~CMa \citep{2010A&A...518L.145R}.
We note that the peak $\tau_{\rmn M}$ obtainable with a 
constant abundance of $3\times 10^{-5}$ is approximately 15 from the peak of Fig.~\ref{f:tauvn}. To
attain larger values of $\tau_{\rmn M}$ requires an increased water fractional abundance in the post-shock gas.
In Model~0, with a compression factor of 9 and pre-shock depth of $\tau_{\rmn M} = 5$, a post-shock
depth of 45 is achieved, or approximately three times the constant abundance limit from
Fig.~\ref{f:tauvn}. It is reasonable
to treat this as a linear multiplier of the fractional water abundance in the saturated regime, implying
a post-shock fractional abundance of $9\times 10^{-5}$ with respect to H$_2$. The Model~0 curves therefore
represent something close to the limiting values achievable for a hydrodynamic shock in a 1\,au cloud.

The C-type models, Models~4-6, have a negligibly small pre-shock maser depth, which follows from
placing the pre-shock ortho-H$_2$O number densities (3-12\,cm$^{-3}$) from Table~\ref{t:shockmods} on the $x$ axis
of Fig.~\ref{f:dense_taumvn}, and reading off the corresponding depth from the $y$ axis for a curve with
$T=672$\,K. These models also have
only one post-shock depth value, since we consider negligible compression during the passage of
the H$_2$O abundance front. Therefore, we plot instead the flux density in the brightest spectral channel
and the specific intensity in the brightest map pixel as a function of shock velocity in
Fig.~\ref{f:mhdfluxdep} for three values of the pre-shock density. Almost all the peak flux densities occur
at the last time sample, when passage of the abundance front is complete. However, the specific
intensities peak at a variety of earlier times, almost certainly indicating competitive action
between rays at the onset of saturation. It is also apparent from Fig.~\ref{f:mhdfluxdep}, that
while the flux density at any speed rises with number density from Model~4 to Model~6, the specific
intensities do so only in a velocity-averaged sense, and the specific intensity, at least in
Models~4 and 5, is not a strong function of speed.

\begin{figure*}
  \includegraphics[scale=0.9,angle=0]{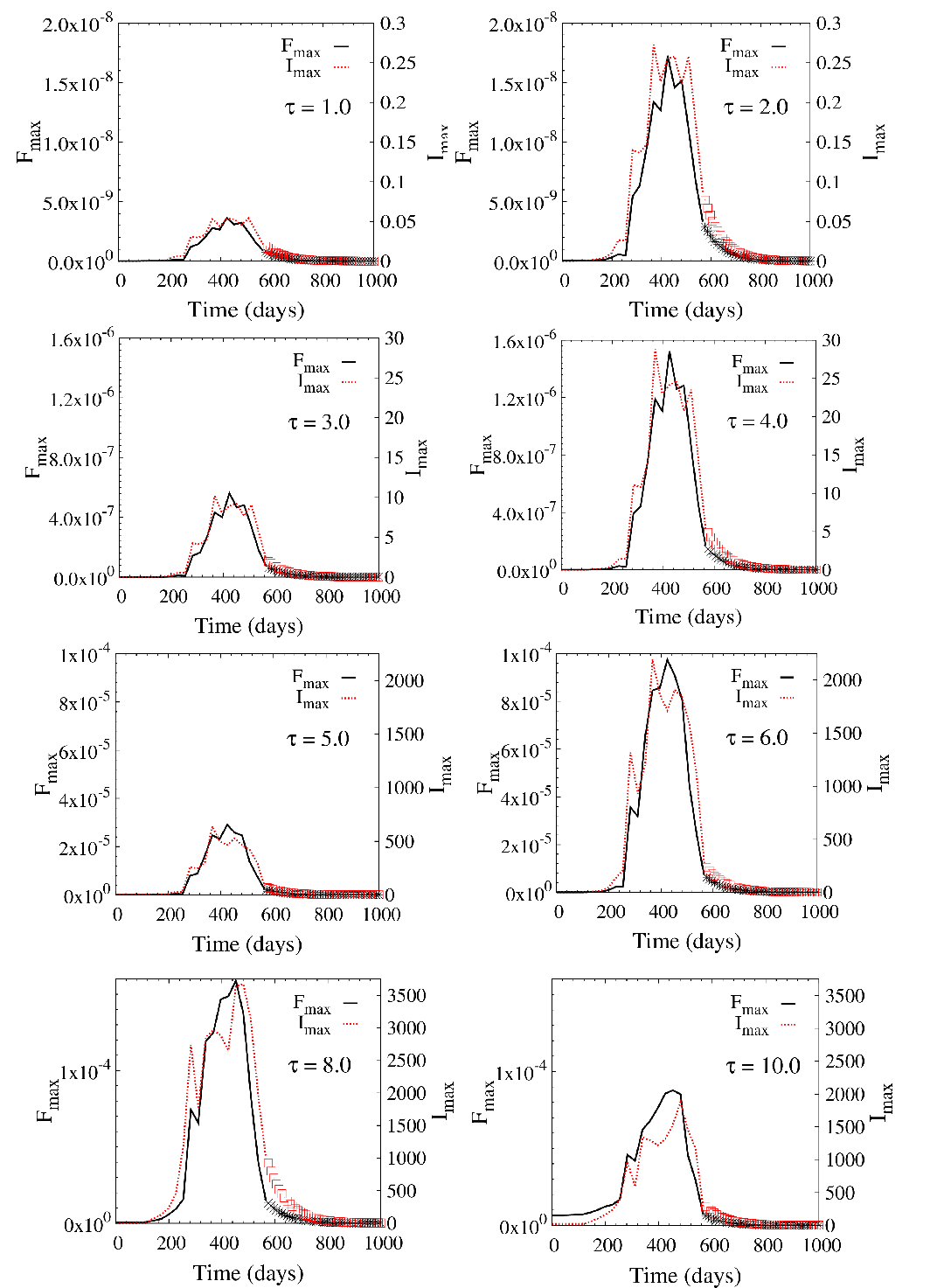}
  \caption{A sequence of light curves, showing the effect of increasing the initial
optical depth of the model. Depth multipliers , $\tau$, in the unshocked cloud, a
measure of density and saturation, are marked in each sub-figure.
The compression factor of the shock, 16,
the shock speed, 6.14\,km\,s$^{-1}$, and the viewpoint of the observer ($\theta = \upi/2$, $\phi=0$)
are the same in all cases; these figures correspond to Model~2 in Table~\ref{t:shockmods}.}
\label{f:low_models}
\end{figure*}

Light curves, with the flux density and peak specific intensity on linear scales, are shown for
Model~2 (hydrodynamic, compression factor of 16) in Fig.~\ref{f:low_models} for all 8 values
of $\tau_{\rmn M}$ studied.  All of the sub-figures are plotted on the same time range
of 1000\,d. To approximate a decay for the flare, we have again plotted exponentials in
symbols with a time constant given by equation (\ref{eq:taudyn}). However, the time constant in
Fig.~\ref{f:low_models} is reduced to 71\,d to allow for increased density, by a factor of $9/16$,
in the post-shock gas, compared to the Model~0 light curve in Fig.~\ref{f:linlight}
The first three panels of Fig.~\ref{f:low_models}, corresponding to pre-shock depth multipliers
of $\tau_{\rmn M} = 1.0,2.0$, and $3.0$ produce flares that would probably not be observable at distances of several
kpc.

\begin{figure}  
   \includegraphics[scale=0.80,angle=0]{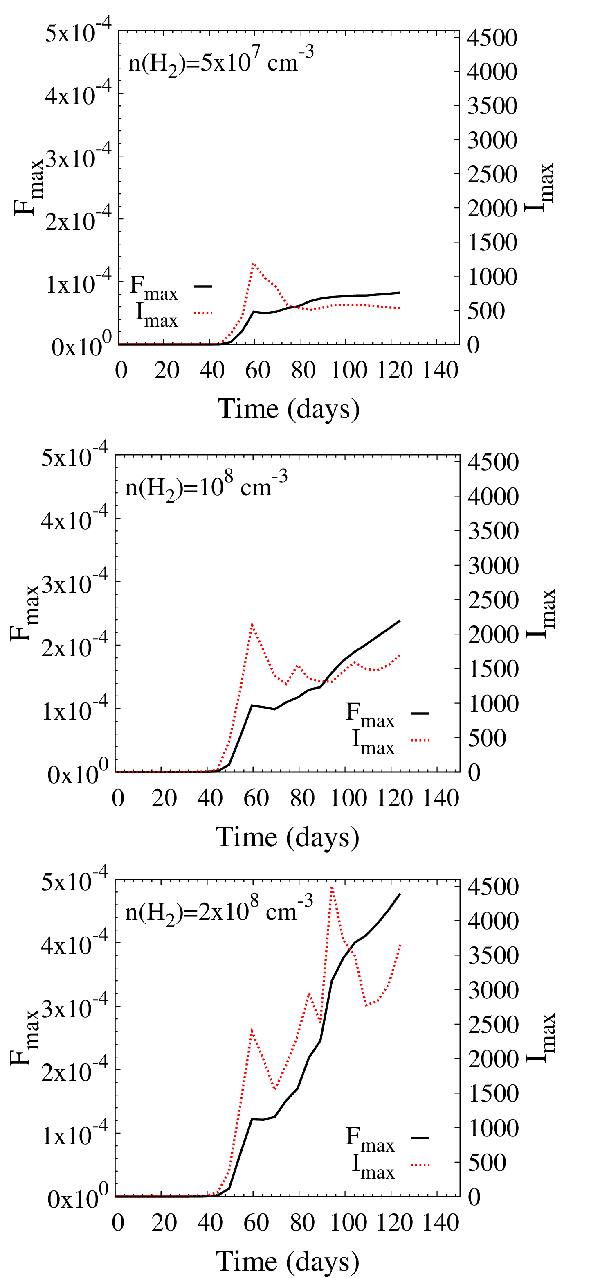}
   \caption{As for Fig.~\ref{f:low_models}, but for the C-type models, Model~4-6 in
Table~\ref{t:shockmods}. The common shock speed is 35\,km\,s$^{-1}$, and each sub-figure
is labelled with the H$_2$ number density of the model.}
\label{f:high_models}
\end{figure}

The counterpart to Fig.~\ref{f:low_models} for the C-type models is Fig.~\ref{f:high_models}, where
we plot the light-curves for a 35\,km\,s$^{-1}$ shock at 3 different number densities of H$_2$. No
decays are shown. The likely outcome, after passage of the H$_2$O abundance front is
complete, is a flat, or slowly rising, segment until the momentum transfer
time is reached, at which point a combination of rapid cooling and high density quenching of the
inversion are likely to cause the light-curve to decay rapidly. If rapid loss of inversion sets in
when the temperature of neutral molecules falls below 400\,K, then the distance over which decay occurs, from cooling
alone, is of order $3\times 10^{13}$\,cm from Fig.~2 of \citet{1996ApJ...456..250K}, or a time
of 99\,d for a shock speed of 35\,km\,s$^{-1}$.

\begin{figure}
  \includegraphics[bb=60 50 460 300, scale=0.72,angle=0]{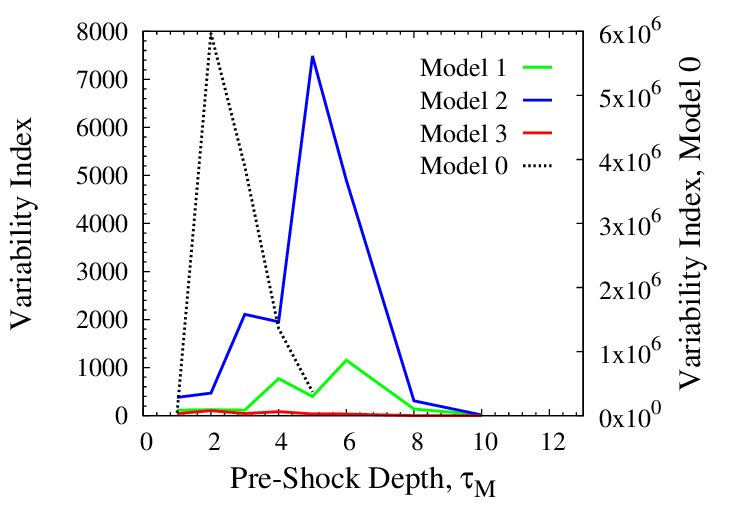}
  \caption{Variability indices in flux density as a function of pre-shock model depth for Model~0 and the
sequence of Models~1-3 with line styles as shown. The left-hand $y$-axis is for the sequence
of models, and the right-hand $y$-axis for Model~0 only.}
\label{f:viplot}
\end{figure}

A powerful flare must be observable, but another criterion of merit is the variability index that
measures how bright the flare is relative to the quiescent state. For these models, this generally
means the flux density from the shocked cloud divided by that from the unshocked original. We
show this for Model~0 to Model~3 in Fig.~\ref{f:viplot}.
All graphs in Fig.~\ref{f:viplot} show a peak, representing the `best' flare in the sense of
a flux-density ratio. However, for all four models, these peaks are at lower values of $\tau_{\rmn M}$ than those corresponding
to strong saturation from Fig.~\ref{f:fluxandepth}. There is therefore generally a compromise to
be found between maximising the variability index or the absolute flux density. We note also
that the Model~0 variability indices can reach several million, whilst the more realistic models
have variability indices that peak at values below 8000. The C-type models have an extremely small optical depth
of H$_2$O in the unshocked gas, and therefore an extremely large variability index.

\subsection{Effect of Viewpoint}
\label{ss:res_viewpoint}

Initially, we consider what happens when
the observer's viewpoint is moved to a selection of positions instead of the view close
to parallel with the shock front, shown in Fig.~\ref{f:viewpoint}. Before shock impact, the 
cloud approximates to a uniform sphere. A large sample of equidistant viewpoints leads to a calculation of
the visibility solid angle, $\overline{\Omega}_{\rmn V}$, of the source: the solid angle of sky
that receives half the flux density of the source.  The visibility solid angle should not be confused with the 
intrinsic beaming pattern that will be considered later. For example, a perfectly spherical maser cloud appears the same 
to all equidistant observers, and so has  $\overline{\Omega}_{\rmn V}=2\upi$\,sr. The same spherical maser, if highly
amplified, might have a very small intrinsic beam solid angle $\ll 2\upi$\,sr, which would make the source
appear much smaller than the physical cloud. For a measurement of $\overline{\Omega}_{\rmn V}$, we choose
the 100 per cent shocked
cloud from Model~0 with a pre-shock model depth of $\tau_{\rmn M}=5$. The result of 1000 formal solutions
in random directions in this case results in a visibility solid angle of
$\overline{\Omega}_{\rmn V} = 0.452 \pm 0.013$\,sr, for the brightest frequency channel, where the uncertainty 
is taken to be the solid angle corresponding to one solution, or 1/1000 of the sky in this example. This uncertainty
estimate follows from the fact that a non-integer number of viewpoints contributes half the flux density, but
we take $\overline{\Omega}_{\rmn V}$ from the nearest integer number.
The $\overline{\Omega}_{\rmn V}$ parameter is useful in estimating the actual number of maser clouds that may exist in
a source. Since the probability of observing a cloud at $>0.5$ of its maximum flux density is $\overline{\Omega}_{\rmn V}/(4\upi)$,
then a randomly placed observer might expect to see one bright cloud in every $4\upi /\overline{\Omega}_{\rmn V}$.

\begin{figure}
  \includegraphics[bb=60 50 460 300, scale=0.72,angle=0]{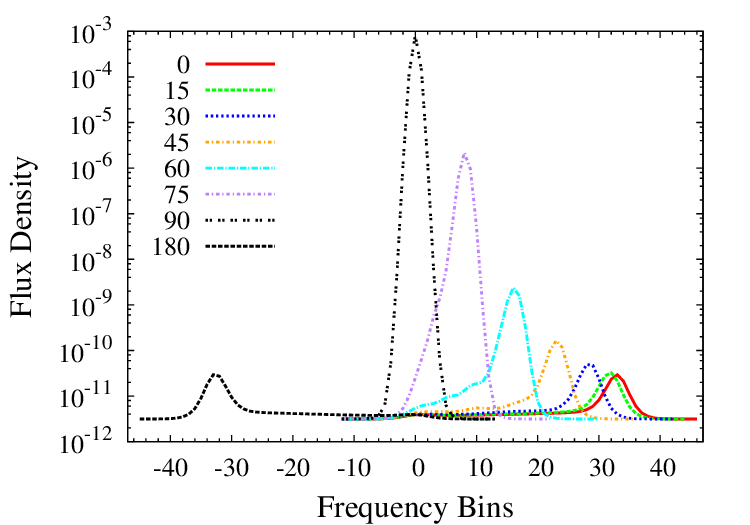}
  \caption{Logarithmic spectra from the Model~0 cloud at the polar viewing angles, $\theta$, shown (in degrees).
The frequency bin number rises with frequency (not velocity). The flux density represents a sum over all
rays (pixels). Each ray has a specific intensity scaled to the saturation intensity, and the observer is
at a standard distance of 1000 domain units.}
\label{f:viewangles}
\end{figure}

We also show sample spectra derived from Model~0 at a compressed fraction of $f=0.75$ 
in Fig.~\ref{f:viewangles} to show the development of the total spread and
the flux density as the viewing position is moved from zero (where the domain $z$-axis and the shock front
are pointing directly towards the observer) through a number of different polar angles, as marked in degrees,
up to 180\degr, where the observer views the cloud from behind the shock front. A logarithmic scale is
used in Fig.~\ref{f:viewangles} to make the spectra viewed significantly away from the 90\degr position
visible. Note that channel number rises with frequency, so the dominant blue-shifted peak from the 
shocked portion of the cloud appears to the right of the figure for a viewing angle of zero degrees, whilst
a very similar red-shifted peak (to the left) appears at 180\degr. The zero of the frequency axis corresponds
to the unshocked portion of the cloud.

\begin{figure}
  \includegraphics[bb=60 50 460 300, scale=0.72,angle=0]{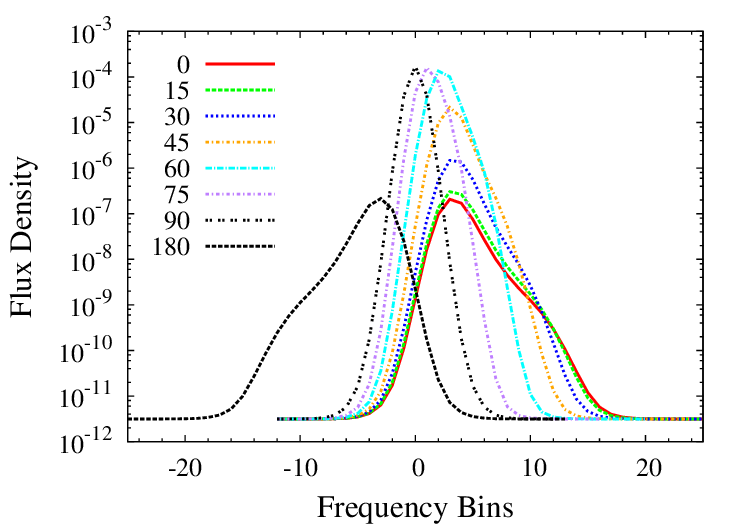}
  \caption{As Fig.~\ref{f:viewangles}, but for the C-type model with a shock speed
of 35\,km\,s$^{-1}$ from Model~5.}
\label{f:viewctype}
\end{figure}

Sample spectra for Model~5 with $v_{\rmn s} = 35$\,km\,s$^{-1}$, as a representative C-type model, are
shown in Fig.~\ref{f:viewctype} for comparison with Fig.~\ref{f:viewangles}. It is immediately
apparent that, without the deformation of the cloud, the bias of high flux density to viewpoints
parallel to the shock front is much less pronounced, making the C-type model significantly less
sensitive to the observer's viewpoint. In this type of model, a bias to higher flux density
when viewed parallel to the shock front is due only to the velocity gradient along the
$z$-axis. The result of the visibility solid angle calculation
in this case is $\overline{\Omega}_{\rmn V} = 3.091 \pm 0.013$\,sr. This much larger value, when compared
to the hydrodynamic case of Model~0 above, confirms numerically the reduced sensitivity to viewpoint
of the C-type models that is suggested by Fig.~\ref{f:viewctype}: a randomly placed observer would
be $6.8$ times more likely to see the flare from Model~5 than from Model~0.

Returning to the situation of a fixed observer, another
great advantage of a 3D model is that it is possible to compute the intrinsic beaming solid angle
of the maser at any point on the surface of the cloud. Operationally, we
choose a target point on the external surface of the domain, along the direct line of sight to the
observer from the domain origin. The maser beaming pattern at this target point can then be computed
by calculating the specific intensities of a large number of rays that converge on the target from
points on a spherical background. If each ray has a very similar solid angle of sky associated with
it at the target, then the beaming angle at that point may be straightforwardly calculated by ordering
the output specific intensities in brightness, and counting the number $N$ that 
generate half the observer's flux density. The half-flux density beaming angle, $\Omega_{\rmn M}$,
is then $\sum_{i=1}^N \delta \Omega_i$ if ray $i$ has an associated beaming solid angle
$\delta \Omega_i$ at the target. If we also
approximate the beaming pattern as an ellipse of solid angle $\Omega_{\rmn M} = \upi \delta \theta_{\rmn M}
\delta \phi_{\rmn M}$, where $\delta \theta_{\rmn M}$ ($\delta \phi_{\rmn M}$) are the intrinsic polar (azimuthal)
beaming angles, then the calculation of $\Omega_{\rmn M}$, and one of the directed beaming angles provides
a complete solution of the beaming-angle problem for one target point. For example, an accurate value of
$\Omega_{\rmn M} = (8.33\pm 0.60 )\times 10^{-5}$\,sr was computed by taking the mean value of 100 similar target points
observed parallel to the shock front for the case of the 100 per cent shocked Model~0 with an initial
depth multiplier of 3.0. Beam solid angles smaller than this have been suggested previously \citep{1991ApJ...367L..63N}.

There is a practical problem in the computation of the beaming angles: the number of rays must be sufficient
to resolve the (very small) beaming solid angle. We increased the order of a HEALPix partition of the sphere
\citep{2005ApJ...622..759G},
measuring the beaming solid angle in each case, and found that the beaming solid angle became significantly
larger than the solid angle associated with an individual ray at order 9, which partitions the sphere
into 3\,145\,728 panels of equal area, each with one ray originating from its centre. Beaming angle computations
were therefore carried out at order 9, with occasional tests at order 10 (12\,582\,912 rays). We note
that our HEALPix algorithm was coded from formulae in the published work cited above, and did not use
any software from the HEALPix website or project.

We now compare the observer's view of the source for a hydrodynamic model in the polar 
and azimuthal directions, where the shocked part of the cloud is highly compressed. The effective geometry
of the compressed cloud, or compressed part of a cloud, is that of a short cylinder, and in the polar direction
the observer looks along the short edge of the cylinder, perpendicular to the shock front
(see Fig.~\ref{f:viewpoint}). In this simple
representation, there is no curvature of the source in this direction, and it therefore approximates to the `screen'
type maser in Fig.~4.1a of \citet{mybook}. Also, although the polar beaming angle will be small in the sense
that its sine and tangent approximate to the angle itself, it will still be
much larger than the typical angular extent of the source, so that we are in the `size-limited' case,
resulting in the flux density equation (4.48) of \citet{mybook}, which is independent of the intrinsic beaming
angle. This is why a model computation of $\delta \theta_{\rmn M}$ is particularly important: it is very difficult
to obtain from observations, but may be estimated from a sequence of VLBI channel images that have a position-velocity
gradient, allowing the cloud size to be estimated from the range of spot positions \citep{2012A&A...546A..16R}.
With a simultaneous measurement of the angular FWHM of each spot, the beaming angle can be calculated. However,
this method needs high sensitvity and assumes an at least approximately spherical cloud surface. Any brightness variation
in the polar direction in images, such as those in Fig.~\ref{f:frames}, is only due to intrinsic brightness variations
as a function of $z$ position along the cylinder edge in Fig.~\ref{f:viewpoint}, and this surface is not 
curved in this direction.

\begin{figure}
  \includegraphics[scale=0.4,angle=0]{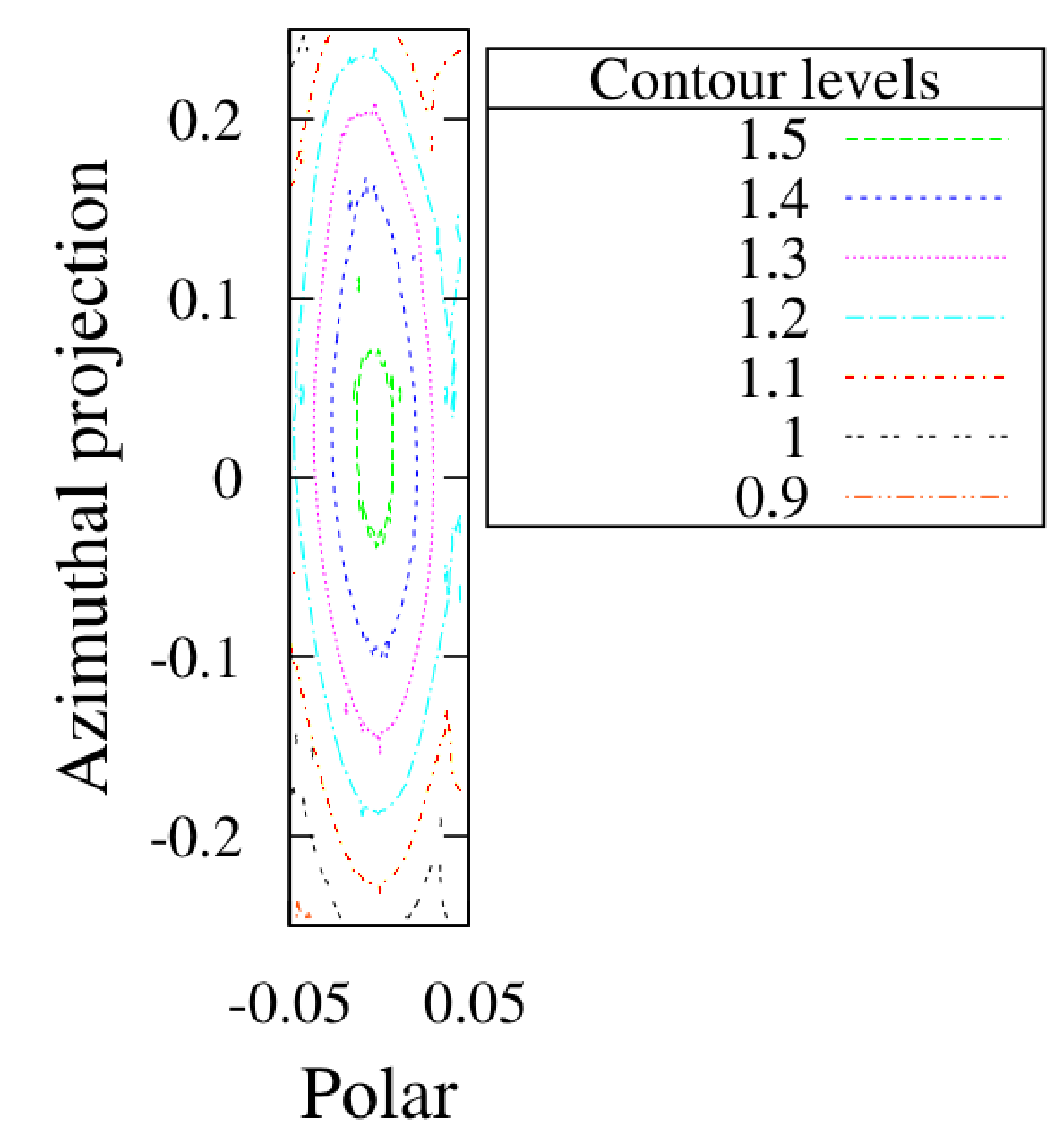}
  \caption{Contours of the intrinsic beaming pattern for Model~0 and initial depth
multiplier 3.0. Levels are in log$_{10}$ of (the specific intensity in multiples of
the saturation intensity). Each ray has an angular offset from the direction of the
brightest, and the $x$ and $y$ axes are the respective projections of this offset
on the polar and azimuthal directions, measured from the centre of the beaming pattern
(the brightest ray). The $x$ and $y$ axis
scales are equal to demonstrate the azimuthal extension of the beaming pattern. A line
towards the observer points out of the page.}
\label{f:beamplot}
\end{figure}

The source surface in the azimuthal direction is curved, following a circular path, parallel to the
shock front, and in or out of the plane of the page in Fig.~\ref{f:viewpoint}.
Unlike the polar direction, we now expect the central intensity of the beaming pattern to vary very
little as the circular path is followed, so that variations in observed intensity for a fixed observer, with
$\sin \theta \simeq 1$ and a particular azimuthal angle, $\phi$, depend upon both the intrinsic 
azimuthal beaming angle, $\delta \phi_{\rmn M}$ and the curvature of the source surface. The situation in the 
azimuthal direction is now rather like Fig.~4.1b in \citet{mybook}, with the flux density given
by a solution of equation (4.52). With symbols, the angular radius of the source in this case can be
written, 
\begin{equation}
\alpha_{\rmn{max}} = R \delta \phi_{\rmn M} / d,
\label{eq:frectangle}
\end{equation}
where $\alpha_{\rmn{max}}$ is the measured angular radius of the source at FWHM, $R$ is the radius of
the short cylinder that represents the shocked region, and
$d$ is the distance of the source from the observer. 
Note that in this case, $\delta \phi_{\rmn M}$ can
be recovered from an interferometric observation of $\alpha_{\rmn{max}}$, provided that $R$ and $d$ are known
independently, and that the angular resolution of the observation is sufficient to determine
$\alpha_{\rmn{max}}$.
From the modeller's perspective, the angular resolution provided by the set of rays in
Fig.~\ref{f:frames} is too coarse to resolve the azimuthal beaming angle, so we compute it using
the HEALPix method introduced above, together with measurements of the asymmetry of the beaming
pattern. The beaming pattern is typically extended in the azimuthal direction, and an example
is shown in Fig.~\ref{f:beamplot}, noting that this diagram shows the true aspect ratio
of the beaming pattern, which is 5.17 for Model~0 and initial $\tau_{\rmn M}=3.0$. Using the ellipse
approximation for the beaming pattern, and the computed value of $8.33\times 10^{-5}$\,sr for
$\Omega_{\rmn M}$, we derive polar and azimuthal beaming angles of
$\delta \theta_{\rmn M} = 2.26\times 10^{-3}$\,rad (0\degr.129)
and $\delta \phi_{\rmn M} = 1.17\times 10^{-2}$\,rad (0\degr.667)
for the 100 per cent compressed cloud from Model~0 with $\tau_{\rmn M}=3.0$.

The ratio $\delta \phi_{\rmn M} / \delta \theta_{\rmn M}$ appears remarkably constant over the range
of depths shown in $\tau_{\rmn M}$. For Model~0, this ratio ranges from 5.2 to 7.6 between depths
of $\tau_{\rmn M} =0.2$ to $4.5$ with no clear trend to increase or decrease. Measurements are
rather inconsistent because the brightness ranges that can be contoured vary so much over the range of
depths used. It is probably significant, however, that $\delta \phi_{\rmn M} / \delta \theta_{\rmn M}$ is larger
for the sub-models of Model~1, ranging from 7.8 to 9.4, again with no clear trend, over the initial depth range of
$1.0-12.5$. This difference must arise from some factor other than the compression factor, which
is the same in Models~0 and 1.

\begin{figure}
  \includegraphics[bb=60 50 460 300, scale=0.72,angle=0]{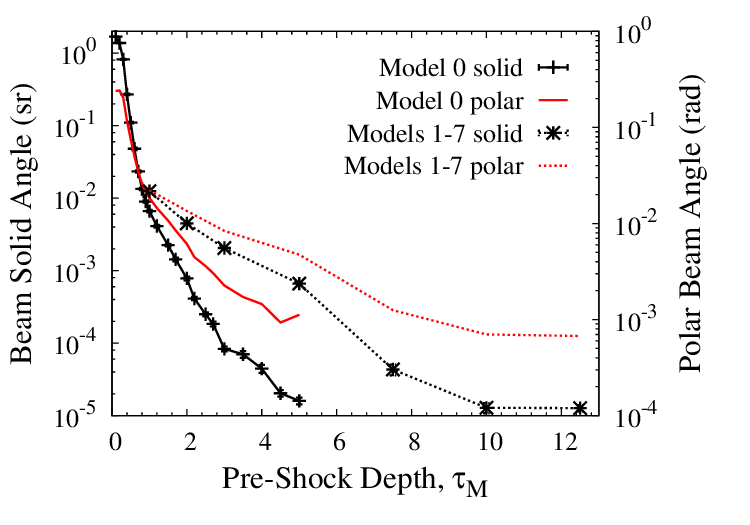}
  \caption{Beaming solid angles (black lines) and beaming angles in the polar direction (perpendicular to the shock front; red lines) 
as a function of the model depth in the unshocked cloud. Solid lines are for Model~0 and dashed lines for Model~1. All
these models have a compression factor of 9. Error bars at points on the solid angle lines show the standard error, based
on the mean of 30 evenly-spaced azimuthal viewing positions.}
\label{f:beams}
\end{figure}

Having considered some particular examples, we plot in Fig.~\ref{f:beams} the beaming
solid angle (left-hand $y$ axis and black lines) and
the polar intrinsic beaming angles (right-hand $y$ axis and red lines), at a similar
range of depths, for Model~0, and for Model~1, as in the figures from Section~\ref{ss:res_density}.
In both black curves, the beaming solid angle falls with rising $\tau_{\rmn M}$. There is no clear residual
level reached for Model~0, but the curve for Model~1 flattens at a level of approximately $1.3\times 10^{-5}$\,sr.
The polar beaming angle does appear to re-broaden at the last point ($\tau_{\rmn M} = 5.0$), 
for Model~0, due to an exceptionally
small value of 4.1 in $\delta \phi_{\rmn M} / \delta \theta_{\rmn M}$. The minimum is $1.12\times 10^{-3}$\,rad
(0\degr.064), at $\tau_{\rmn M} =4.5$, before the apparent re-broadening under strong saturation. The red dotted curve,
representing $\delta \theta_{\rmn M}$ for Model~1, follows the solid angle in apparently flattening out at
the highest depths used. In this case, values of $\delta \phi_{\rmn M} / \delta \theta_{\rmn M}$ remain typical for
the last pair of points, with $\delta \theta_{\rmn M}$ flattening at a value of approximately $7\times 10^{-4}$\,rad
(0\degr.04). Azimuthal beaming angles can be recovered from all depth in  Fig.~\ref{f:beams} from the polar
angles, beaming solid angles and the ellipse assumption for the beaming pattern.

\begin{figure}
   \includegraphics[bb=70 70 410 262, scale=0.85,angle=0]{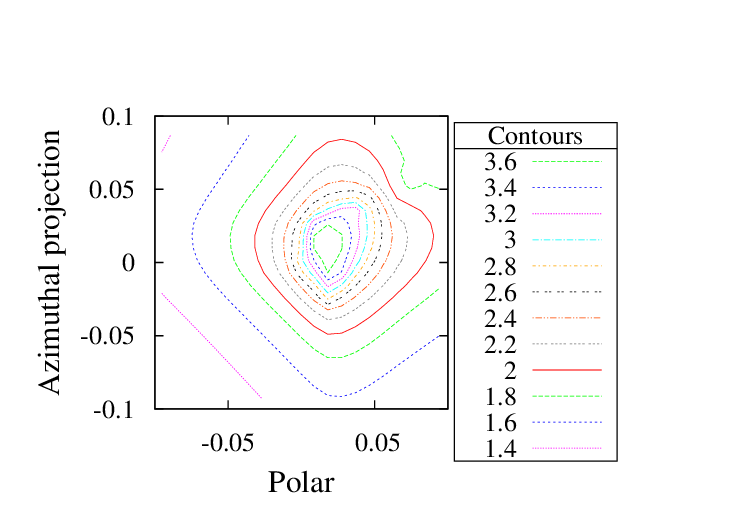}
   \caption{As for Fig.~\ref{f:beamplot}, but for the C-type model, Model~5 with a
shock speed of 35\,km\,s$^{-1}$}
\label{f:mhd_beam}
\end{figure}

In the case of MHD models, there is again no set of pre-shock $\tau_{\rmn M}$ values to use as the
$x$ axis of a graph similar to Fig.~\ref{f:beams}. However, we computed a one-off value of the
beaming solid angle, and the polar intrinsic beaming angle, for Model~5, $v=35$\,km\,s$^{-1}$. The
respective values were $\Omega_{\rmn M} = 1.43 \times 10^{-4}$\,sr and $\delta \phi_{\rmn M} = 8.33\times 10^{-3}$\,rad
(0\degr.48). These beaming angles
are substantially larger than those found for the hydrodynamic models under similar levels of
saturation. We also found that the beaming pattern is less extended in the azimuthal direction
in the MHD models with a ratio of $\delta \phi_{\rmn M} / \delta \theta_{\rmn M} = 1.54$. The beaming pattern for
Model~5, $v=35$\,km\,s$^{-1}$ is shown in Fig.~\ref{f:mhd_beam}.

\begin{figure}
  \includegraphics[bb=60 50 460 300, scale=0.72,angle=0]{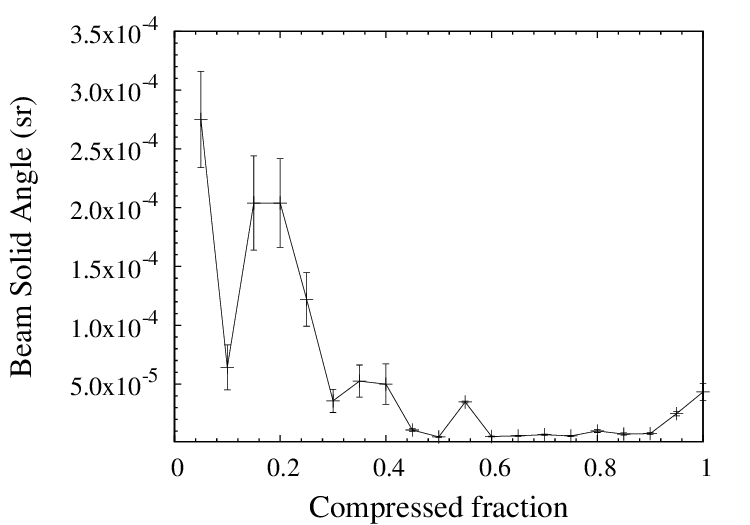}
  \caption{Variation of $\Omega_{\rmn M}$ with compressed fraction, $f$, for Model~1 (pre-shock
maser depth = 7.5). Error
bars are shown for the standard error on the mean of 30 azimuthal observation angles.}
\label{f:beamcompf}
\end{figure}

It is possibly also useful to consider the variation of beaming angles with the shock fraction, $f$, in the
cloud, and an example is plotted in Fig.~\ref{f:beamcompf}. An assumption made for this plot is that the
beaming angle is unaffected by the radiation from the unshocked part of the cloud. This is a reasonable assumption
once a significant fraction of the cloud is compressed, since the radiation from the unshocked
part of the cloud is usually much weaker.
The data is from Model~1, and the sub-model has a  pre-shock depth of $\tau_{\rmn M}=7.5$.
There are almost certainly larger uncertainties arising from the exact position of the line of sight
and the discretization of the domain than the error bars shown in Fig.~\ref{f:beamcompf}, which correspond only to the standard
error on a mean of 30 azimuthal angles, and this is particularly true for low values of $f$, where the
fraction of nodes in the compressed region is small. There is a trend of falling $\Omega_{\rmn M}$ until
the compressed fraction reaches $0.4-0.6$, with little change thereafter, though there is
possibly a small, but significant, broadening at $f=0.9-1.0$.

\subsection{Effect of Shock Speed}
\label{ss:res_speed}

In this section, it is straightforward to study a full set of C-type models, since these
have sub-models identified by shock speed, varying from
15--40\,km\,s$^{-1}$, and each model has a fixed density. In the
case of the hydrodynamic models, Models~1-3, there is one velocity only for each complete
model, and we choose sub-models that have a depth parameter that gives close to
maximum output (plotted in Fig.~\ref{f:fluxandepth}): 
we use $\tau_{\rmn M}=8.0$ for Models~1 and 2, but $\tau_{\rmn M}=6.0$ for Model~3,
so as not to be too greatly affected by quenching in Model~3. With these settings, post-shock
conditions in all 3 models are  
close to the peak of Fig.~\ref{f:tauvn}. For each shock speed, we plot
a light curve. Owing to the considerable range of shock speeds,
we do not plot the $x$ axis in time, but in fraction of the cloud swept. This parameter
exceeds 1.0 for the C-type models to allow for the passage of a transition zone, in
which the H$_2$O abundance rises from its pre-shock value to its full post-shock value ($4\times 10^{-4}$
with respect to H$_2$ in Models~4-6) after passage of the abundance front.

\begin{figure}
  \includegraphics[bb=60 50 460 300, scale=0.72,angle=0]{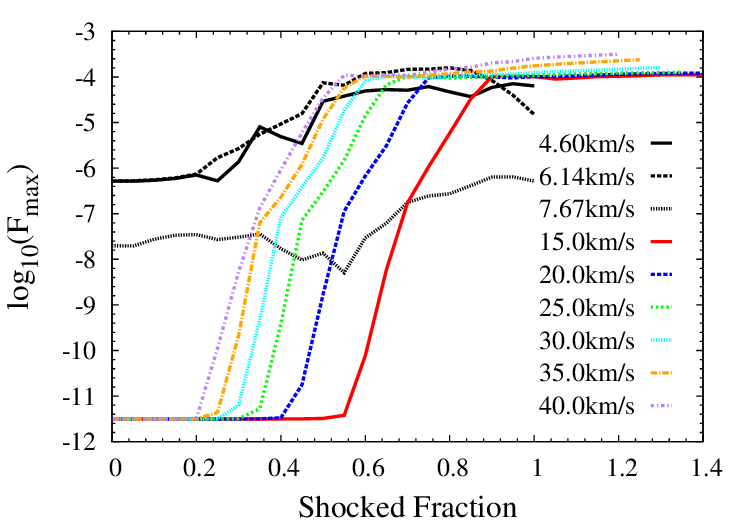}
  \caption{Light curves for nine different shock speeds, as marked in the key. Black lines
correspond to hydrodynamic models (Models~1,2 and 3 in order of increasing speed) and other 
colours, to the C-type shocks of all speeds from Model~5. The $x$-axis
is the fraction of the $z$-axis extent of the cloud traversed by the shock (or the H$_2$O
abundance front in the C-type cases) and the $y$-axis
is the base-10 logarithm of the flux density.}
\label{f:velfig}
\end{figure}

We plot light curves for the nine shock speeds in Fig.~\ref{f:velfig}.
For the C-type shocks, the general trend is for the flux density achieved at the peak of the flare to rise with
shock velocity, and therefore with the velocity shift present in the cloud. However, the
final flux densities achieved differ by only a factor of $\sim$ 3 between velocities of $15$ and
$40$\,km\,s$^{-1}$. As expected, the
rise towards the peak is progressively more delayed as the shock speed is reduced, and the
transition zone becomes wider. There appears to be a significant
secondary effect in that the C-type light curves are smoother than those of the hydrodynamic
models. The hydrodynamic models also tend to have light curves that peak at a lower shocked fraction
than 1.0 at the higher shock speeds, so that there is already a significant decay in flux density 
of Models~2 and 3 by the time the shock passage through the cloud is complete. The
shocked fractions yielding the highest flux density in Models~2 and 3 are typically 80-90 per cent
(see Section~\ref{ss:res_density}). The
optimum amount of overall compression for a bright flare appears to be somewhere between 16 (Model~2) and
25 (Model~3).

\begin{figure*}
  \includegraphics[scale=0.9,angle=0]{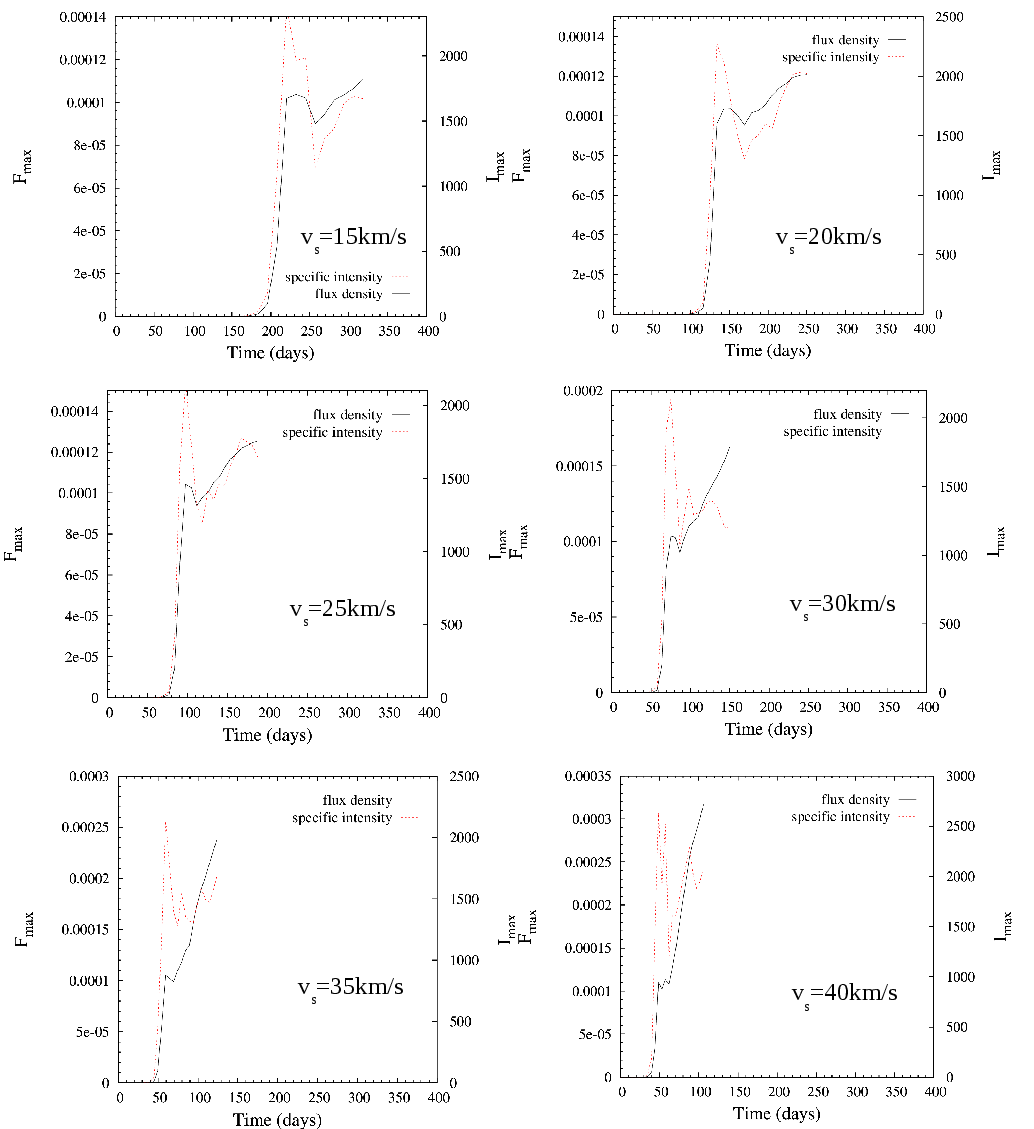}
  \caption{A sequence of flare light curves, showing the effect of increasing shock
speed, as marked for each sub-figure. All these speeds conform to the labelling
of sub-models from Table~\ref{t:shockmods}. The H$_2$ number density in
cloud is 10$^8$\,cm$^{-3}$ in all cases, and the viewpoint ($\theta=\upi/2$,$\phi=0.0$) is
as in Fig.~\ref{f:low_models}. All sub-figures are shown over the same time
range of 400\,d for comparison.}
\label{f:speed_models}
\end{figure*}

As a complement to Fig.~\ref{f:velfig}, we plot light curves for Model~5 at all shock speeds
on a linear scale in Fig.~\ref{f:speed_models}, with a standard time range of 400\,d, so
that the shortening of the shock duration with rising shock speed can
be appreciated. In these light-curves, we have not followed the model evolution through the decays which would be expected
to be of  duration similar to the momentum-transfer time, due to
a combination of compression quenching and cooling of the cloud.

\section{Discussion}
\label{discuss}

The question to be discussed is, 'how good is a shock mechanism for generating flares?'. The
answer is `very good', provided that there is no requirement of periodicity, in the sense
of single clouds, rather than whole source regions, and that the
transition concerned has a pumping scheme that is dominated by collisions. Variability indices
can reach the order of millions for shocks that approximate to ideal hydrodynamic, isothermal
types, but hundreds to thousands are probably more typical given the work 
presented in Section~\ref{ss:res_density}.
Rise timescales of order 30\,d are achievable for au-radius clouds, but the overall flare, particularly
in the case of C-type shocks, with a long momentum transfer time,
can take an order of magnitude longer, and the behaviour after the shock passage is complete,
including the time for relaxation, is not addressed in any detail by the current models.
We consider our results for achievable flux density and variability index to be somewhat
conservative, since the parameter $\Gamma$ in equation (\ref{eq:isat}) has been considered a constant,
but is likely to increase somewhat in the post-shock gas because it contains density-dependent
collision rates. Any increase in $\Gamma$ raises the saturation intensity, allowing higher
intensities to develop before the onset of saturation.

Much of the work
reported above has concentrated on au-scale clouds, and a single one of these, at kpc distance,
can produce a flare of order 100\,Jy under the best conditions of viewing angle and shock
properties (peak flux density of Model~0 from Fig.~\ref{f:fluxandepth} and equation (\ref{eq:fkpc})). 
The more realistic models are somewhat less effective, largely because of the
quenching effect on the maser inversion of very high densities. An important point that enhances
the potential of shocked clouds to explain very large flares is the result that the bright
part of a shocked cloud appears substantially smaller than its pre-shock counterpart and, given
the limited dynamic range of telescopes, a measured, single-channel, spot size of order 1\,au may imply that
the original cloud had a radius of 25-40\,au. With this sort of logic, an observed 1\,au
feature could flare to some thousands of Jy at kpc distances. However, there are limits
to this consideration of `invisible' pre-shock material imposed by the need to keep timescales
short enough. In terms of sensitivity to the viewing angle, C-type shocks are considerably
more visible because the cloud is not significantly compressed during the rise time of
the flare, and there is consequently less restriction to lines of sight close to parallel
with the shock front.

Flares of the highest flux density, for a given cloud size, require a density enhancement in the post-shock
gas that places the number density of the maser species close to the quenching density
(for o-H$_2$O parameters, close to the peak of one of the curves in Fig.~\ref{f:tauvn},
or Fig.~\ref{f:dense_taumvn}). Pre-shock conditions that include
a number density well to the low side of the quenching density, and therefore to a high
shock speed (in hydrodynamic models) with a pre-shocked cloud of low, probably unsaturated, maser depth
are also pre-requisites for a very high variability index. This
view is supported by the observation that almost all strongly saturated nodes (those
with remaining inversion $<$0.9) are found in the shocked part of the domain (see
Fig.~\ref{f:histogs}). This saturation distribution follows from the fact that only
the shocked gas can contain rays that pass diametrically through material that is all
at the high, post-shock density. C-type models have no difficulty in this respect, since the pre-shock
material has a very low value of $\tau_{\rmn M}$ (or inversion), and the variability index is consequently very high.
It is quite possible that an additional mechanism is required to explain the very
brightest H$_2$O masers flares, for example line-of-sight overlap between clouds for the
140\,kJy outburst in W51 \citep{2023ApJ...955...10V}.

Models~4-6, based on MHD shocks, lead to flares with rise times of shorter duration
than pure hydrodynamic sytems (see Fig.~\ref{f:speed_models}). This is partly due to the fairly obvious point that an
MHD shock needs to be faster than a hydrodynamic type to achieve the same ultimate compression
factor. However, this is offset to some extent by the delay introduced by a transition zone in which
the abundance of H$_2$O rises towards its final value. A much longer delay in decay, that is necessary
for the momentum coupling of the ionic and neutral fluids, may make the overall flare from
a C-type shock at least as long as that from a significantly slower hydrodynamic type with
prompt compression. Decay times in the C-type case are difficult to predict with the current
version of the code, and may include contributions from cooling, compression beyond the quenching
limit for the inversion and dispersion of the final, compressed cloud on a timescale based on the 
post-shock Alfv\'{e}n speed, which is considerably larger than the isothermal sound speed for Models~0-3.
The fastest reasonable C-shock speeds can probably produce a flare that is complete within
$\sim$800\,d for an original cloud of radius 1\,au, at least an order of magnitude longer than the
rapid initial rise times in Fig.~\ref{f:speed_models}.
It is apparent from Fig~\ref{f:low_models} and Fig~\ref{f:high_models} that the rise time of
flares is rather insensitive to the optical depth of the medium. However, increasing saturation
smooths the light curve during the brightest part of the flare, where it has become
comparable to the peak value after a rise time of order a few $\times 10$\,d.
Overall, the form of the light curve from a
flare is an important observational diagnostic.

The present work has concentrated on deriving the parameters of maser flares that can
be generated by the impact of a shockwave on an individual cloud or condensation. It is
perhaps interesting to speculate on how many such objects might be able to contribute to
a flare in a real source. An estimate of 100 `hot-spots' for H$_2$O masers in a typical
star-forming region source was made by \citet{2007ASPC..365..196S}, some or all of which
might take part in a maser flare. Such a number makes it quite easy to achieve flux
densities of thousands of Jy from groups of au-scale clouds at kpc distances. Groups
of maser, or potential maser, clouds also allow us to consider a shock mechanism as
a source of periodic or quasi-periodic maser flares: if a periodic event generates
a sequence of shockwaves, then even if shock passage destroys individual clouds, a new
group may be excited and compressed by the next shock, leading to periodicity in a
statistical sense.

Beaming solid angles of the size derived in Section~\ref{ss:res_viewpoint} certainly make
it possible to achieve the very high brightness temperatures discussed in Section~\ref{ss:samples}.
Also, measured azimuthal beaming angles are potentially useful for determining the saturation
state of a shock-driven maser flare, particularly if there is an independent measurement
of the shock speed, which likely controls the asymmetry of the beaming pattern. Given that
the models discussed in Section~\ref{ss:res_viewpoint} always show a signifcant asymmetry,
then for the same beaming solid angle, a shocked cloud will always exhibit an azimuthal
(polar) beaming angle that is larger (smaller) than the beaming angle derived for a spherical
cloud. It is not clear how this beam structure, based on a model approximating to
 the `short cylinder' shocked
slab in Fig.~\ref{f:viewpoint}, relates to the long filamentary cylinder approximations (length
to diameter ratio 5-50) described in \citet{1989ApJ...346..983E}. Perhaps in the most
general terms we can consider the filament diameter to correspond to the thickness of the
compressed layer in our hydrodynamic  models, and the filament length to the original cloud diameter.
This then leads to aspect ratios (length/diameter) of 4-16. These figures are consistent
with the lower part of the range in \citet{1989ApJ...346..983E}. The upper range is perhaps
not accessible because of our lower shock speeds.

\begin{figure}
  \includegraphics[bb=110 40 410 292, scale=1.0,angle=0]{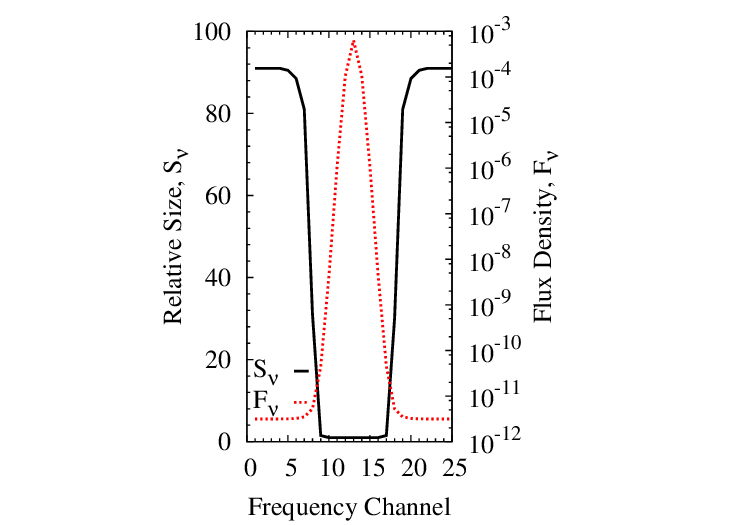}
  \caption{Apparent area of the maser cloud at the half flux-density level as a
function of frequency (black line), and the logarithmic spectrum (red line) for
comparison of widths. The area scale sets the minimum value to 1.0, and the
line centre of this figure
corresponds to the final panel of Fig.~\ref{f:frames}: Model~0, initial depth 3.0
and compressed fraction 1.0. The 25 frequency channels shown cover 7 Doppler widths.
The viewpoint is $(\theta,\phi) = (\upi/2,0.0)$.}
\label{f:sizeplot}
\end{figure}

In terms of the definitions used in \citet{2011A&A...525A..56R}, we expect the shock-generated
masers from our hydrodynamic models to be `matter bounded' \citep{1992ApJ...394..221E} in the polar direction, so that we should see
something close to the full thickness of the compressed region of the cloud. We expect
amplification bounded emission in the azimuthal direction because of the curvature of the
surface in this direction. C-type models are closer to amplification bounded in both directions,
but still have a substantial brightness bias to polar angles near $\upi/2$.
Beaming-angle asymmetry yields a relatively large beaming angle in the azimuthal  direction, so that 
the predicted extension in this direction is larger than would be expected for a circular beaming 
pattern. We therefore expect images of shocked clouds to be a better approximation to 
their true size  when measured along the minor axis, that is perpendicular
to the shock front, compared with
spherical clouds of constant velocity, where the maser image at half flux density
is significantly smaller than the actual cloud radius
\citep{1992ApJ...394..221E,2018MNRAS.477.2628G}. This expectation appears to be substantially
borne out in the images in Fig.~\ref{f:frames}. 
As a further check, \citet{2011A&A...525A..56R}
predict that the observed size of a shock generated, matter-bounded, maser should be rather insensitive to the
frequency channel at which the observation is made, but that an amplification-bounded maser should
appear much larger in the line wings, for example 69 times larger comparing the large and small box
areas in their Fig.~9 (lower panel). When we plot object area, at the half flux-density level, relative
to the value at line centre in Fig.~\ref{f:sizeplot}, we see a broad minimum, covering about 2.3
Doppler widths. We suggest this region is dominated by the brightest, matter-bounded emission, before
the very faint azimuthally-extended emission, that is amplification bounded, produces much larger
areas in the extreme line wings. However, this increase in area would not be detectable to a telescope
with a dynamic range of $\lesssim 10^6$, as can be seen from the overplotted spectrum in Fig.~\ref{f:sizeplot}.

\subsection{Relevance to Accretion Sources}
\label{ss:rel_accn}

It is instructive to discuss how our models of shock-excited maser flares from individual au-scale
clouds fits into the broader picture of a global source geometry, particularly a source of the classic
type combining an accretion disc and bipolar outflow. A source of this type that has been observed in
particular detail is IRAS21078+5211, in which 22-GHz H$_2$O masers exhibit outflow-oriented proper-motions,
and have been analysed as having a disc-wind origin \citep{2022NatAs...6.1068M}. Scatter in $v_{\rmn{LSR}}$
of several km\,s$^{-1}$ is said to be inconsistent with much smaller positional scatter of VLBI
maser features, and therefore
evidence for excitation by weak C-type shocks travelling through the maser clouds. In this source, we
identify our maser clouds with the observational radio knots in \citet{2022NatAs...6.1068M}: these are
clearly resolved into small clusters of distinct objects at the 0.7\,milliarcsec resolution, giving
a linear scale of 1.14\,au at 1630\,pc that is consistent with our model cloud scale. Physically, the
maser clouds would be regions of enhanced density in the MHD disc wind resulting from variations in
the physical conditions near the disc surface. A variability timescale of order 1 month
\citep{2022NatAs...6.1068M,2018ARep...62..200K} related to individal spectral features is consistent
with, for example, our Model~4 (top panel, Figure~\ref{f:high_models}) in which more than half the maximum flux density, and
the brightest specific intensity are achieved within 20\,d of flare onset. The shock speed in our
figure is 35\,km\,s$^{-1}$, though two-thirds of this speed, 23\,km\,s$^{-1}$, would still yield a rise time of 30\,d.
which is also consistent with month-scale variability. These modelled speeds are rather difficult to
match to any precise values from observations: \citet{2018ARep...62..200K} suggest $\sim$15\,km\,s$^{-1}$, whilst fitted
flow velocities, of which the shock velocity is presumably a significant fraction, in Table~2 of
\citet{2022NatAs...6.1068M} range from 9 to 85\,km\,s$^{-1}$.

The Introduction to the present work refers to a couple of accretion-burst sources. It is not obvious that a shock-based
variability mechanism should apply in such objects, since the main result of an accretion event is
a burst of radiation, often powerful in the IR, that favours radiatively-pumped maser transitions.
A case in point is the MM1 source in the massive star-forming region G358.93-0.03, where the progress
of the `heat wave', or radiation burst, was tracked across the nearly face-on accretion disc via its
effect on 6.7-GHz CH$_3$OH masers \citep{2023NatAs...7..557B}. A clear link between spiral sub-structures
in the accretion disc of MM1 and accretion bursts has been demonstrated by \citet{2020NatAs...4.1170C}.
Several radiatively-pumped maser species, particularly 3 newly discovered transitions, traced the
spiral sub-structure in the disc, and flared following the IR radiation enhancement from the accretion
burst. However, the 22-GHz H$_2$O masers in this source are more mysterious, because of a lack of
interferometer data during the H$_2$O flare, which was delayed by $\sim$100\,d with respect to 6.7-GHz
CH$_3$OH. There are two e-VLA images of G358.93-0.03 before and after the H$_2$O maser flare, showing
drastic change in the source structure \citep{2022A&A...664A..44B}. Whilst the 22-GHz masers in MM1
itself are significantly altered, spectrally and spatially, the distribution is broadly aligned with
the outflow in the before and after images. However, maser clusters in the more distant sources MM2, MM4
and MM5, present in the `before' image became undetectable, and a new source appears in the `after'
image between MM4 and MM5, though probably associated with the latter. \citet{2022A&A...664A..44B}
specluate that the H$_2$O maser flare is associated with this new source, noting that the vacuum
light-travel time from MM1 to the new source (in the sky plane) of 78\,d is comparable to the 100\,d
delay in the maser flares. If this association is correct, we offer two possibilities for the generation
of the new H$_2$O maser cluster. The first is that the radiation from the accretion event in MM1 has
a spectral energy distribution (SED) that is IR-dominated, and very similar to those in
\citet{2021A&A...646A.161S}. This radiation impinges on gas containing very dense ($\sim10^{10}$\,H$_2$
\,cm$^{-3}$) and cold ($<$50\,K) water molecules. Under these conditions, a {\em radiative} pumping
scheme can operate for the 22-GHz transition \citep{2022MNRAS.513.1354G}. This possibility is 
attractive in that the IR burst is destructive for the existing masers in hotter, more rarefied
conditions, but can explain the rapid rise of the flare 
(over 20-30\,d, \citealt{2022A&A...664A..44B}) in a previously maser-free site.
  The second possibility is that the dust shrouding of the young stellar object in MM1 is patchy,
allowing radiation approximating to its raw SED to escape along certain lines of sight. Assuming most radiation
with $\lambda < 111$\,nm, is lost in dissociating or ionizing hydrogen, there is still a band $145-111$\,nm
that can heat gas efficiently due to photoelectric emission from grains. The
9 best burst models in \citet{2021A&A...646A.161S} (their Table~A6) have a mean stellar temperature, $T_*$ of 22\,570\,K and
mean radius of 10.31\,R$_{\sun}$. From formulae in Section~30.2 of \citet{2011piim.book.....D}
such photoelectric emission
deposits energy at an approximate  rate of $9.28\times 10^{-27} n_{\rmn H} (n_\gamma /3\times 10^{-3} \rmn{cm^{-3}})$\,erg\,cm$^{-3}$\,s$^{-1}$,
where $n_\gamma$ is the number density of the UV photons and $n_{\rmn H}$ is the number density of H-nuclei. The
radiation field, at $T_*$, diluted to a distance of $13\,500$\,au, has $n_\gamma \simeq 270$\,cm$^{-3}$ of
photons in the 145-111\,nm band, so for a number density $n_{\rmn H}=2\times 10^8$\,cm$^{-3}$, typical of the pre-shock
gas in the C-type models used in the present work, the photoelectric heating rate is $1.64\times 10^{-14}$\,J\,m$^{-3}$\,s$^{-1}$.
An increase, $\Delta T$, in the gas thermal energy due to this heating rate, $\Gamma_{\rmn pe}$, follows
$\Delta T= 2\Gamma_{\rmn pe} t/(3 k_{\rmn B} n_{\rmn H_2})$ after $t$ seconds. From this formula, a temperature
increase of 800\,K requires only a little over 100\,s. There remains the question of how quickly this
energy can be transferred from the photoelectrons to the H$_2$ gas. For electron knietic energies of
a few eV, the elastic collision cross-section for collisions with H$_2$ are typically $>$10$^{-19}$\,m$^2$
\citep{2008JPCRD..37..913Y}, so the mean-free time for the electrons is $1.7\times 10^5/E_{\rm eV}^{1/2}$\,s. Energy can therefore
be transferred from the radiation field to the H$_2$ gas within a few days. The rate of energy deposition
will, of course, fall as the radiation progresses into the gas. We use an extinction cross section
due to dust of $8.16\times 10^{-26}$\,m$^2$ at 100\,nm to calculate an optical depth at this wavelength of
$\tau_{1000}=2.441 L_{\rmn{AU}}$, so the energy is mostly deposited a layer of astronomical-unit scale, linking
ultraviolet extinction by dust to the scale of model clouds studied in the present work.
Since the time required for heating is much less than the sound crossing time of the heated region, the heated gas
will become hot and overpressured with respect to its surroudings. Under these conditions it is
very likely that shock expansion will follow into cooler gas further from the radiation source.
In this way, an initial radiative event caused by accretion can ultimately drive a collision-based
pumping scheme after a delay little longer than the light travel time to the new maser zone. 

\section{Conclusions}
\label{conclusion}

Shock-wave modification of gaseous clouds of scale a few au to a few tens of au provides a
mechanism that can explain maser flares with the largest observed brightness temperatures
($\gtrsim 10^{17}$\,K). The mechanism is limited to transitions with a mainly collisional
pumping mechanism, and the output is highly directional, dominated by directions parallel
to the shock front, but this effect is less pronounced for C-type shocks. 
Magnetically dominated (C-type) shocks can generate masers with similar flux
densities to the ideal hydrodynamic variety, and the magnetic types typically produce
flares with an extremely large variability index, since these models have a minimal
inversion in the 22-GHz transition of the maser species (ortho-H$_2$O) in the pre-shock fluid. C-type shocks also produce 
flares of shorter duration, particular during the rise to peak, because faster shocks are 
required to achieve the same (ultimate) compression.
Density of the maser species in the post-shock gas is important, but only in the sense that it needs to be
in a `sensible' range that is sufficient to achieve saturation, but low enough to avoid
quenching the pump. Maser beaming pattens show a strong asymmetry, being considerably
more extended parallel to the shock front. We find consistency with earlier work that
suggests that the observed image, perpendicular to the shock front, in a shock-generated maser 
is much closer to the true size of the supporting cloud than in the quiescent spherical case.

\section*{Acknowledgments}

MDG and SE acknowledge funding from the UK Science and Technology Facilities
Council (STFC) as part of the consolidated grant ST/P000649/1 to the Jodrell Bank
Centre for Astrophysics at the University of Manchester. MDG acknowledges
financial support from the National Astronomical Research Institute of Thailand (NARIT)
whilst on sabbatical at their HQ in Chiang Mai, Thailand.
This work was performed, in part, using the DiRAC Data Intensive service at Leicester, operated 
by the University of Leicester IT Services, which forms part of the STFC 
DiRAC HPC Facility (www.dirac.ac.uk). The equipment was funded by BEIS capital 
funding via STFC capital grants ST/K000373/1 and ST/R002363/1 and 
STFC DiRAC Operations grant ST/R001014/1. DiRAC is part of the National e-Infrastructure.
Data used in this work was generated under DiRAC award dp124. The authors would
also like to thank the management of the Chalawan supercomputer at NARIT.

\section*{Data Availability}

The data underlying this article will be shared on reasonable re-
quest to the corresponding author.

\bibliographystyle{mn2e}

\begin{thebibliography}{}

\bibitem[\protect\citeauthoryear{{Abramowitz} \& {Stegun}}{{Abramowitz} \&
  {Stegun}}{1972}]{absteg}
{Abramowitz} M.,  {Stegun} I.~A.,  1972, {Handbook of Mathematical Functions}

\bibitem[\protect\citeauthoryear{{Araya}, {Hofner}, {Goss}, {Kurtz},
  {Richards}, {Linz}, {Olmi} \& {Sewi{\l}o}}{{Araya}
  et~al.}{2010}]{2010ApJ...717L.133A}
{Araya} E.~D.,  {Hofner} P.,  {Goss} W.~M.,  {Kurtz} S.,  {Richards} A.~M.~S.,
  {Linz} H.,  {Olmi} L.,    {Sewi{\l}o} M.,  2010, \apjl, 717, L133

\bibitem[\protect\citeauthoryear{{Asanok}, {Etoka}, {Gray}, {Thomasson},
  {Richards} \& {Kramer}}{{Asanok} et~al.}{2010}]{2010MNRAS.404..120A}
{Asanok} K.,  {Etoka} S.,  {Gray} M.~D.,  {Thomasson} P.,  {Richards} A.~M.~S.,
     {Kramer} B.~H.,  2010, \mnras, 404, 120

\bibitem[\protect\citeauthoryear{{Ashimbaeva}, {Colom}, {Krasnov}, {Lekht},
  {Pashchenko}, {Rudnitskii} \& {Tolmachev}}{{Ashimbaeva}
  et~al.}{2020a}]{2020ARep...64..586A}
{Ashimbaeva} N.~T.,  {Colom} P.,  {Krasnov} V.~V.,  {Lekht} E.~E.,
  {Pashchenko} M.~I.,  {Rudnitskii} G.~M.,    {Tolmachev} A.~M.,  2020a,
  Astronomy Reports, 64, 586

\bibitem[\protect\citeauthoryear{{Ashimbaeva}, {Colom}, {Krasnov}, {Lekht},
  {Pashchenko}, {Rudnitskii} \& {Tolmachev}}{{Ashimbaeva}
  et~al.}{2020b}]{2020ARep...64..839A}
{Ashimbaeva} N.~T.,  {Colom} P.,  {Krasnov} V.~V.,  {Lekht} E.~E.,
  {Pashchenko} M.~I.,  {Rudnitskii} G.~M.,    {Tolmachev} A.~M.,  2020b,
  Astronomy Reports, 64, 839

\bibitem[\protect\citeauthoryear{{Ashimbaeva}, {Colom}, {Lekht}, {Pashchenko},
  {Rudnitskii} \& {Tolmachev}}{{Ashimbaeva} et~al.}{2016}]{2016AstL...42..652A}
{Ashimbaeva} N.~T.,  {Colom} P.,  {Lekht} E.~E.,  {Pashchenko} M.~I.,
  {Rudnitskii} G.~M.,    {Tolmachev} A.~M.,  2016, Astronomy Letters, 42, 652

\bibitem[\protect\citeauthoryear{{Ashimbaeva}, {Colom}, {Lekht}, {Pashchenko},
  {Rudnitskii} \& {Tolmachev}}{{Ashimbaeva} et~al.}{2018}]{2018ARep...62..609A}
{Ashimbaeva} N.~T.,  {Colom} P.,  {Lekht} E.~E.,  {Pashchenko} M.~I.,
  {Rudnitskii} G.~M.,    {Tolmachev} A.~M.,  2018, Astronomy Reports, 62, 609

\bibitem[\protect\citeauthoryear{{Ashimbaeva}, {Krasnov}, {Lekht},
  {Pashchenko}, {Rudnitskii} \& {Tolmachev}}{{Ashimbaeva}
  et~al.}{2020}]{2020ARep...64...15A}
{Ashimbaeva} N.~T.,  {Krasnov} V.~V.,  {Lekht} E.~E.,  {Pashchenko} M.~I.,
  {Rudnitskii} G.~M.,    {Tolmachev} A.~M.,  2020, Astronomy Reports, 64, 15

\bibitem[\protect\citeauthoryear{{Barlow}, {Nguyen-Q-Rieu} \& et al.}{{Barlow}
  et~al.}{1996}]{1996A&A...315L.241B}
{Barlow} M.~J.,  {Nguyen-Q-Rieu}   et al. T.,  1996, \aap, 315, L241

\bibitem[\protect\citeauthoryear{{Bayandina}, {Brogan}, {Burns}, {Caratti o
  Garatti}, {Chibueze}, {van den Heever}, {Kurtz}, {MacLeod}, {Moscadelli},
  {Sobolev}, {} \& {\em et al.}}{{Bayandina}
  et~al.}{2022}]{2022A&A...664A..44B}
{Bayandina} O.~S.,  {Brogan} C.~L.,  {Burns} R.~A.,  {Caratti o Garatti} A.,
  {Chibueze} J.~O.,  {van den Heever} S.~P.,  {Kurtz} S.~E.,  {MacLeod} G.~C.,
  {Moscadelli} L.,  {Sobolev} A.~M.,  {}   {\em et al.} 2022, \aap, 664, A44

\bibitem[\protect\citeauthoryear{{Bayandina}, {Burns}, {Kurtz},
  {Shakhvorostova} \& {Val'tts}}{{Bayandina}
  et~al.}{2019}]{2019ApJ...884..140B}
{Bayandina} O.~S.,  {Burns} R.~A.,  {Kurtz} S.~E.,  {Shakhvorostova} N.~N.,
  {Val'tts} I.~E.,  2019, \apj, 884, 140

\bibitem[\protect\citeauthoryear{{Bergman} \& {Humphreys}}{{Bergman} \&
  {Humphreys}}{2020}]{2020A&A...638A..19B}
{Bergman} P.,  {Humphreys} E.~M.~L.,  2020, \aap, 638, A19

\bibitem[\protect\citeauthoryear{{Burns}, {Handa}, {Nagayama}, {Sunada} \&
  {Omodaka}}{{Burns} et~al.}{2016}]{2016MNRAS.460..283B}
{Burns} R.~A.,  {Handa} T.,  {Nagayama} T.,  {Sunada} K.,    {Omodaka} T.,
  2016, \mnras, 460, 283

\bibitem[\protect\citeauthoryear{{Burns}, {Uno}, {Sakai}, {Blanchard}, {Rosli},
  {Orosz}, {Yonekura}, {Tanabe}, {Sugiyama}, {Hirota}, {} \& {\em et
  al.}}{{Burns} et~al.}{2023}]{2023NatAs...7..557B}
{Burns} R.~A.,  {Uno} Y.,  {Sakai} N.,  {Blanchard} J.,  {Rosli} Z.,  {Orosz}
  G.,  {Yonekura} Y.,  {Tanabe} Y.,  {Sugiyama} K.,  {Hirota} T.,  {}   {\em et
  al.} 2023, Nature Astronomy, 7, 557

\bibitem[\protect\citeauthoryear{{Chen}, {Sobolev}, {Ren}, {Parfenov}, {Breen},
  {Ellingsen}, {Shen}, {Li}, {MacLeod}, {Baan}, {} \& {\em et al.}}{{Chen}
  et~al.}{2020}]{2020NatAs...4.1170C}
{Chen} X.,  {Sobolev} A.~M.,  {Ren} Z.-Y.,  {Parfenov} S.,  {Breen} S.~L.,
  {Ellingsen} S.~P.,  {Shen} Z.-Q.,  {Li} B.,  {MacLeod} G.~C.,  {Baan} W.,  {}
    {\em et al.} 2020, Nature Astronomy, 4, 1170

\bibitem[\protect\citeauthoryear{{Chen} \& {Cai}}{{Chen} \&
  {Cai}}{2001}]{2001ApMCo.124..351C}
{Chen} Y.,  {Cai} D.,  2001, App. Maths. \& Computation, 124, 351

\bibitem[\protect\citeauthoryear{{Chibueze}, {MacLeod}, {Vorster}, {Hirota},
  {Brogan}, {Hunter} \& {van Rooyen}}{{Chibueze}
  et~al.}{2021}]{2021ApJ...908..175C}
{Chibueze} J.~O.,  {MacLeod} G.~C.,  {Vorster} J.~M.,  {Hirota} T.,  {Brogan}
  C.~L.,  {Hunter} T.~R.,    {van Rooyen} R.,  2021, \apj, 908, 175

\bibitem[\protect\citeauthoryear{{Colom}, {Ashimbaeva}, {Lekht}, {Pashchenko},
  {Rudnitskii}, {Krasnov} \& {Tolmachev}}{{Colom}
  et~al.}{2019}]{2019ARep...63..814C}
{Colom} P.,  {Ashimbaeva} N.~T.,  {Lekht} E.~E.,  {Pashchenko} M.~I.,
  {Rudnitskii} G.~M.,  {Krasnov} V.~V.,    {Tolmachev} A.~M.,  2019, Astronomy
  Reports, 63, 814

\bibitem[\protect\citeauthoryear{{Colom}, {Ashimbaeva}, {Lekht}, {Pashchenko},
  {Rudnitskij}, {Krasnov} \& {Tolmachev}}{{Colom}
  et~al.}{2021}]{2021MNRAS.507.3285C}
{Colom} P.,  {Ashimbaeva} N.~T.,  {Lekht} E.~E.,  {Pashchenko} M.~I.,
  {Rudnitskij} G.~M.,  {Krasnov} V.~V.,    {Tolmachev} A.~M.,  2021, \mnras,
  507, 3285

\bibitem[\protect\citeauthoryear{{Colom}, {Lekht}, {Pashchenko}, {Rudnitskii}
  \& {Tolmachev}}{{Colom} et~al.}{2016}]{2016ARep...60..730C}
{Colom} P.,  {Lekht} E.~E.,  {Pashchenko} M.~I.,  {Rudnitskii} G.~M.,
  {Tolmachev} A.~M.,  2016, Astronomy Reports, 60, 730

\bibitem[\protect\citeauthoryear{{Coppola}, {Mizzi}, {Bruno}, {Esposito},
  {Galli}, {Palla} \& {Longo}}{{Coppola} et~al.}{2016}]{2016MNRAS.457.3732C}
{Coppola} C.~M.,  {Mizzi} G.,  {Bruno} D.,  {Esposito} F.,  {Galli} D.,
  {Palla} F.,    {Longo} S.,  2016, \mnras, 457, 3732

\bibitem[\protect\citeauthoryear{{Cragg}, {Johns}, {Godfrey} \&
  {Brown}}{{Cragg} et~al.}{1992}]{1992MNRAS.259..203C}
{Cragg} D.~M.,  {Johns} K.~P.,  {Godfrey} P.~D.,    {Brown} R.~D.,  1992,
  \mnras, 259, 203

\bibitem[\protect\citeauthoryear{{De Ceuster}, {Ceulemans}, {Srivastava},
  {Homan}, {Bolte}, {Yates}, {Decin}, {Boyle} \& {Hetherington}}{{De Ceuster}
  et~al.}{2022}]{2022JOSS....7.3905D}
{De Ceuster} F.,  {Ceulemans} T.,  {Srivastava} A.,  {Homan} W.,  {Bolte} J.,
  {Yates} J.,  {Decin} L.,  {Boyle} P.,    {Hetherington} J.,  2022, The
  Journal of Open Source Software, 7, 3905

\bibitem[\protect\citeauthoryear{{De Ceuster}, {Homan}, {Yates}, {Decin},
  {Boyle} \& {Hetherington}}{{De Ceuster} et~al.}{2020}]{2020MNRAS.492.1812D}
{De Ceuster} F.,  {Homan} W.,  {Yates} J.,  {Decin} L.,  {Boyle} P.,
  {Hetherington} J.,  2020, \mnras, 492, 1812

\bibitem[\protect\citeauthoryear{{de Jong}}{{de
  Jong}}{1973}]{1973A&A....26..297D}
{de Jong} T.,  1973, \aap, 26, 297

\bibitem[\protect\citeauthoryear{{Deguchi} \& {Watson}}{{Deguchi} \&
  {Watson}}{1989}]{1989ApJ...340L..17D}
{Deguchi} S.,  {Watson} W.~D.,  1989, \apjl, 340, L17

\bibitem[\protect\citeauthoryear{{Draine}}{{Draine}}{2011}]{2011piim.book.....D}
{Draine} B.~T.,  2011, {Physics of the Interstellar and Intergalactic Medium}

\bibitem[\protect\citeauthoryear{{Elitzur} \& {Asensio Ramos}}{{Elitzur} \&
  {Asensio Ramos}}{2006}]{2006MNRAS.365..779E}
{Elitzur} M.,  {Asensio Ramos} A.,  2006, \mnras, 365, 779

\bibitem[\protect\citeauthoryear{{Elitzur}, {Hollenbach} \& {McKee}}{{Elitzur}
  et~al.}{1989}]{1989ApJ...346..983E}
{Elitzur} M.,  {Hollenbach} D.~J.,    {McKee} C.~F.,  1989, \apj, 346, 983

\bibitem[\protect\citeauthoryear{{Elitzur}, {Hollenbach} \& {McKee}}{{Elitzur}
  et~al.}{1992}]{1992ApJ...394..221E}
{Elitzur} M.,  {Hollenbach} D.~J.,    {McKee} C.~F.,  1992, \apj, 394, 221

\bibitem[\protect\citeauthoryear{{Elitzur}, {McKee} \& {Hollenbach}}{{Elitzur}
  et~al.}{1991}]{1991ApJ...367..333E}
{Elitzur} M.,  {McKee} C.~F.,    {Hollenbach} D.~J.,  1991, \apj, 367, 333

\bibitem[\protect\citeauthoryear{{Faure} \& {Josselin}}{{Faure} \&
  {Josselin}}{2008}]{2008A&A...492..257F}
{Faure} A.,  {Josselin} E.,  2008, \aap, 492, 257

\bibitem[\protect\citeauthoryear{{Field}, {Gray} \& {de St.~Paer}}{{Field}
  et~al.}{1994}]{1994A&A...282..213F}
{Field} D.,  {Gray} M.~D.,    {de St.~Paer} P.,  1994, \aap, 282, 213

\bibitem[\protect\citeauthoryear{{Frisch}}{{Frisch}}{1988}]{1988rmgm.conf..339F}
{Frisch} H.,  1988, in Saas-Fee Advanced Course 18: Radiation in Moving Gaseous
  Media {Radiative Transfer with Frequency Redistribution - Asymptotic methods
  scaling laws and approximate solutions}.
p.~339

\bibitem[\protect\citeauthoryear{{Goedhart}, {Gaylard} \& {van der
  Walt}}{{Goedhart} et~al.}{2004}]{2004MNRAS.355..553G}
{Goedhart} S.,  {Gaylard} M.~J.,    {van der Walt} D.~J.,  2004, \mnras, 355,
  553

\bibitem[\protect\citeauthoryear{{Goldreich} \& {Kwan}}{{Goldreich} \&
  {Kwan}}{1974}]{1974ApJ...190...27G}
{Goldreich} P.,  {Kwan} J.,  1974, \apj, 190, 27

\bibitem[\protect\citeauthoryear{{G{\'o}rski}, {Hivon}, {Banday}, {Wandelt},
  {Hansen}, {Reinecke} \& {Bartelmann}}{{G{\'o}rski}
  et~al.}{2005}]{2005ApJ...622..759G}
{G{\'o}rski} K.~M.,  {Hivon} E.,  {Banday} A.~J.,  {Wandelt} B.~D.,  {Hansen}
  F.~K.,  {Reinecke} M.,    {Bartelmann} M.,  2005, \apj, 622, 759

\bibitem[\protect\citeauthoryear{{Gray}}{{Gray}}{2012}]{mybook}
{Gray} M.~D.,  2012, {Maser Sources in Astrophysics}.
Cambridge University Press, Cambridge, UK

\bibitem[\protect\citeauthoryear{{Gray}, {Baggott}, {Westlake} \&
  {Etoka}}{{Gray} et~al.}{2019}]{2019MNRAS.486.4216G}
{Gray} M.~D.,  {Baggott} J.,  {Westlake} J.,    {Etoka} S.,  2019, \mnras, 486,
  4216 (Paper~2)

\bibitem[\protect\citeauthoryear{{Gray}, {Baudry}, {Richards}, {Humphreys},
  {Sobolev} \& {Yates}}{{Gray} et~al.}{2016}]{2016MNRAS.456..374G}
{Gray} M.~D.,  {Baudry} A.,  {Richards} A.~M.~S.,  {Humphreys} E.~M.~L.,
  {Sobolev} A.~M.,    {Yates} J.~A.,  2016, \mnras, 456, 374

\bibitem[\protect\citeauthoryear{{Gray}, {Etoka} \& {Pimpanuwat}}{{Gray}
  et~al.}{2020}]{2020MNRAS.498L..11G}
{Gray} M.~D.,  {Etoka} S.,    {Pimpanuwat} B.,  2020, \mnras, 498, L11

\bibitem[\protect\citeauthoryear{{Gray}, {Etoka}, {Richards} \&
  {Pimpanuwat}}{{Gray} et~al.}{2022}]{2022MNRAS.513.1354G}
{Gray} M.~D.,  {Etoka} S.,  {Richards} A.~M.~S.,    {Pimpanuwat} B.,  2022,
  \mnras, 513, 1354

\bibitem[\protect\citeauthoryear{{Gray}, {Etoka}, {Travis} \&
  {Pimpanuwat}}{{Gray} et~al.}{2020}]{2020MNRAS.493.2472G}
{Gray} M.~D.,  {Etoka} S.,  {Travis} A.,    {Pimpanuwat} B.,  2020, \mnras,
  493, 2472 (Paper~3)

\bibitem[\protect\citeauthoryear{{Gray}, {Mason} \& {Etoka}}{{Gray}
  et~al.}{2018}]{2018MNRAS.477.2628G}
{Gray} M.~D.,  {Mason} L.,    {Etoka} S.,  2018, \mnras, 477, 2628 (Paper~1)

\bibitem[\protect\citeauthoryear{{Gwinn}, {Moran} \& {Reid}}{{Gwinn}
  et~al.}{1992}]{1992ApJ...393..149G}
{Gwinn} C.~R.,  {Moran} J.~M.,    {Reid} M.~J.,  1992, \apj, 393, 149

\bibitem[\protect\citeauthoryear{{Hirota}, {Tsuboi}, {Kurono}, {Fujisawa},
  {Honma}, {Kim}, {Imai} \& {Yonekura}}{{Hirota}
  et~al.}{2014}]{2014PASJ...66..106H}
{Hirota} T.,  {Tsuboi} M.,  {Kurono} Y.,  {Fujisawa} K.,  {Honma} M.,  {Kim}
  M.~K.,  {Imai} H.,    {Yonekura} Y.,  2014, \pasj, 66, 106

\bibitem[\protect\citeauthoryear{{Hollenbach}, {Elitzur} \&
  {McKee}}{{Hollenbach} et~al.}{2013}]{2013ApJ...773...70H}
{Hollenbach} D.,  {Elitzur} M.,    {McKee} C.~F.,  2013, \apj, 773, 70

\bibitem[\protect\citeauthoryear{{Imai}, {Iwata} \& {Miyoshi}}{{Imai}
  et~al.}{1999}]{1999PASJ...51..473I}
{Imai} H.,  {Iwata} T.,    {Miyoshi} M.,  1999, \pasj, 51, 473

\bibitem[\protect\citeauthoryear{{Inayoshi}, {Sugiyama}, {Hosokawa}, {Motogi}
  \& {Tanaka}}{{Inayoshi} et~al.}{2013}]{2013ApJ...769L..20I}
{Inayoshi} K.,  {Sugiyama} K.,  {Hosokawa} T.,  {Motogi} K.,    {Tanaka} K.
  E.~I.,  2013, \apj, 769, L20

\bibitem[\protect\citeauthoryear{{Kaufman} \& {Neufeld}}{{Kaufman} \&
  {Neufeld}}{1996}]{1996ApJ...456..250K}
{Kaufman} M.~J.,  {Neufeld} D.~A.,  1996, \apj, 456, 250

\bibitem[\protect\citeauthoryear{{Kim}, {Kim}, {Kurayama}, {Honma}, {Sasao},
  {Surcis}, {Cant{\'o}}, {Torrelles} \& {Kim}}{{Kim}
  et~al.}{2013}]{2013ApJ...767...86K}
{Kim} J.-S.,  {Kim} S.-W.,  {Kurayama} T.,  {Honma} M.,  {Sasao} T.,  {Surcis}
  G.,  {Cant{\'o}} J.,  {Torrelles} J.~M.,    {Kim} S.~J.,  2013, \apj, 767, 86

\bibitem[\protect\citeauthoryear{{King} \& {Florance}}{{King} \&
  {Florance}}{1964}]{1964ApJ...139..397K}
{King} J.~I.~F.,  {Florance} E.~T.,  1964, \apj, 139, 397

\bibitem[\protect\citeauthoryear{{Kobayashi}, {Shimoikura}, {Omodaka} \&
  {Diamond}}{{Kobayashi} et~al.}{2000}]{2000aprs.conf..109K}
{Kobayashi} H.,  {Shimoikura} T.,  {Omodaka} T.,    {Diamond} P.~J.,  2000, in
  {Hirabayashi} H.,  {Edwards} P.~G.,   {Murphy} D.~W.,  eds, Astrophysical
  Phenomena Revealed by Space VLBI {Monitoring of the Orion-KL Water Maser
  Outburst}.
pp 109--112

\bibitem[\protect\citeauthoryear{{Krasnov}, {Lekht}, {Minnebaev}, {Pashchenko},
  {Rudnitskii} \& {Tolmachev}}{{Krasnov} et~al.}{2018}]{2018ARep...62..200K}
{Krasnov} V.~V.,  {Lekht} E.~E.,  {Minnebaev} V.~M.,  {Pashchenko} M.~I.,
  {Rudnitskii} G.~M.,    {Tolmachev} A.~M.,  2018, Astronomy Reports, 62, 200

\bibitem[\protect\citeauthoryear{{Krasnov}, {Lekht}, {Rudnitskii}, {Pashchenko}
  \& {Tolmachev}}{{Krasnov} et~al.}{2015}]{2015AstL...41..517K}
{Krasnov} V.~V.,  {Lekht} E.~E.,  {Rudnitskii} G.~M.,  {Pashchenko} M.~I.,
  {Tolmachev} A.~M.,  2015, Astronomy Letters, 41, 517

\bibitem[\protect\citeauthoryear{{Lees}}{{Lees}}{1973}]{1973ApJ...184..763L}
{Lees} R.~M.,  1973, \apj, 184, 763

\bibitem[\protect\citeauthoryear{{MacLeod}, {Smits}, {Goedhart}, {Hunter},
  {Brogan}, {Chibueze}, {van den Heever}, {Thesner}, {Banda} \&
  {Paulsen}}{{MacLeod} et~al.}{2018}]{2018MNRAS.478.1077M}
{MacLeod} G.~C.,  {Smits} D.~P.,  {Goedhart} S.,  {Hunter} T.~R.,  {Brogan}
  C.~L.,  {Chibueze} J.~O.,  {van den Heever} S.~P.,  {Thesner} C.~J.,  {Banda}
  P.~J.,    {Paulsen} J.~D.,  2018, \mnras, 478, 1077

\bibitem[\protect\citeauthoryear{{MacLeod}, {Smits}, {Green} \& {van den
  Heever}}{{MacLeod} et~al.}{2021}]{2021MNRAS.502.5658M}
{MacLeod} G.~C.,  {Smits} D.~P.,  {Green} J.~A.,    {van den Heever} S.~P.,
  2021, \mnras, 502, 5658

\bibitem[\protect\citeauthoryear{{McCarthy}, {Breen}, {Kaczmarek}, {Chen},
  {Parfenov}, {Sobolev}, {Ellingsen}, {Burns}, {MacLeod}, {Sugiyama}, {} \&
  {\em et al.}}{{McCarthy} et~al.}{2023}]{2023MNRAS.522.4728M}
{McCarthy} T.~P.,  {Breen} S.~L.,  {Kaczmarek} J.~F.,  {Chen} X.,  {Parfenov}
  S.,  {Sobolev} A.~M.,  {Ellingsen} S.~P.,  {Burns} R.~A.,  {MacLeod} G.~C.,
  {Sugiyama} K.,  {}   {\em et al.} 2023, \mnras, 522, 4728

\bibitem[\protect\citeauthoryear{{Melnick}, {} \& {\em et al.}}{{Melnick}
  et~al.}{2000}]{2000ApJ...539L..87M}
{Melnick} G.~J.,  {}   {\em et al.} 2000, \apjl, 539, L87

\bibitem[\protect\citeauthoryear{{Moscadelli}, {Sanna}, {Beuther}, {Oliva} \&
  {Kuiper}}{{Moscadelli} et~al.}{2022}]{2022NatAs...6.1068M}
{Moscadelli} L.,  {Sanna} A.,  {Beuther} H.,  {Oliva} A.,    {Kuiper} R.,
  2022, Nature Astronomy, 6, 1068

\bibitem[\protect\citeauthoryear{{Moscadelli}, {Sanna}, {Goddi}, {Krishnan},
  {Massi} \& {Bacciotti}}{{Moscadelli} et~al.}{2019}]{2019A&A...631A..74M}
{Moscadelli} L.,  {Sanna} A.,  {Goddi} C.,  {Krishnan} V.,  {Massi} F.,
  {Bacciotti} F.,  2019, \aap, 631, A74

\bibitem[\protect\citeauthoryear{{Nakamura}, {Motogi} \& {Fujisawa}}{{Nakamura}
  et~al.}{2021}]{2021eavw.workE...3M}
{Nakamura} M.,  {Motogi} K.,    {Fujisawa} K.,  2021, in 13th East Asian VLBI
  Workshop 2021 {A kinematic study of the disk-outflow system around a
  high-mass protostar G59.783+0.065 probed by CH3OH and H2O masers}.
p.~3

\bibitem[\protect\citeauthoryear{{Nedoluha} \& {Watson}}{{Nedoluha} \&
  {Watson}}{1991}]{1991ApJ...367L..63N}
{Nedoluha} G.~E.,  {Watson} W.~D.,  1991, \apjl, 367, L63

\bibitem[\protect\citeauthoryear{{Olech}, {Szymczak}, {Wolak}, {G{\'e}rard} \&
  {Bartkiewicz}}{{Olech} et~al.}{2020}]{2020A&A...634A..41O}
{Olech} M.,  {Szymczak} M.,  {Wolak} P.,  {G{\'e}rard} E.,    {Bartkiewicz} A.,
   2020, \aap, 634, A41

\bibitem[\protect\citeauthoryear{{Parfenov} \& {Sobolev}}{{Parfenov} \&
  {Sobolev}}{2014}]{2014MNRAS.444..620P}
{Parfenov} S.~Y.,  {Sobolev} A.~M.,  2014, \mnras, 444, 620

\bibitem[\protect\citeauthoryear{{Prozesky} \& {Smits}}{{Prozesky} \&
  {Smits}}{2020}]{2020MNRAS.491.2536P}
{Prozesky} A.,  {Smits} D.~P.,  2020, \mnras, 491, 2536

\bibitem[\protect\citeauthoryear{{Rajabi} \& {Houde}}{{Rajabi} \&
  {Houde}}{2017}]{2017SciA....3E1858R}
{Rajabi} F.,  {Houde} M.,  2017, Science Advances, 3, e1601858

\bibitem[\protect\citeauthoryear{{Rajabi}, {Houde}, {Bartkiewicz}, {Olech},
  {Szymczak} \& {Wolak}}{{Rajabi} et~al.}{2019}]{2019MNRAS.484.1590R}
{Rajabi} F.,  {Houde} M.,  {Bartkiewicz} A.,  {Olech} M.,  {Szymczak} M.,
  {Wolak} P.,  2019, \mnras, 484, 1590

\bibitem[\protect\citeauthoryear{{Richards}, {Cohen}, {Crocker}, {Lekht},
  {Mendoza} \& {Samodurov}}{{Richards} et~al.}{2005}]{2005Ap&SS.295...19R}
{Richards} A.~M.~S.,  {Cohen} R.~J.,  {Crocker} M.,  {Lekht} E.~E.,  {Mendoza}
  E.,    {Samodurov} V.~A.,  2005, \apss, 295, 19

\bibitem[\protect\citeauthoryear{{Richards}, {Elitzur} \& {Yates}}{{Richards}
  et~al.}{2011}]{2011A&A...525A..56R}
{Richards} A.~M.~S.,  {Elitzur} M.,    {Yates} J.~A.,  2011, \aap, 525, A56

\bibitem[\protect\citeauthoryear{{Richards}, {Etoka}, {Gray}, {Lekht},
  {Mendoza-Torres}, {Murakawa}, {Rudnitskij} \& {Yates}}{{Richards}
  et~al.}{2012}]{2012A&A...546A..16R}
{Richards} A.~M.~S.,  {Etoka} S.,  {Gray} M.~D.,  {Lekht} E.~E.,
  {Mendoza-Torres} J.~E.,  {Murakawa} K.,  {Rudnitskij} G.,    {Yates} J.~A.,
  2012, \aap, 546, A16

\bibitem[\protect\citeauthoryear{{Royer}, {Decin} \& {Wesson}}{{Royer}
  et~al.}{2010}]{2010A&A...518L.145R}
{Royer} P.,  {Decin} L.,    {Wesson} R. e.~a.,  2010, \aap, 518, L145

\bibitem[\protect\citeauthoryear{{Salii}, {Zinchenko}, {Liu}, {Sobolev},
  {Aberfelds} \& {Su}}{{Salii} et~al.}{2022}]{2022MNRAS.512.3215S}
{Salii} S.~V.,  {Zinchenko} I.~I.,  {Liu} S.-Y.,  {Sobolev} A.~M.,  {Aberfelds}
  A.,    {Su} Y.-N.,  2022, \mnras, 512, 3215

\bibitem[\protect\citeauthoryear{{Sanna}, {Moscadelli}, {Cesaroni}, {Tarchi},
  {Furuya} \& {Goddi}}{{Sanna} et~al.}{2010}]{2010A&A...517A..78S}
{Sanna} A.,  {Moscadelli} L.,  {Cesaroni} R.,  {Tarchi} A.,  {Furuya} R.~S.,
  {Goddi} C.,  2010, \aap, 517, A78

\bibitem[\protect\citeauthoryear{{Sch{\"o}ier}, {van der Tak}, {van Dishoeck}
  \& {Black}}{{Sch{\"o}ier} et~al.}{2005}]{2005A&A...432..369S}
{Sch{\"o}ier} F.~L.,  {van der Tak} F.~F.~S.,  {van Dishoeck} E.~F.,    {Black}
  J.~H.,  2005, \aap, 432, 369

\bibitem[\protect\citeauthoryear{{Shakhvorostova}, {Vol'vach}, {Vol'vach},
  {Dmitrotsa}, {Bayandina}, {Val'tts}, {Alakoz}, {Ashimbaeva} \&
  {Rudnitskii}}{{Shakhvorostova} et~al.}{2018}]{2018ARep...62..584S}
{Shakhvorostova} N.~N.,  {Vol'vach} L.~N.,  {Vol'vach} A.~E.,  {Dmitrotsa}
  A.~I.,  {Bayandina} O.~S.,  {Val'tts} I.~E.,  {Alakoz} A.~V.,  {Ashimbaeva}
  N.~T.,    {Rudnitskii} G.~M.,  2018, Astronomy Reports, 62, 584

\bibitem[\protect\citeauthoryear{{Shimoikura}, {Kobayashi}, {Omodaka},
  {Diamond}, {Matveyenko} \& {Fujisawa}}{{Shimoikura}
  et~al.}{2005}]{2005ApJ...634..459S}
{Shimoikura} T.,  {Kobayashi} H.,  {Omodaka} T.,  {Diamond} P.~J.,
  {Matveyenko} L.~I.,    {Fujisawa} K.,  2005, \apj, 634, 459

\bibitem[\protect\citeauthoryear{{Slysh}, {Alakoz} \& {Migenes}}{{Slysh}
  et~al.}{2010}]{2010MNRAS.404.1121S}
{Slysh} V.~I.,  {Alakoz} A.~V.,    {Migenes} V.,  2010, \mnras, 404, 1121

\bibitem[\protect\citeauthoryear{{Sobolev}, {Moran}, {Gray}, {Alakoz}, {Imai},
  {Baan}, {Tolmachev}, {Samodurov} \& {Ladeyshchikov}}{{Sobolev}
  et~al.}{2018}]{2018ApJ...856...60S}
{Sobolev} A.~M.,  {Moran} J.~M.,  {Gray} M.~D.,  {Alakoz} A.,  {Imai} H.,
  {Baan} W.~A.,  {Tolmachev} A.~M.,  {Samodurov} V.~A.,    {Ladeyshchikov}
  D.~A.,  2018, \apj, 856, 60

\bibitem[\protect\citeauthoryear{{Stecklum}, {Wolf}, {Linz}, {Caratti o
  Garatti}, {Schmidl}, {Klose}, {Eisl{\"o}ffel}, {Fischer}, {Brogan}, {Burns},
  {} \& {\em et al.}}{{Stecklum} et~al.}{2021}]{2021A&A...646A.161S}
{Stecklum} B.,  {Wolf} V.,  {Linz} H.,  {Caratti o Garatti} A.,  {Schmidl} S.,
  {Klose} S.,  {Eisl{\"o}ffel} J.,  {Fischer} C.,  {Brogan} C.,  {Burns} R.~A.,
   {}   {\em et al.} 2021, \aap, 646, A161

\bibitem[\protect\citeauthoryear{{Strelnitski}}{{Strelnitski}}{2007}]{2007ASPC..365..196S}
{Strelnitski} V.,  2007, in {Haverkorn} M.,  {Goss} W.~M.,  eds, SINS - Small
  Ionized and Neutral Structures in the Diffuse Interstellar Medium Vol.~365 of
  Astronomical Society of the Pacific Conference Series, {AU Structures in
  Turbulent Outflows from YSOs Revealed by H$_{2}$O Masers}.
p.~196

\bibitem[\protect\citeauthoryear{{Strelnitski}, {Holder}, {Shishov} \&
  {Nezhdanova}}{{Strelnitski} et~al.}{2017}]{2017A&AT...30..173S}
{Strelnitski} V.~S.,  {Holder} B.~P.,  {Shishov} V.~I.,    {Nezhdanova} N.~I.,
  2017, Astronomical and Astrophysical Transactions, 30, 173

\bibitem[\protect\citeauthoryear{{Strelnitskii}}{{Strelnitskii}}{1982}]{1982SvAL....8...86S}
{Strelnitskii} V.~S.,  1982, Soviet Astronomy Letters, 8, 86

\bibitem[\protect\citeauthoryear{{Surcis}, {Vlemmings}, {Curiel}, {Hutawarakorn
  Kramer}, {Torrelles} \& {Sarma}}{{Surcis} et~al.}{2011}]{2011A&A...527A..48S}
{Surcis} G.,  {Vlemmings} W.~H.~T.,  {Curiel} S.,  {Hutawarakorn Kramer} B.,
  {Torrelles} J.~M.,    {Sarma} A.~P.,  2011, \aap, 527, A48

\bibitem[\protect\citeauthoryear{{Surcis}, {Vlemmings}, {van Langevelde},
  {Goddi}, {Torrelles}, {Cant{\'o}}, {Curiel}, {Kim} \& {Kim}}{{Surcis}
  et~al.}{2014}]{2014A&A...565L...8S}
{Surcis} G.,  {Vlemmings} W.~H.~T.,  {van Langevelde} H.~J.,  {Goddi} C.,
  {Torrelles} J.~M.,  {Cant{\'o}} J.,  {Curiel} S.,  {Kim} S.~W.,    {Kim}
  J.~S.,  2014, \aap, 565, L8

\bibitem[\protect\citeauthoryear{{Tennyson}, {Hulme}, {Naim} \&
  {Yurchenko}}{{Tennyson} et~al.}{2016}]{2016JPhB...49d4002T}
{Tennyson} J.,  {Hulme} K.,  {Naim} O.~K.,    {Yurchenko} S.~N.,  2016, Journal
  of Physics B Atomic Molecular Physics, 49, 044002

\bibitem[\protect\citeauthoryear{{Uscanga}, {Cant{\'o}}, {Curiel}, {Anglada},
  {Torrelles}, {Patel}, {G{\'o}mez} \& {Raga}}{{Uscanga}
  et~al.}{2005}]{2005ApJ...634..468U}
{Uscanga} L.,  {Cant{\'o}} J.,  {Curiel} S.,  {Anglada} G.,  {Torrelles} J.~M.,
   {Patel} N.~A.,  {G{\'o}mez} J.~F.,    {Raga} A.~C.,  2005, \apj, 634, 468

\bibitem[\protect\citeauthoryear{{van der Walt}}{{van der
  Walt}}{2011}]{2011AJ....141..152V}
{van der Walt} D.~J.,  2011, \aj, 141, 152

\bibitem[\protect\citeauthoryear{{van Dishoeck}, {Kristensen}, {Mottram},
  {Benz}, {Bergin}, {Caselli}, {Herpin}, {Hogerheijde}, {Johnstone} \&
  {Liseau}}{{van Dishoeck} et~al.}{2021}]{2021A&A...648A..24V}
{van Dishoeck} E.~F.,  {Kristensen} L.~E.,  {Mottram} J.~C.,  {Benz} A.~O.,
  {Bergin} E.~A.,  {Caselli} P.,  {Herpin} F.,  {Hogerheijde} M.~R.,
  {Johnstone} D.,    {Liseau} e.~a.,  2021, \aap, 648, A24

\bibitem[\protect\citeauthoryear{{Volvach}, {Volvach} \& {Larionov}}{{Volvach}
  et~al.}{2023}]{2023ApJ...955...10V}
{Volvach} A.~E.,  {Volvach} L.~N.,    {Larionov} M.~G.,  2023, \apj, 955, 10

\bibitem[\protect\citeauthoryear{{Vol'vach}, {Vol'vach}, {Larionov}, {MacLeod},
  {van den Heever}, {Wolak}, {Olech}, {Ipatov}, {Ivanov} \&
  {Mikhailov}}{{Vol'vach} et~al.}{2019}]{2019ARep...63...49V}
{Vol'vach} L.~N.,  {Vol'vach} A.~E.,  {Larionov} M.~G.,  {MacLeod} G.~C.,  {van
  den Heever} S.~P.,  {Wolak} P.,  {Olech} M.,  {Ipatov} A.~V.,  {Ivanov}
  D.~V.,    {Mikhailov} A.~G.,  2019, Astronomy Reports, 63, 49

\bibitem[\protect\citeauthoryear{{Volvach}, {Volvach}, {Larionov}, {MacLeod},
  {Wolak}, {Olech}, {Kramer}, {Menten}, {Kraus}, {Brand}, {Zanicelli}, {Poppi}
  \& {Rigini}}{{Volvach} et~al.}{2019}]{2019AstL...45..321V}
{Volvach} L.~N.,  {Volvach} A.~E.,  {Larionov} M.~G.,  {MacLeod} G.~C.,
  {Wolak} P.,  {Olech} M.,  {Kramer} B.,  {Menten} K.,  {Kraus} A.,  {Brand}
  J.,  {Zanicelli} A.,  {Poppi} S.,    {Rigini} S.,  2019, Astronomy Letters,
  45, 321

\bibitem[\protect\citeauthoryear{{Volvach}, {Volvach}, {Larionov}, {Wolak},
  {Kramer}, {Menten}, {Kraus}, {Brand}, {Zanichelli}, {Poppi}, {Rigini},
  {Ipatov}, {Ivanov}, {Mikhailov} \& {Mel'nikov}}{{Volvach}
  et~al.}{2019}]{2019ARep...63..652V}
{Volvach} L.~N.,  {Volvach} A.~E.,  {Larionov} M.~G.,  {Wolak} P.,  {Kramer}
  B.,  {Menten} K.,  {Kraus} A.,  {Brand} J.,  {Zanichelli} A.,  {Poppi} S.,
  {Rigini} S.,  {Ipatov} A.~V.,  {Ivanov} D.~V.,  {Mikhailov} A.~G.,
  {Mel'nikov} A.,  2019, Astronomy Reports, 63, 652

\bibitem[\protect\citeauthoryear{{Voronkov}, {Sobolev}, {Ellingsen} \&
  {Ostrovskii}}{{Voronkov} et~al.}{2005}]{2005MNRAS.362..995V}
{Voronkov} M.~A.,  {Sobolev} A.~M.,  {Ellingsen} S.~P.,    {Ostrovskii} A.~B.,
  2005, \mnras, 362, 995

\bibitem[\protect\citeauthoryear{{Xu}, {Li}, {Reid}, {Menten}, {Zheng},
  {Brunthaler}, {Moscadelli}, {Dame} \& {Zhang}}{{Xu}
  et~al.}{2013}]{2013ApJ...769...15X}
{Xu} Y.,  {Li} J.~J.,  {Reid} M.~J.,  {Menten} K.~M.,  {Zheng} X.~W.,
  {Brunthaler} A.,  {Moscadelli} L.,  {Dame} T.~M.,    {Zhang} B.,  2013, \apj,
  769, 15

\bibitem[\protect\citeauthoryear{{Yates}, {Field} \& {Gray}}{{Yates}
  et~al.}{1997}]{1997MNRAS.285..303Y}
{Yates} J.~A.,  {Field} D.,    {Gray} M.~D.,  1997, \mnras, 285, 303

\bibitem[\protect\citeauthoryear{{Yoon}, {Song}, {Han}, {Hwang}, {Chang}, {Lee}
  \& {Itikawa}}{{Yoon} et~al.}{2008}]{2008JPCRD..37..913Y}
{Yoon} J.-S.,  {Song} M.-Y.,  {Han} J.-M.,  {Hwang} S.~H.,  {Chang} W.-S.,
  {Lee} B.,    {Itikawa} Y.,  2008, Journal of Physical and Chemical Reference
  Data, 37, 913

\end{thebibliography}

\appendix

\section[]{Variable Velocity Development}
\label{a:vvd}

With reference to equation (\ref{eq:formsol0}), we define here the symbols used. The
velocity width parameter is $W(\tau) = \sqrt{2 k_{\rmn B} T_{\rmn K}(\tau) / m_X}$, and
its domain average by node is $\bar{W}$, where $k_{\rmn B}$ is Boltzmann's constant, $T_{\rmn K}$ is the kinetic
temperature and $m_X$ is the molecular mass of species $X$. Scaled distance along a ray, 
propagating in direction $\hat{\vec{n}}$, is
represented by the optical depth $\tau$. The parameter $\eta (\tau)$ is $\bar{W}/W(\tau)$, and
dimensionless frequency and velocity are respectively denoted by $\bar{\nu}$ and $\vec{u}$. The
background dimensionless specific intensities, $i_{\rmn{BG}}(\Omega)$, are also multiples of the saturation
intensity, and here we allow for the possibility that the background level may vary with
direction ($\Omega$) on the sky surrounding the domain.

As in Papers~1-3, a nodal solution for the inversion is found through a saturation term that
involves a frequency-and-angle averaged intensity that is itself eliminated in favour of an
expression in terms of the inversions in nodes that border the ray paths converging on the
node in question. In the earlier papers, it proved possible to perform the frequency averaging
analytically, but this cannot be applied with variable velocity: although the analytical
frequency integral can still be carried out, the resulting expression cannot be factored
to restore multiple integrals (from the power-series expansion of an exponential) to the
product of identical integrals, each in a single spatial variable. It is therefore not
possible to reduce the expression to a single power series with one spatial integral as the
argument. We therefore adopt the rather different strategy described here.

A formal frequency average of equation (\ref{eq:formsol0}) over a gaussian molecular response
generates the averaged intensity,
\begin{align}
i (\tau) & = \frac{i_{\rmn{BG}}(\Omega)}{\mathrm{\upi}^{1/2}}
\int_{-\infty}^{\infty} d\bar{\nu} e^{-(\bar{\nu} - \hat{\vec{n}} \cdot \vec{u}(\tau))^2} \nonumber \\
         & \times \exp \left\{
    \int_{\tau_0}^\tau d\tau' \delta'(\tau')
    e^{-(\bar{\nu} - \hat{\vec{n}} \cdot \vec{u}(\tau'))^2 }
                                          \right\},
\label{eq:formsol1}
\end{align}
where we have now assumed that $\eta$, the width parameter, is 1 everywhere, equivalent
to assuming that the domain is isothermal. The frequency integral in equation (\ref{eq:formsol1}) is
now modified by defining the new dimensionless frequency 
$\varpi = \bar{\nu} - \hat{\vec{n}}_q \cdot \vec{u}(\vec{r})$. This change of variable is
applied to make the gaussian in equation (\ref{eq:formsol1}) zero-centered, as convenient for
numerical quadrature. With the further definition $\Delta \vec{u}(\tau') = \vec{u}(\vec{r}) - \vec{u}(\tau')$,
and specifying the intensity of a particular ray, $q$, equation (\ref{eq:formsol1}) becomes
\begin{align}
i_q (\tau) & = \frac{i_{\rmn{BG},q}}{\mathrm{\upi}^{1/2}}
\!\! \int_{-\infty}^{\infty} \!\! \exp \left\{ 
       \int_{\tau_0}^\tau \delta'(\tau') e^{-(\varpi + \hat{\vec{n}}_q \cdot \Delta \vec{u}(\tau'))^2} d\tau'
                                \right\} \nonumber \\ 
           & \times e^{-\varpi^2} d\varpi.
\label{eq:formsol1a}
\end{align}
The exponential in braces in equation (\ref{eq:formsol1a}) contains a spatial integral that is
discretized according to a finite element scheme similar to that used in
\citetalias{2019MNRAS.486.4216G} and \citetalias{2020MNRAS.493.2472G}. We
first define
\begin{equation}
Y_q  = \sum_{j=1}^{J(q)} \int_{\tau_{\rmn{in},j}}^{\tau_{\rmn{out},j}} \delta'(\tau')
    e^{-(\varpi + \hat{\vec{n}}_q \cdot \vec{u}(\tau'))^2 }  d\tau',
\label{eq:ykthing}
\end{equation}
where the expression in equation (\ref{eq:ykthing}) shows the integral along ray $q$
separated into the $J(q)$ elemental contributions that appear between the entry
point to the domain and arrival at a target node. The integral in equation (\ref{eq:ykthing})
is now over the path through a single element, $j$, with entry point
$\tau_{\rmn{in},j}$, and the corresponding exit point $\tau_{\rmn{out},j}$. The expression for
$Y_q$ is equivalent to the argument of the main exponential in equation (\ref{eq:formsol1a}), so
this can now be written in the much simpler form,
\begin{equation}
i_q (\tau) = (i_{\rmn{BG},q} /\mathrm{\upi}^{1/2}) \int_{-\infty}^\infty 
    e^{-\varpi^2}  \exp \left\{ 
                         Y_q (\varpi)
                      \right\} d\varpi .
\label{eq:formsol2}
\end{equation}

If the $j$th element has a shape function for its $i$th node of the form,
\begin{equation}
   f_i(x,y,z) = a_i + b_i x + c_i y + d_i z ,
\end{equation}
where $x,y,z$ are dimensionless coordinates within the element, and the quantities with
the subscript $i$ are shape-function coefficients, then $f_i$ can be parameterized in terms
of a single variable $0 \leq t \leq 1$, for example,
\begin{equation}
f_i(t) = a_i + b_i (x_{\rmn{in}} + t \Delta x ) + c_i (y_{\rmn{in}} + t \Delta y) + d_i (z_{\rmn{in}} + t \Delta z),
\end{equation}
where $(x_{\rmn{in}},y_{\rmn{in}},z_{\rmn{in}})$ are the coordinates of the entry point $\tau_{\rmn{in},j}$, and
$(\Delta x, \Delta y, \Delta x)$ are the coordinate differences between the exit and
entry points. Summing terms that do not multiply $t$ as
$A_i = a_i + b_i x_{\rmn{in}} + c_i y_{\rmn{in}} + d_i z_{\rmn{in}}$, and those that do, as
$B_i = b_i \Delta x + c_i \Delta y + d_i \Delta z$, the $i$th shape function becomes
simply,
\begin{equation}
f_i (t) = A_i + B_i t .
\label{eq:sfcoi}
\end{equation}
Any spatially variable quantity can then be found, in the interior of element $j$, by summing 
equation (\ref{eq:sfcoi}), multiplied by nodal values of the quantity, over the nodes. For
example, for some ray distance $\tau_{\rmn{in},j} \leq t \leq \tau_{\rmn{out},j}$, the dimensionless bulk
velocity is approximated by
\begin{equation}
\vec{u}(t) = \sum_{m=1}^M \vec{u}_m (A_m + B_m t) ,
\label{eq:usigma}
\end{equation}
where $\vec{u}_m$ is the velocity at the $m$th mode, and each element has $M$ nodes.
A similar expansion to equation (\ref{eq:usigma}) is made for the unknown, $\delta'(t)$.

We now use equation (\ref{eq:usigma}) and its analogue for $\delta'(t)$ to write $Y_q$ in terms of
nodal values of quantities that vary across the domain. With the aid of the further
definitions,
\begin{equation}
\alpha_{qj}^{(\varpi)} = \varpi + \hat{\vec{n}}_q \cdot \vec{u}_i - \sum_{m=1}^M A_m \hat{\vec{n}}_q \cdot \vec{u}_m ,
\label{eq:alphaqjvarp}
\end{equation}
where $\vec{u}_i$ is the velocity at the target node, and
\begin{equation}
\beta_{qj} = \sum_{m=1}^M B_m \hat{\vec{n}}_q \cdot \vec{u}_m ,
\label{eq:betaqj}
\end{equation}
equation (\ref{eq:ykthing}) reduces to the form, with $\tau_j$ as the distance covered
in passing through element $j$,
\begin{equation}
Y_q = \sum_{j=1}^{J(q)} \tau_j \sum_{\mu = 1}^M \delta'_\mu \int_0^1 (A_{j\mu} + B_{j\mu} t) e^{-(\alpha_{qj}^{(\varpi)} - \beta_{qj} t)^2} dt .
\label{eq:book7}
\end{equation}
There are various ways of carrying out the integral in equation (\ref{eq:book7}): perhaps the most
satisfactory is to extract a factor of $\exp -[\alpha_{qj}^{(\varpi)}]^2$ from the integral, while
defining new constants $a = \beta_{qj}^2$ and $b= \alpha_{qj}^{(\varpi)} \beta_{qj}$, reducing equation (\ref{eq:book7})
to the form
\begin{equation}
Y_q = \! \sum_{j=1}^{J(q)} \tau_j e^{   -[   \alpha_{qj}^{(\varpi)}    ]^2     } 
           \!\!\sum_{\mu = 1}^M \delta'_\mu \! \! \int_0^1 \!\! (A_{j\mu} + B_{j\mu} t) 
                    e^{- ( a t^2 - 2 b t   )   }   dt, 
\label{eq:ykstandard}
\end{equation}
that is tractable in terms of standard integrals. Of particular interest is
the formula from \citet{absteg}, reproduced here as
\begin{equation}
\int e^{-(ax^2 + 2 b x + c)} dx = \frac{\mathrm{\upi}^{1/2}e^{-c }}{2 a^{1/2}} e^{b^2/a} \mathrm{erf} 
      \left\{
        \sqrt{a} x + \frac{b}{\sqrt{a}}
      \right\} + \kappa ,
\end{equation}
where $\kappa$ is a constant of integration. One integral is exactly of this form, and
the other is reducible to this same integral through integration by parts. Results are in terms
of exponentials and error functions, and the integrated form of equation (\ref{eq:ykstandard}) is
\begin{align}
Y_q & = \sum_{j=1}^{J(q)} \sum_{\mu=1}^{M} \frac{\tau_j \delta'_\mu B_{j\mu}}{2\beta_{qj}} \left\{
         \frac{e^{-(\alpha_{qj} + \varpi )^2 }  - e^{-( \gamma_{qj} + \varpi )^2}}{\beta_{qj}} \right. \nonumber \\
   & \left. +
    \frac{\mathrm{\upi}^{1/2}}{\beta_{qj}} (\varepsilon_{qj} + \varpi)
        \left[ \mathrm{erf}(\alpha_{qj} + \varpi ) - \mathrm{erf}(\gamma_{qj} + \varpi ) \right] 
                                                             \right\},
\label{eq:ykint}
\end{align}
where further new constants are $\alpha_{qj} = \alpha_{qj}^{(\varpi)} - \varpi$,
 $\gamma_{qj} = \alpha_{qj} - \beta_{qj}$ and $\varepsilon_{qj} = \alpha_{qj} + A_{j\mu} \beta_{qj} / B_{j\mu}$.
If the double sum over elements and local nodes in equation (\ref{eq:ykint}) is replaced by a single
sum over the global node numbers of those nodes that bound ray $q$, noting that such nodes may be members of more than
one element in $J(q)$, then we can pre-compute the coefficients, $\Phi_{j,k}^{q,i}$, where $j$ is now the
global node index. These coefficients include everything in equation (\ref{eq:ykint}) except the nodal
values of the inversion itself, $\delta'_\mu$. It is now possible to write
equation (\ref{eq:ykint}) in the simple form $Y_q = \sum_{j=1}^{J(q)} \delta'_j \Phi_{j,k}^{q,i}$.
We substitute this expression for $Y_q$ in equation (\ref{eq:formsol2}), noting that $Y_q$ is a
function of $\varpi$ through the cofficients $\Phi_{j,k}^{q,i}$. We now replace the frequency
integral with a Gauss-Hermite numerical quadrature with $K$ abcissae, $\varpi_k$, and corresponding
weights $\zeta_k$, with $k$ in the range 1 to $K$. We generally use the Gauss-Hermite quadrature
scheme, as this is designed to approximate integrals that are of a form where the integrand
is a product of a gaussian and another function, such that
\begin{equation}
\int_{-\infty}^\infty e^{-\varpi^2} f(\varpi) d\varpi \simeq \sum_{k=1}^K \zeta_k f(\varpi_k) .
\label{eq:ghquad}
\end{equation}
The resulting form of equation (\ref{eq:formsol2}), discretized in frequency, is
\begin{equation}
i_q (\tau) = (i_{\rmn{BG},q} /\mathrm{\upi}^{1/2}) \sum_{k=1}^K \zeta_k
                   \exp \left\{ 
                        \sum_{j=1}^{J(q)} \delta'_j \Phi_{j,k}^{q,i}
                      \right\} .
\label{eq:formsol3}
\end{equation}
A formal average of equation (\ref{eq:formsol3}) over the $q$ rays with their associated solid angles
results in the mean intensity, $\bar{j}(\vec{r})$, at the target node, which is given by
\begin{equation}
\bar{j}(\vec{r})\! = \! \frac{1}{4\mathrm{\upi}^{3/2}} \! \sum_{q=1}^Q
                          \frac{i_{\rmn{BG},q} {\cal A}_q}{l_q^2}
                            \! \times \! \sum_{k=1}^K \zeta_k 
                       \exp \! \left[ \tau_{\rmn{M}} \! \sum_{j=1}^{J(q)}
                           \! \delta'_j \Phi_{j,k}^{q,i}
                          \right] ,
\label{eq:formsol4}
\end{equation}
where a global depth multiplier, $\tau_{\rmn M}$, has been extracted from every saturation coefficient. 

\section[]{Absolute Saturation Parameters}
\label{a:satparms}

We begin with the definition of the saturation intensity from equation (\ref{eq:isat}) of the
main text. As in Appendix~A of \citetalias{2020MNRAS.493.2472G}, we use a
form of the definition where we assume that the A-value of the maser and
collisional rates across the maser transition itself contribute negligibly compared to the
overall loss rate, represented by the parameter $\Gamma$.

The main difficulty in equation (\ref{eq:isat}) is evaluating $\Gamma$ to a reasonable accuracy,
since the other parameters, comprising the line-centre frequency (21.5549\,GHz), the
Einstein A-value of the maser transition ($A_{ij} = 1.835 \times 10^{-9}$\,s$^{-1}$) and
the statistical weight ratio $g_i/g_j = 13/11$, are all easily available from the 
ortho-H$_2$O {\sc radex} datafile provided by the {\it LAMBDA} database
\citep{2005A&A...432..369S}. The necessary numbers to calculate $\Gamma$ can be extracted from the same
database, noting that $\Gamma$ is essentially the non-maser loss rate from the upper level
of the maser transition, $J_{K_a,K,c} = 6_{2,6}$, in the vibrational ground state.

There is one radiative transition from the upper maser level (to $5_{1,5}$) with a significant
A-value of 0.758\,Hz. However, it is clear that with number densities typical of the shocked
gas in the models used in the present work, significant contributions to $\Gamma$ will also come
from collisional rate coefficients. We ignore upward radiative transitions from $6_{2,6}$ on
the grounds that there is negligible external pumping radiation. The file from the {\it LAMBDA}
database contains two tables of downward second-order collisional rate coefficients: the
first is for collisions with H$_2$, the second, with electrons. These tables are derived
from work by \citet{2008A&A...492..257F}. Although the rate coefficients in the electron table
are typically much stronger than those from H$_{2}$, by approximately 3 orders of magnitude,
we do not consider electron collisions further, since their effect is marginal, even with
an electron abundance as high as 10$^{-4}$ with respect to H$_2$. The sum of the downward
rate coefficients from $6_{2,6}$ is $3.168\times 10^{-10}$\,cm$^3$\,s$^{-1}$, using the figures
for 800\,K. A similar total of $3.379\times 10^{-10}$\,cm$^3$\,s$^{-1}$ arises from summing
the first 20 upward rate coefficients. Transitions to higher levels add little due to a general
decay in the magnitude of the coefficients and increasingly adverse Boltzmann exponentials.
Summing these collisional terms and the single strong A-value yields
\begin{equation}
\Gamma = 0.758 + 0.06547 n_8 \;\; \mathrm{s^{-1}}
\label{eq:genz}
\end{equation}
where $n_8$ is the number density of H$_2$ in multiples of 10$^8$\,cm$^{-3}$.

Equation~\ref{eq:genz} has the slightly unfortunate effect of making the loss rate density,
and therefore model, dependent. As an example, we consider the model described in Section~\ref{ss:samples}.
In this case $\tau_{\rmn M}=3$ and the corresponding number density of o-H$_2$O in the unshocked gas from Fig.~\ref{f:tauvn}
is 440\,cm$^{-3}$. As the models in \citet{2016MNRAS.456..374G} use a standard ratio of $3 \times 10^{-5}$ for o-H$_2$O
to H$_2$, we obtain a value of $n_8 = 0.147$. However, as saturation is only important in the shocked gas we
multiply this figure by 9, the compression factor, noting that the post-shock inversion follows the
density in this model, but not in the others. Therefore, we arrive at $n_8 = 1.32$. With 
the aid of equation (\ref{eq:genz})
we then obtain $\Gamma = 0.844$\,s$^{-1}$ and a saturation intensity from equation (\ref{eq:isat}) of 
$3.11 \times 10^{-11}$\,W\,m$^{-2}$\,Hz$^{-1}$.

\subsection{Conversion of flux density to Jy}
\label{ss:convjy}

The starting point here is equation (A3) of \citetalias{2020MNRAS.493.2472G}, which may be
written $f_{\rmn{kpc}} = f_{\nu} (R_{\rmn{AU}} / 206265 d_{\rmn{kpc}})^2$, where $f_{\rmn{kpc}}$ is the dimensionless
flux density of a source of size $R_{\rmn{AU}}$ in astronomical units at a distance $d_{\rmn{kpc}}$
kiloparsec, and $f_{\nu}$ is the flux density at the standard distance of 1000 domain
units and scaled by the saturation intensity of the model. To obtain a dimensioned flux
density, we need only multiply the above equation by the saturation intensity
of a given model and divide by $10^{-26}$ to put the result in Jy. For the example given
above where $I_{\rmn s} = 3.11 \times 10^{-11}$\,W\,m$^{-2}$\,Hz$^{-1}$, the dimensioned flux
density becomes
\begin{equation}
F_{\rmn kpc} = 7.31 \times 10^4 f_{\nu} (R_{\rmn{AU}} / d_{\rmn{kpc}})^2 \;\; \mathrm{Jy} ,
\end{equation}
which is equation (\ref{eq:fkpc}) of the main text,
noting that the leading constant is significantly larger for the 22-GHz H$_2$O
transition in the present work than for the 6.7-GHz methanol transition considered
in \citetalias{2020MNRAS.493.2472G} because of the higher saturation intensity in the water
transition.

\clearpage

\end{document}